\documentstyle[12pt,prd,aps,epsf]{revtex}
\def\tstrut{\vrule height2.5ex depth0pt width0pt} 
\newcommand{\comp}{{\rm C}\hspace{-1ex}\rule{0.1mm}{1.5ex}\hspace{1ex}}
\newcommand{\reales}{{\rm R}\hspace{-1ex}\rule{0.1mm}{1.5ex}\hspace{1ex}}
\newcommand{\naturales}{{\rm N}\hspace{-1ex}\rule{0.1mm}{1.5ex}\hspace{1ex}}
\def\er#1#2{\relax\ifmmode{}^{+#1}_{-#2}\else$^{+#1}_{-#2}$\fi}
\begin{document}
\draft
\tighten
\def\footnoterule{\kern-3pt \hrule width\hsize \kern3pt}

\title{
Bethe-Salpeter Approach for Unitarized Chiral Perturbation Theory.}

\author{
J. Nieves and E. Ruiz Arriola
} 

\address{
{~} \\
Departamento de F\'{\i}sica Moderna \\
Universidad de Granada \\
E-18071 Granada, Spain
}

\date{\today}
\maketitle

\thispagestyle{empty}

\begin{abstract}
The Bethe-Salpeter equation restores exact elastic unitarity in the
$s-$ channel by summing up an infinite set of chiral loops. We use
this equation to show how a chiral expansion can be undertaken in the
two particle irreducible amplitude and the propagators accomplishing
exact elastic unitarity at any step. Renormalizability of the
amplitudes  can be achieved by allowing for an
infinite set of counter-terms as it is the case in ordinary Chiral
Perturbation Theory. Crossing constraints can be imposed on the
parameters to a given order. Within this framework, we calculate the
leading and next-to-leading contributions to the elastic $\pi \pi$
scattering amplitudes, for all isospin channels, and to the vector and
scalar pion form factors in several renormalization schemes.  A
satisfactory description of amplitudes and form factors is
obtained. In this latter case, Watson's theorem is automatically
satisfied. From such studies we obtain a quite accurate determination
of some of the ChPT $SU(2)-$low energy parameters (${\bar l}_1 - {\bar
l}_2 = -6.1\er{0.1}{0.3} $ and ${\bar l}_6= 19.14 \pm 0.19$). We also
compare the two loop piece of our amplitudes to recent two--loop
calculations.

\end{abstract}


\vspace*{1cm}
\centerline{\it PACS: 11.10.St;11.30.Rd; 11.80.Et; 13.75.Lb;
14.40.Cs; 14.40.Aq\\}
\centerline{\it Keywords: Bethe-Salpeter Equation, Chiral Perturbation
Theory, Unitarity, $\pi\pi$-Scattering, Resonances. }


\newpage
\setcounter{page}{1}

\section{Introduction}

The dynamical origin of resonances in $\pi\pi$ scattering has been a
recurrent subject in low energy particle physics \cite{Rev}.
Analyticity, unitarity, crossing and chiral symmetry have provided
main insights into this subject. It is well known that so far no
solution is available exactly fulfilling all these requirements. In
practical calculations some of the properties mentioned above have to
be given up. Standard Chiral Perturbation Theory (ChPT) furnishes
exact crossing and restores unitarity order by order in the chiral
expansion. In typical calculations the unitarity limit is reached at
about a center of mass (CM) energy $\sqrt{s} \sim 4 \sqrt{\pi} f \sim
670 $ MeV\footnote{Through this paper $m$ is the pion mass, for which
we take 139.57 MeV, and $f$ the pion decay constant, for which we take
93.2 MeV.} still a low scale compared to the known resonances. Simply
because of this reason standard ChPT is unable to describe the
physically observed resonances, namely the $\rho$ and the
$\sigma$. But even if the unitarity limit was much larger, and thus
resonances would appear at significantly smaller scales than it,
standard ChPT would be unable to generate them since its applicability
requires the existence of a gap between the pion states and the
hadronic states next in energy, which for $\pi\pi$ scattering are
precisely the resonances. Resonances clearly indicate the presence of
non perturbative physics, and  thus a pole on the second Riemann sheet
(signal of the resonance) cannot be obtained in perturbation theory to
finite order. Hence, the energy regime for which ChPT holds has to
satisfy $ s << m_R^2 $ regardless of the unitarity limit. It so
happens that both the resonances and the unitarity limit lead to
similar scales, but there is so far no compelling reason to ascribe
this coincidence to some underlying dynamical feature or
symmetry\footnote{ To see this, consider for instance the large $N_c$
limit where $f \sim \sqrt{N_c} $ but $m_R \sim N_c^0 $ and $ \Gamma_R
\sim 1/N_c $, so in this limit $f >> m_R$, which suggests a
possible scenario where resonances can appear well below the unitarity
limit.}.

The desire to describe resonances using standard ChPT as a guide
has led some authors to propose several approaches which favor some
of the properties, that the exact scattering amplitude should satisfy, 
respect to others. Thus, Pade
Re-summation (PR)~\cite{dht90}, Large $N_f-$Expansion
(LNE)~\cite{ei84}, Inverse Amplitude Method (IAM)~\cite{dp93} , Current
Algebra Unitarization (CAU)~\cite{bbo97}, Dispersion Relations
(DP)~\cite{pb97}, Roy Equations~\cite{W97}, Coupled Channel
Lippmann-Schwinger Approach
(CCLS)~\cite{O97} and hybrid approaches~\cite{O98} have been
suggested. Besides their
advantages and success to describe the data in the low-lying resonance
region, any of them has specific drawbacks. In all above approaches
except by LNE and CCLS it is not clear which is the ChPT series of
diagrams which has been summed up. This is not the case for the CCLS
approach, but there a three momentum cut-off is introduced, hence
breaking translational Lorentz invariance and therefore the scattering
amplitude can be only evaluated in the the CM 
frame. On the other hand, though the LNE and CAU approaches
preserve crossing symmetry, both of them violate unitarity. Likewise,
those approaches which preserve exact unitarity violate crossing
symmetry.

A clear advantage of maintaining elastic
unitarity lies in the unambiguous identification of
the phase shifts. This is particularly useful to describe resonances
since the modulus of the partial wave amplitude reaches at 
the resonant energy the maximum value allowed by
unitarity. On the other hand, a traditional objection to any
unitarization scheme is provided by the non-uniqueness of the
procedure; this ambiguity is related to our lack of knowledge of an
appropriate expansion parameter for resonant energy physics. This drawback
does not invalidate the systematics and
predictive power of unitarization methods, although it is true that
improvement has a different meaning for different schemes. In addition,
unitarization by itself is not sufficient to predict a resonance, some
methods work while others do not. Against unitarization it is also
argued that since crossing symmetry is violated, 
the connection to a Lagrangian framework is lost, 
and hence there is no predictive power for other processes in
terms of a few phenomenological constants. 

The Bethe-Salpeter equation (BSE) provides a natural framework beyond
perturbation theory to treat the relativistic two body problem from a
Quantum Field Theory (QFT) point of view~\cite{BS51}.  This approach
allows to treat both the study of the scattering and of the bound
state properties of the system. In practical applications, however,
approximations have to be introduced which, generally speaking,
violate some known properties of the underlying QFT. This failure only
reflects our in-capability of guessing the exact solution. The problem
with the truncations is that almost always the micro-causality
requirement is lost, and hence the local character of the theory. As a
consequence, properties directly related to locality such as CPT,
crossing and related ones are not exactly satisfied. This has led some
authors to use the less stringent framework of relativistic quantum
mechanics as a basis to formulate the few body problem~\cite{kp91}.
It would be, of course, of indubitable interest the formulation of a
chiral expansion within such a framework.

Despite of these problems, even at the lowest order approximation, or
ladder approximation, the BSE sums an infinite set of diagrams,
allowing for a manifest implementation of elastic unitarity. This is
certainly very relevant in the scattering region and more specifically
if one aims to describe resonances, as we have discussed above. At higher
orders, to implement unitarity one needs to include  inelastic
processes due to particle production ( $\pi\pi \to K\bar{K}$, 
$\pi\pi \to \pi\pi\pi\pi $,$\cdots$).

In general, the renormalization of the BSE is a difficult problem for
QFT's in the continuum. The complications arise because typical local
features, i.e.  properties depending on micro-causality, like crossing
are broken by the approximate nature of the solution. This means that
there is no a renormalized Lagrangian which {\it exactly} reproduces
the amplitude, but only up to the approximate level of the solution.
This is the price payed for manifest unitarity. Fortunately, from the
point of view of the Effective Field Theory (EFT) idea, this problem
can be tackled in a manageable and analytical way. Since this is a low
energy expansion in terms of the appropriate relevant degrees of
freedom, interactions are suppressed in powers of momentum and thus
the BSE equation can be solved and renormalized explicitly by
expanding the iterated two particle irreducible amplitude in a power
series of momentum. This requires the introduction of a finite number
of counter-terms for a given order in the expansion.  The higher the
order in this expansion, the larger the number of parameters, and thus
the predictive power of the expansion diminishes. Thus an acceptable
compromise between predictive power and degree of accuracy has to be
reached. Within the BSE such a program, although possible, has the
unpleasant feature of arbitrariness in the renormalization scheme,
although this is also the case in standard ChPT\footnote{There, higher
order terms than the computed ones are exactly set to zero. This
choice is as arbitrary as any other one where these higher order terms
are set different from zero. The advantages of taking them different
from zero as dictated by unitarity will become clear along the
paper.}.

In this paper we  study 
the scattering of pseudoscalar mesons
by means of the BSE in the context of ChPT. Most useful information can be 
extracted from chiral symmetry, which dictates the energy dependence
of the scattering amplitude as a power series expansion of $1 /f^2$.  
The unknown coefficients of the
expansion increase with the order.  By using the BSE we can
also predict the energy dependence of the scattering amplitude in terms of
unknown coefficients and in agreement with the unitarity requirement.
This can be done in a way to comply with the known behavior in ChPT.

One important outcome of our calculation is the justification of
several methods based on algebraic manipulations of the on-shell
scattering amplitude, and which make no reference to the set of diagrams
which are summed up. Moreover, some off-shell quantities
are obtained.  Although it is true that physical quantities only make
sense when going to the mass shell there is no doubt that off-shell
quantities do enter into few body calculations.

Among others, the main results of the present investigation are the
following ones:
\begin{itemize}
\item A quantitative accurate description of both pion form factors
and $\pi\pi$ elastic scattering amplitudes is achieved in an energy
region wider than the one in which ChPT works. We implement exact
elastic unitarity whereas crossing symmetry is perturbatively
restored. For the form factors, the approach presented here
automatically satisfies Watson's theorem and it goes beyond the
leading order Omn\`es representation which is traditionally used.
\item A meticulous error treatment is undertaken. Our investigations
on the pion electric form factor together with our statistical and
systematic error analysis provide a very accurate determination 
of some of the ChPT $SU(2)-$low energy parameters,
\[
{\bar l}_1 - {\bar l}_2 = -6.1\er{0.1}{0.3}, \,\,\,\, {\bar l}_6 = 19.14
\pm 0.19 
\]
\item The current approach reproduces ChPT to one loop. However and
mainly due to the precise determination of the difference ${\bar l}_1
- {\bar l}_2 $ quoted above, our one loop results have much smaller
errors than most of the previously published ones. As a consequence,
we generate some of the two-loop corrections more precisely than
recent two loop calculations do.
\item  The present framework allows for an improvement of
any computed order of ChPT. 
\end{itemize}

The paper is organized as follows. In Sect.~\ref{sec:bse} we discuss
the BSE for the case of scattering of pseudoscalar mesons together
with our particular conventions and definitions. In
Sect.~\ref{sec:off} we review and improve the lowest order solution
for the off-shell amplitudes, found in a previous work~\cite{ej99}. We
also show that, already at this level, a satisfactory description of
data can be achieved when a reasonable set of ChPT low energy
parameters is used. This is done in
Subsect.~\ref{sec:num-res-off}. Technical details are postponed to
Appendix~\ref{sec:appea}. The computation of the next-to-leading order
corrections, within this off-shell scheme, turns out to be rather
cumbersome. The difficulties can be circumvented by introducing what
we call the ``on-shell'' scheme, as we do in Sect.~\ref{sec:on}. After
having discussed some renormalization issues, the decoupling of the
off-shell amplitudes is achieved in Subsect.~\ref{sec:pwd} by means of
a generalized partial wave expansion. In this context, the
renormalization conditions can also be stated in a more transparent
way. The findings of Subsect.~\ref{sec:pwd} allow us to write an exact
$T-$matrix if an exact two particle irreducible amplitude was
known. In Subsect.~\ref{sec:uni}, we show that the found solution
satisfies off-shell unitarity and in the next two subsections we
present several systematic expansions of the two particle irreducible
amplitude. In Subsect.~\ref{sec:form}, we show how our amplitudes can
be fruitfully employed for the calculation of pseudoscalar meson form
factors, in harmony with Watson's theorem. Next-to-leading order
numerical results for both $\pi\pi$ scattering and form factors
(vector and scalar) are presented in
Subsect.~\ref{sec:num-res-on}. This study allows us to determine, very
precisely, the parameters ${\bar l}_1, {\bar l}_2$ (specially their
difference) and ${\bar l}_6$ from experimental data. Besides, we also
compare with recent two-loop calculations. In
Subsect.~\ref{sec:compari} we compare some results obtained at leading
and next-to-leading accuracy, within the BSE on--shell scheme. Such a
study tests the convergence of the approach presented in this work. In
Sect.~\ref{sec:c_non} we try to understand qualitatively the origin of
some renormalization constants which naturally appear within the BSE
approach and in particular a remarkable formula for the width of the
$\rho$ meson, very similar to the celebrated KSFR, is deduced. In
Sect.~\ref{sec:comp} we briefly analyze alternative unitarization
methods on the light of the BSE approach. Conclusions are presented in
Sect.~\ref{sec:concl}. Finally, in Appendix~\ref{sec:appeb} the
leading and next-to-leading elastic $\pi\pi$ scattering amplitudes in
ChPT are compiled and we also give analytical expressions for the $u-$
and $t-$ unitarity chiral corrections projected over both isospin and
angular momentum.

\section{The Bethe-Salpeter Equation.}
\label{sec:bse}

\begin{figure}[t]
\vspace{-10.5cm}
\hbox to\hsize{\hfill\epsfxsize=0.75\hsize
\epsffile[52 35 513 507]{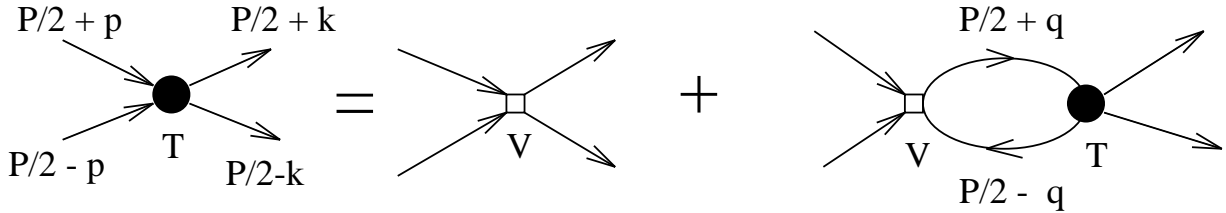}\hfill}
\vspace{1.5cm}
\caption[pepe]{\footnotesize Diagrammatic representation of the BSE
equation. It is also 
sketched the used kinematics.}
\vspace{0.5cm}
\label{fig:kin}
\end{figure}

Let us consider the scattering of two identical mesons of 
mass, $m$. The BSE can, with the kinematics described in
Fig.~\ref{fig:kin}, be written as
\begin{eqnarray}
T_P(p,k) &=& V_P(p,k) + {\rm i}\int\frac{d^4
q}{(2\pi)^4}T_P(q,k)\Delta(q_+) \Delta(q_-) V_P(p,q)\label{eq:bs}
\end{eqnarray}
where $q_{\pm} = (P/2\pm q)$ and $T_P(p,k)$ and $V_P(p,k)$ are the
total scattering amplitude\footnote{The normalization of the amplitude
$T$ is determined by its relation with the differential cross section
in the CM system of the two identical mesons and it is given by
$d\sigma /d\Omega = |T_P(p,k)|^2 / 64\pi^2 s$, where $s=P^2$. The
phase of the amplitude $T$ is such that the optical theorem reads
${\rm Im} T_P(p,p) = - \sigma_{\rm tot} (s^2-4s\,m^2)^{1/2}$, with
$\sigma_{\rm tot} $ the total cross section. The contribution to the
amputated Feynman diagram is $(-{\rm i} T_P(p,k) )$ in the
Fig.~\protect\ref{fig:kin}.} and the two particle irreducible
amplitude ({\it potential}) respectively. Besides, $\Delta$ is the
exact pseudoscalar meson propagator. Note that the previous equation
requires knowledge about the off-shell potential and the off-shell
amplitude\footnote{We will see later that this off-shellness can be
disregarded within the framework of EFT's and using an appropriate
regularization scheme.}. Clearly, for the exact  potential $V$
and propagator $\Delta$ the BSE provides an exact solution of the
scattering amplitude $T$~\cite{BS51}. Obviously an exact solution for
$T$ is not accessible, since $V$ and $\Delta$ are not exactly known.
An interesting property, direct consequence of the two particle
irreducible character of the potential, is that in the elastic
scattering region $ s > 4m^2 $, $V$ is a real function, and that it
also has a discontinuity for $s < 0 $, i.e. $ t > 4m^2 $. We will also
see below that, because of the inherent freedom of the renormalization
program of an EFT, the definition of the potential is ambiguous, and
some reference scale ought to be introduced in general. The exact
amplitude, of course, will be scale independent.

Probably, the most 
appealing feature of the BSE is that it accomplishes the two particle
unitarity requirement\footnote{Cutkosky's rules lead to the
substitution 
$\Delta(p) \to = (-2\pi{\rm i}) \delta^+ (p^2-m^2) $}
\begin{eqnarray}
T_P(p,k) - T_P(k,p)^* &=&
-{\rm i}(2\pi)^2 \int\frac{d^4 q}{(2\pi)^4}
T_P(q,k)\,   \delta^+ \left( q_+^2-m^2 \right)
                        \delta^+ \left( q_-^2-m^2 \right) T_P(q,p)^* 
\label{eq:off-uni}
\end{eqnarray}
as can be deduced from the real character of the potential above
threshold $ s > 4m^2 $, with 
$\delta^+(p^2-m^2) = \Theta(p^0)\delta(p^2-m^2)$. It is important to
stress that although the above unitarity condition  is considered
most frequently for the on-shell amplitude, it is also fulfilled even 
for off-shell amplitudes. We will see below
that these conditions are verified in practice by our amplitudes.

Isospin invariance leads to the following decomposition of the 
two identical isovector meson scattering amplitude for the 
process $ (P/2+p,a)+(P/2-p,b) \to (P/2+k,c)+(P/2-k,d) $,
\begin{eqnarray}
T_P (p,k)_{ab;cd} &=& A_P (p,k) \delta_{ab} \delta_{cd}
                  + B_P (p,k) \delta_{ac} \delta_{bd}
                  + C_P (p,k) \delta_{ad} \delta_{bc} \label{eq:defTabcd}
\end{eqnarray}
with $a,b,c$ and $d$ Cartesian isospin indices. With our conventions, 
the Mandelstam variables are defined as $s=P^2$,
$t=(p-k)^2$ and finally $u=(p+k)^2$ and the  isospin projection
operators, $P^I$, are given by~\cite{BD65}
\begin{eqnarray}
P^0_{ab;cd} &=& {2\over 3} \delta_{ab} \delta_{cd} \nonumber\\
P^1_{ab;cd} &=& \left( \delta_{ac} \delta_{bd}
                     -\delta_{ad} \delta_{bc} \right) \nonumber\\
P^2_{ab;cd} &=&  \left( \delta_{ac} \delta_{bd}
                     +\delta_{ad} \delta_{bc}
-{2\over 3} \delta_{ab} \delta_{cd} \right)\label{eq:iso-pro}
\end{eqnarray}
The isospin decomposition of the amplitude is then
$ T_P (p,k)_{ab;cd} = \sum_I P^I_{ab;cd} T_P^I (p,k) $, where
\begin{eqnarray}
T_P^0 (p,k) &=& {1\over 2} \left(3 A_P (p,k) + B_P (p,k) + C_P (p,k)\right)
\nonumber 
\\ T_P^1 (p,k) &=& {1\over 2} \left( B_P (p,k) - C_P (p,k) \right) \nonumber\\
T_P^2 (p,k) &=& {1\over 2} \left( B_P (p,k) 
+ C_P(p,k) \right) \label{eq:def_ti}
\end{eqnarray}
Based on the identical particle character of the scattered particles we
have the following symmetry properties for the isospin amplitudes
\begin{equation}
T_P^I (p,k) = (-)^I T_P^I (p,-k) = (-)^I T_P^I (-p,k) \label{eq:iden0}
\end{equation}
The above relations imply: 
\begin{eqnarray}
A_P (p,k) &=& A_P (p,-k)= A_P (-p,k) \nonumber\\
B_P (p,k)&=& C_P (p,-k)=C_P (-p,k) \label{eq:iden}
\end{eqnarray}
On the other hand crossing symmetry requires  
\begin{equation}
T_P (p,k)_{ab;cd} = T_{(p-k)}
\left(\frac{p+k+P}{2},\frac{p+k-P}{2}\right)_{ac;bd}
\end{equation}
which implies
\begin{eqnarray}
B_P(p,k) &=& A_{(p-k)}\left(\frac{p+k+P}{2},\frac{p+k-P}{2}\right)\nonumber\\
&&\nonumber\\
C_P(p,k) &=& C_{(p-k)}\left(\frac{p+k+P}{2},\frac{p+k-P}{2}
\right)\label{eq:cross}
\end{eqnarray}
Finally crossing, together with rotational invariance also implies:
\begin{equation}
T_P (p,k)_{ab;cd} = T_{-P} (-k,-p)_{cd;ab} = T_{-P} (-k,-p)_{ab;cd} =
T_{P} (k,p)_{ab;cd} \label{eq:cross0}
\end{equation}
which forces to all three functions $A_P,B_P$ and $C_P$ to be
symmetric under the exchange of $k$ and $p$. 
The relations of Eqs.~(\ref{eq:iden}) and~(\ref{eq:cross}) can be
combined to obtain the standard parameterization on the mass shell:
\begin{eqnarray}
A_P(p,k) &=& A(s,t,u)=A(s,u,t)\nonumber\\
B_P(p,k) &=& A(t,s,u)\nonumber\\ 
C_P(p,k) &=& A(u,t,s)\label{eq:cross1}
\end{eqnarray}
On the other hand the relations in Eq.~(\ref{eq:def_ti}) can be
inverted, and thus one gets
\begin{eqnarray}
A_P (p,k) &=& {2\over 3} \Big(  T_P^0 (p,k) - T_P^2 (p,k)\Big) \nonumber\\
B_P (p,k) &=& T_P^2 (p,k) + T_P^1 (p,k)  \nonumber\\
C_P (p,k) &=& T_P^2 (p,k) - T_P^1 (p,k) \label{eq:cross3}
\end{eqnarray}
Hence, the crossing conditions stated in Eq.~(\ref{eq:cross1}),
impose a set of non trivial relations between the amplitudes $T_P^0 (p,k)$, 
$T_P^1 (p,k)$ and $T_P^2 (p,k)$.  

For comparison with the experimental CM phase shifts, $\delta_{IJ}(s)$,
we define the on-shell amplitude for each isospin channel as
\begin{eqnarray}
T^I (s,t) = T_P^I (p,k)\,\,;  \qquad p^2=k^2=m^2-s/4\,\,; \qquad P \cdot p = P \cdot
k = 0
\end{eqnarray}
and then the projection over each partial wave $J$ in the CM frame, 
$T_{IJ}(s)$, is given by 
\begin{eqnarray}
T_{IJ}(s) = \frac12 \int_{-1}^{+1}
P_J\left(\cos\theta\right) T_P^I(p,k)~
d(\cos\theta) &=& \frac{{\rm i} 8\pi
s}{\lambda^{\frac12}(s,m^2,m^2)}
\left [ e^{2{\rm i}\delta_{IJ}(s)} -1 \right ]
\end{eqnarray}
where $\theta$ is the angle between $\vec{p}$ and $\vec{k}$ in the CM
frame, $P_J$ the Legendre polynomials and $\lambda(x,y,z) =
x^2+y^2+z^2 -2xy-2xz-2yz$. Notice that in our normalization the
unitarity limit implies $ |T_{IJ}(s)| < 16 \pi s / \lambda^{1/2}
(s,m^2,m^2) $.

\section{ Off-shell BSE Scheme }
\label{sec:off}

\subsection{\bf Chiral expansion of the potential and propagator}
\label{sec:off-amp}

We propose an expansion along  the lines of ChPT both for
the exact potential ($V$) and the exact propagator ($\Delta$),
\begin{eqnarray}
\Delta (p) &=& \Delta^{(0)}(p) + \Delta^{(2)}(p) + \dots \nonumber\\
V_P (p,k)  &=& ^{(0)} V_P (p,k) + ^{(2)} V_P (p,k) + \dots
\end{eqnarray}
Thus, at lowest order in this expansion, $V$ should be replaced by the
${\cal O}(p^2)$ chiral amplitude ($^{(2)}T$) and $\Delta$ by the free
meson propagator, $\Delta^{(0)}(r)=(r^2-m^2+{\rm i}\epsilon)^{-1}$.
Even at lowest order, by solving Eq.~(\ref{eq:bs}) we sum up an
infinite set of diagrams. This approach, at lowest order and in the
chiral limit reproduces the bubble re-summation undertaken in
Ref.~\cite{bc93}. This expansion is related to the approach recently
pursued for low energy $NN-$scattering where higher order $t-$ and
$u-$channel contributions to the potential are suppressed in the heavy
nucleon mass limit~\cite{k97}.

To illustrate the procedure, let us consider elastic $\pi \pi$
scattering in the $I=0,1,2$ channels.  One word of caution should be
said, before proceeding to practical calculations. As one sees the
solution of the BSE requires knowledge about off--shell quantities,
like the potential, $V_P (p,k)$ . This is generally an ambiguous
quantity which depends on the particular choice of the pion field. On--
shell quantities should however be independent of such a choice. The
calculations carried out below correspond to a particular ansatz. A
more detailed discussion on other possible choices is postponed to
Appendix~\ref{sec:parapi}. The main result is that a judicious choice of
the renormalization scheme, leads to on-shell quantities free from
this arbitrariness.

\subsubsection{I=0 $\pi \pi$ scattering. }

At lowest order, the off-shell potential $V$ in this channel
is given by
\begin{eqnarray}
V_P^0(p,k) \approx
^{(2)}T_P^0 (p,k) & = & \frac{5m^2-3s-2(p^2+k^2)}{2f^2}
\end{eqnarray}
To solve Eq.~(\ref{eq:bs}) with the above potential we proceed by
iteration. The second Born approximation suggests a solution of the
form
\begin{eqnarray}
T_P^0(p,k) &=& A(s) + B(s) (p^2+k^2) + C(s) p^2 k^2 \label{eq:i0}
\end{eqnarray}
where $A(s)$,$B(s)$ and $C(s)$ are functions to be determined.
Note that,
as a simple one loop calculation shows, there appears a new off-shell
dependence ($p^2k^2$) not present in the ${\cal O}(p^2)$ potential
$^{(2)}T$. That is similar to what happens in standard ChPT~\cite{GL84}.

The above ansatz reduces the BSE to a linear algebraic system of four
equations with three unknowns (Eq.~(\ref{eq:sys}) in Appendix~\ref{sec:appea}).
The system turns out to be compatible and the solution of it is given 
in Eqs.~(\ref{eq:i0_1}) and~(\ref{eq:i0_2})  of the
Appendix~\ref{sec:appea}.

At the lowest order in the chiral expansion examined here, the isoscalar
amplitude on the mass shell and in the CM frame ($\vec{P}=0, p^0 = k^0
= 0, P^0 = \sqrt{s}$) is purely $s-$wave, and its inverse  can be
obtained from Eqs.~(\ref{eq:i0_1}) and~(\ref{eq:i0_2}). It reads 
\begin{eqnarray}
T^{-1}_{00}(s) &=& -I_0(s) +
\frac{2\left(f^2+I_2(4m^2)\right)^2}{2I_4(4m^2)+(m^2-2s)f^2
+ (s-4m^2)I_2(4m^2) }\label{eq:i_0}
\end{eqnarray}
where $I_0(s)$ is a logarithmically divergent integral, which explicit
expression is given in Eq.~(\ref{eq:i2n}). Similarly $I_2(4m^2)$ and
$I_4(4m^2)$ are divergent quantities, which are defined in terms of
the quadratic and quartic divergent integrals $I_2(s)$ and
$I_4(s)$ also introduced in Eq.~(\ref{eq:i2n}). Thus the above
expression for the inverse amplitude requires renormalization, we will
back to this point in Subsect.~\ref{sec:reno}.

\subsubsection{I=1 $\pi \pi$ scattering}

At lowest order, the off-shell potential $V$ in this channel
is approximated by
\begin{eqnarray}
V_P^1(p,k) \approx
^{(2)}T_P^1(p,k) & = & \frac{2p \cdot k}{f^2}
\end{eqnarray}
As before, to solve Eq.~(\ref{eq:bs}) with the above potential we propose a
solution of the form
\begin{eqnarray}
T_P^1(p,k) &=& M(s) p\cdot k + N(s)  (p\cdot P)( k\cdot P)\label{eq:i1ans}
\end{eqnarray}
where $M$ and $N$ are functions to be determined. Note that, as
expected from our previous discussion for the isoscalar case,
there appears a new off-shell dependence
($(p\cdot P)( k\cdot P) $) not present in the ${\cal O}(p^2)$ potential.
Again, this ansatz reduces the BSE to a linear algebraic  system of
equations which provides the full off-shell scattering
amplitude, which is given in Eq.~(\ref{eq:i1_1}).

At the lowest order presented here, we have only $p-$wave contribution
and the resulting inverse CM amplitude on the mass shell, after angular
momentum projection, reads
\begin{eqnarray}
T^{-1}_{11}(s) =  &=&
-I_0(s) + \frac{ 2I_2(4m^2)-6f^2}{s-4m^2}\label{eq:i_1}
\end{eqnarray}
Similarly to the isoscalar case discussed
previously, the above equation presents divergences which need to be
consistently renormalized, this issue will be addressed in
Subsect.~\ref{sec:reno}.

\subsubsection{I=2 $\pi \pi$ scattering. }

At lowest order, the off-shell potential $V$ in this channel
is given by
\begin{eqnarray}
V_P^2(p,k) \approx
^{(2)}T_P^2 (p,k) & = & \frac{m^2-(p^2+k^2)}{f^2}
\end{eqnarray}
This resembles very much the potential for the isoscalar case, and we search
for a solution of the BSE of the form
\begin{eqnarray}
T_P^2(p,k) &=& A(s) + B(s) (p^2+k^2) + C(s) p^2 k^2 \label{eq:i2ans}
\end{eqnarray}
Similarly to the case $I=0$, the functions $A(s)$, $B(s)$ and $C(s)$
can be readily determined and are given in the Appendix~\ref{sec:appea}.

At the lowest order in the chiral expansion examined here, the $I=2$
amplitude on the mass shell and in the CM frame  
is purely $s-$wave, and from Eqs.~(\ref{eq:i2_1}) and~(\ref{eq:i2_2}))
we find its inverse  reads
\begin{eqnarray}
T^{-1}_{20}(s) &=& -I_0(s) +
\frac{2\left(f^2+I_2(4m^2)\right)^2}{2I_4(4m^2)+(s-2m^2)f^2
+ (s-4m^2)I_2(4m^2) }\label{eq:i_2}
\end{eqnarray}
Once again, the above equation has to be renormalized. 

\subsection{On-shell and off-shell unitarity}

As we have already anticipated, the solutions of the BSE must satisfy
on-shell and off-shell unitarity. This is an important check for our
amplitudes. This implies in turn conditions on the discontinuity
(${\rm Disc}\,[f(s)] \equiv f(s+i\epsilon)-f(s-i\epsilon),\,s> 4m^2) $
of the functions $A(s)$, $B(s)$ and $C(s)$ for the $ I=0 $ and $I=2$
cases and $N(s)$ and $M(s) $ for the isovector one. Going through the
unitarity conditions implicit in Eq.~(\ref{eq:off-uni}), we get a
set of constraints which are compiled in Eqs.~(\ref{eq:dis_1}) 
and~(\ref{eq:dis_2}) of the  Appendix~\ref{sec:appea}.

After a little of algebra, one can readily check that the
discontinuity conditions of Eqs.~(\ref{eq:dis_1})-(\ref{eq:dis_2}) are
satisfied by the off-shell amplitudes found in 
Subsect.~\ref{sec:off-amp}.  That guarantees that the solutions of the BSE
found in that subsection satisfy both off-shell and on-shell unitarity.

For the on-shell case, elastic unitarity can be checked in a much
simpler manner than that presented up to now. The on-shell amplitudes
can be expressed in the following suggestive form which, as we will
see, can be understood in terms of dispersion relations,
\begin{equation}
T^{-1}_{IJ}(s) = - {\bar I}_0 (s) - C_{IJ} + {1\over V_{IJ} (s)}\label{eq:tinv}
\end{equation}
where $C_{IJ}$ is a constant and the potentials are trivially read off
from the on shell amplitudes in Eqs~(\ref{eq:i_0}),(\ref{eq:i_1})
and~(\ref{eq:i_2}).  These potentials contain an infinite power series
of $1/f^2$. In the on-shell limit, the unitarity condition for the
partial waves is more easily expressed in terms of the inverse
amplitude, for which the optical theorem for $s>4m^2$ reads
\begin{equation}
{\rm Im }T^{-1}_{IJ}(s+{\rm i}\epsilon)
= \frac{\lambda^{\frac12}(s,m^2,m^2)}{16\pi s}
= {1\over 16\pi }  \sqrt{1-{4m^2 \over s}} = - 
{\rm Im }{\bar I}_0 (s+{\rm i}\epsilon)
\label{eq:uni} 
\end{equation}
thus,  the on-shell amplitudes found  in
Subsect.~\ref{sec:off-amp}  for the several isospin-angular momentum channels 
trivially meet this requirement.

\subsection{Crossing properties}

On the mass-shell and at the lowest order in the chiral expansion
presented in Subsect.~\ref{sec:off-amp}, the Eq.~(\ref{eq:cross3})
leads to
\begin{eqnarray}
A_P (p,k) &=& {2\over 3} \Big(  T_{00} (s) - T_{20} (s) \Big) \nonumber\\
B_P (p,k) &=& T_{20} (s) - 3 {u-t \over s - 4m^2 } T_{11}(s) \nonumber\\
C_P (p,k) &=& T_{20} (s) + 3 {u-t \over s - 4m^2 } T_{11}(s) \label{eq:cross4}
\end{eqnarray}
Obviously, the crossing conditions stated in Eq.~(\ref{eq:cross1}) are
not satisfied by the functions defined in Eq.~(\ref{eq:cross4}),
although they are fulfilled at lowest order in $1/f^2$. This is a
common problem in all unitarization schemes. In the BSE the origin
lies in the fact that the kernel of the equation, $\int d^4 q \cdots
\Delta(q_+)\Delta(q_-) \cdots$ breaks explicitly crossing, and given a
potential $V$ it only sums up all $s-$channel loop contributions
generated by it.  Thus this symmetry is only recovered when an exact
{\it potential} $V$, containing $t-$ and $u-$channel loop
contributions, is used.

Actually, in our treatment of the channels $ I=0,1,2$ we
require $3+2+3=8$ undetermined constants\footnote{As we will discuss in
the next subsection, a consistent renormalization program allows one
to take the divergent integrals
$I_0(4m^2), I_2(4m^2), I_4(4m^2)$  independent of each other in each
isospin channel.} , whereas crossing imposes
to order $ 1/f^4 $ only 4, namely ${\bar l}_{1,2,3,4}$. This is not as
severe as one might think since we are summing up an infinite series in
$1/f^2$. Nevertheless, we will show below how this information on
crossing can be implemented at the level of partial waves.

\subsection{Renormalization of the amplitudes}
\label{sec:reno}

To renormalize the on-shell amplitudes given in
Eqs.~(\ref{eq:i_0}),(\ref{eq:i_1}) and ~(\ref{eq:i_2}), or the
corresponding ones for the case  of off-shell scattering, we
note that in the spirit of an EFT  all
possible counter-terms should be considered. This can be achieved in
our case in a perturbative manner, making use of the formal
expansion of the bare amplitude
\begin{eqnarray}
T = V + VG_0V + VG_0VG_0V+ \cdots \, ,
\end{eqnarray}
where $G_0$ is the two particle propagator. Thus, a counter-term
series should be added to the bare amplitude such that the sum of both
becomes finite. At each order in the perturbative expansion, the
divergent part of the counter-term series is completely
determined. However, the finite piece remains arbitrary. Our
renormalization scheme is such that the renormalized amplitude can be
cast, again, as in Eqs.~(\ref{eq:i_0}),(\ref{eq:i_1}) and
~(\ref{eq:i_2}).  This amounts in practice, to interpret the
previously divergent quantities $I_{2n}(4m^2)$ as renormalized free
parameters. After having renormalized, we add a superscript $R$ to
differentiate between the previously divergent, $I_{2n}(4m^2)$, and
now finite quantities, $I^R_{2n}(4m^2)$. These parameters and
therefore the renormalized amplitude can be expressed in terms of
physical (measurable) magnitudes. In principle, these quantities
should be understood in terms of the underlying QCD dynamics, but in
practice it seems more convenient so far to fit $I^R_{2n}(4m^2)$ to
the available data.  The threshold properties of the amplitude
(scattering length, effective range, etc..) can then be determined
from them.  Besides the pion properties $m$ and $f$, at this order in
the expansion we have 8 parameters.  The appearance of 8 new
parameters is not surprising because the highest divergence we find is
quartic ($I_4(s)$) for the channels $I=0$ and $I=2$ and quadratic
($I_2(s)$) for the isovector channel and therefore to make the
amplitudes convergent we need to perform 3+3+2 subtractions
respectively. This situation is similar to what happens in standard
ChPT where one needs to include low-energy parameters ($\bar l$'s). In
fact, if $t-$ and $u-$ channel unitarity corrections are neglected, a
comparison of our (now) finite amplitude,
Eqs.~(\ref{eq:i_0}),(\ref{eq:i_1}) and ~(\ref{eq:i_2}), to the $ {\cal
O} (p^4) \,\, \pi\pi$ amplitude in terms of some of these ${\bar l}$'s
becomes possible. Such a comparison will be discussed in the next
subsection.

This renormalization scheme leads to a renormalized amplitude which
does not derive from a renormalized Lagrangian. This is again closely
linked to the violation of exact crossing symmetry, and it is detailed 
discussed in the Sect.~\ref{sec:appea1} of the Appendix~\ref{sec:appea}.

\subsection{Crossing Symmetry Restoration and  Comparison 
with one loop ChPT} \label{sec:csr}

Our amplitudes contain undetermined parameters which, as stated
previously outnumber those allowed by crossing symmetry at ${\cal
O}(p^4)$. Nevertheless, by imposing suitable constraints on our
parameters we can fulfill crossing symmetry approximately. We do this
at the level of partial wave amplitudes. For completeness, we
reproduce here a discussion from Ref.~\cite{ej99}, where this issue
was first addressed. At the lowest order in the chiral expansion
proposed in this section, we approximate, in the scattering region
$s>4m^2$, the ${\cal O}(1/f^4)$ $t-$ and $u-$ channel unitarity
corrections (function $h_{IJ}$ in Eq.~(\ref{eq:gl}) of the
Appendix~\ref{sec:appeb}) 
by a Taylor expansion around threshold to order $(s-4m^2)^2$. At next
order in our expansion (when the full ${\cal O}(p^4)$-corrections are
included both in the {\it potential} and in the pion propagator) we
will recover the full $t-$ and $u-$ channel unitarity logs at ${\cal
O}(1/f^4)$, and at the next order (${\cal O}(1/f^6)$), we will be
approximating these logs by a Taylor expansion to order
$(s-4m^2)^3$. Thus, the analytical structure of the amplitude derived
from the left hand cut is only recovered perturbatively. This is in
common to other approaches (PR, DP, IAM $\cdots$) fulfilling exact
unitarity in the $s$-channel, as discussed
in~\cite{dp93},~\cite{pb97},\cite{GM91}. Thus, this approach violates
crossing symmetry. At order ${\cal O}(1/f^4)$ our isoscalar $s-$,
isovector $p-$ and isotensor $s-$wave amplitudes are polynomials of
degree two in the variable $(s-4m^2)$, with a total of eight (3+2+3)
arbitrary coefficients ($I^{R,I=0}_{0,2,4}(4m^2)$,
$I^{R,I=1}_{0,2}(4m^2)$, $I^{R,I=2}_{0,2,4}(4m^2)$ ), and there are no
logarithmic corrections to account for $t-$ and $u-$channel unitarity
corrections. Far from the left hand cut, these latter corrections can
be expanded in a Taylor series to order $(s-4m^2)^2$, but in that case
the one loop $SU(2)$ ChPT amplitudes can be cast as second order
polynomials in the variable $(s-4m^2)$, with a total of four (${\bar
l}_{1,2,3,4}$) arbitrary coefficients~\cite{GL84}.  To restore, in
this approximation, crossing symmetry in our amplitudes requires the
existence of four constraints between our eight undetermined
parameters. These relations can be found in Eq.~(\ref{eq:cons}) in the
Appendix~\ref{sec:appea}.

Once these constraints are implemented in our model, there exists a
linear relation between our remaining four undetermined parameters
($I^{R,I=0}_{0,4}(4m^2)$, $I^{R,I=1}_{0,2}(4m^2)$) and the most
commonly used ${\bar l}_{1,2,3,4}$ parameters. Thus, all eight parameters 
($I^{R,I=0,2}_{0,2,4}(4m^2)$, $I^{R,I=1}_{0,2}(4m^2)$) can be
expressed in terms of ${\bar l}_{1,2,3,4}$ (see Eq.~(\ref{eq:li}) of
Appendix~\ref{sec:appea}). 
\subsection{Numerical results for $I=0,1,2$}
\label{sec:num-res-off}
After the above discussion, it is clear that at this order we have
four independent parameters $I^{R,I=0}_0$, $I^{R,I=0}_4$,
$I^{R,I=1}_0$ and $I^{R,I=1}_2$ which can be determined either from a
combined $\chi^2-$fit to the isoscalar and isotensor $s-$ and
isovector $p-$wave elastic $\pi\pi$ phase shifts or through,
Eq.~(\ref{eq:li}), from the Gasser-Leutwyler or other estimates of the
${\bar l}_{1,2,3,4}$ low energy parameters. In a previous
work,~\cite{ej99}, we have already discussed the first procedure
($\chi^2-$fit) and thus we will follow here the second one. Therefore,
we will try to address the following question: Does the lowest order
of the off-shell BSE approach together with reasonable values for the
${\bar l}_{1,2,3,4}$ parameters describe the observed $\pi\pi$
phase-shifts in the intermediate energy region?  To answer the
question, we will consider two sets of parameters:
\begin{eqnarray}
{\rm set\,\, {\bf A}:\,\,\,} && {\bar l}_1 = -0.62 \pm
0.94,\,\,{\bar l}_2 = 6.28 \pm 0.48,\,\, {\bar l}_3 = 2.9 \pm 2.4,\,\,
{\bar l}_4 = 4.4 \pm 0.3 \nonumber\\
{\rm set\,\, {\bf B}:\,\,\,} && {\bar l}_1 = -1.7\phantom{0} \pm
1.0\phantom{0},\,\,{\bar l}_2 = 6.1\phantom{0} \pm 0.5\phantom{0},\,\,
{\bar l}_3 = 2.9 \pm 2.4,\,\, {\bar l}_4 = 4.4 \pm 0.3
\label{eq:elesab} 
\end{eqnarray}
In both sets ${\bar l}_3$ and ${\bar l}_4$ have been determined from
the $SU(3)$ mass formulae and the scalar radius as suggested
in~\cite{GL84} and in~\cite{bct98}, respectively. On the other hand
the values of ${\bar l}_{1,2}$  come from the analysis of
Ref.~\cite{rdgh91} of the data on $K_{l4}-$decays (set {\bf A})
and from the combined study of  
$K_{l4}-$decays and $\pi\pi$ with some unitarization
procedure (set {\bf B}) performed in Ref.~\cite{bcg94}.

\begin{figure}


\begin{center}                                                                
\leavevmode
\epsfysize = 600pt
\epsfbox{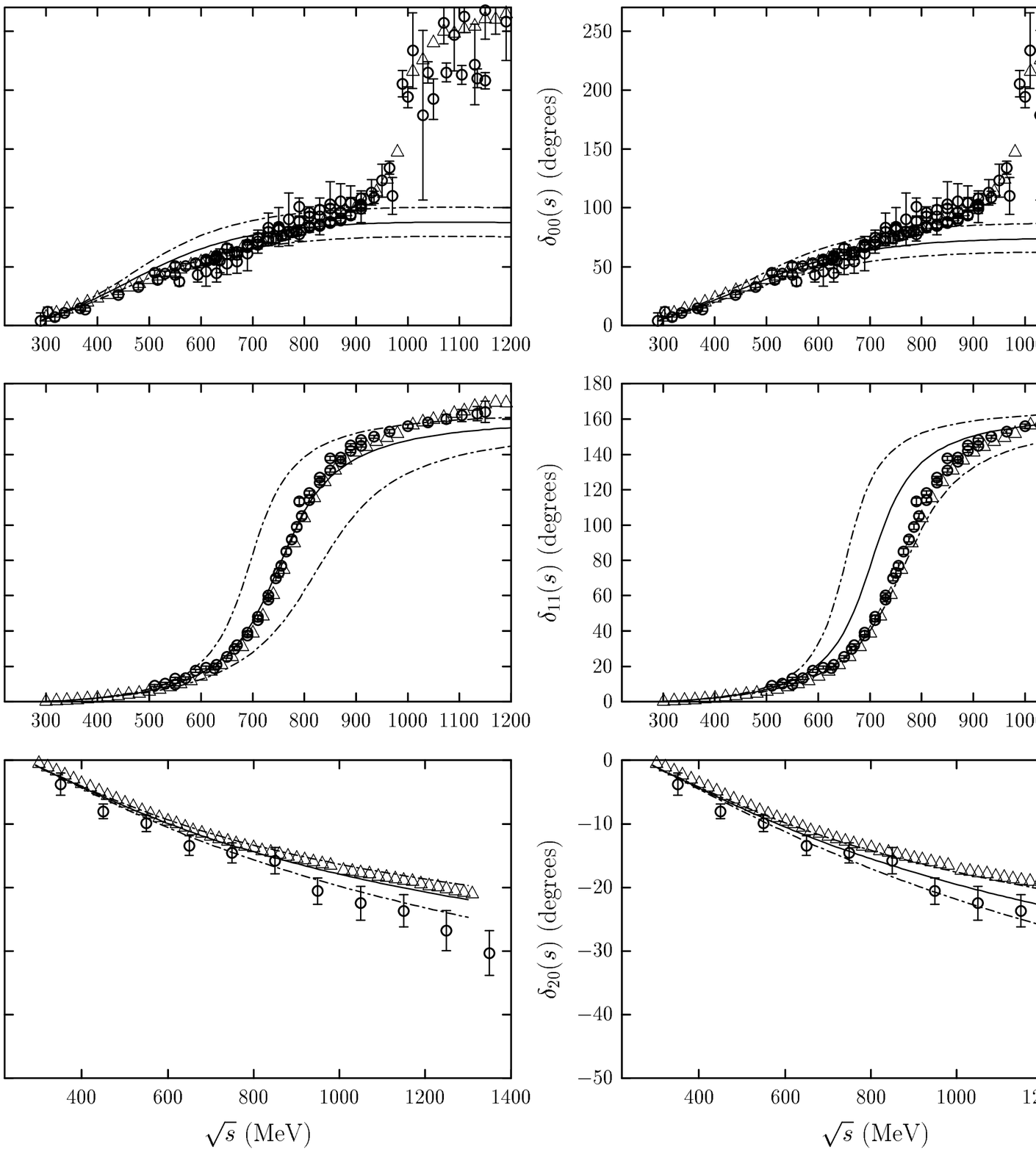}\phantom{tonto}
\end{center}
\vspace{-2.5cm}
\caption[pepe]{\footnotesize Several $\pi\pi$ phase shifts as a function
of the total CM energy $\protect\sqrt s$ for both sets of ${\bar
l}$'s quoted in Eq.~(\protect\ref{eq:elesab}). Left (right) panels
have been obtained with the set {\bf A} ({\bf B}) of parameters. Solid
lines are the predictions of the off-shell BSE approach, at lowest
order, for the different $IJ-$channels. Dashed lines are the 68\%
confidence limits.  Top panels ($I=0$, $J=0$): circles stand for the
experimental analysis of Refs.~\protect\cite{pa73} -
\cite{klr97}. Middle panels ($I=1$, $J=1$): circles stand for the
experimental analysis of Refs.~\protect\cite{pa73} and
~\protect\cite{em74}. Bottom panels ($I=2$, $J=0$): circles stand for the
experimental analysis of Ref.~\protect\cite{ho77}.  In all plots the
triangles are the Frogatt and Petersen phase-shifts
(Ref.~\protect\cite{fp77}) with no errors due to the lack of error
estimates in the original analysis.}
\label{fig:off-res}
\end{figure}

Results are presented in Fig.~\ref{fig:off-res} and in
Table~\ref{tab:off-res}.  In the figure we show the prediction (solid
lines) of the off-shell BSE approach, at lowest order, for the $s-$
and $p-$wave $\pi \pi$ scattering phase-shifts for all isospin
channels and for both sets {\bf A} (left panels) and {\bf B} (right
panels) of the ${\bar l}$'s parameters. We assume Gauss distributed
errors for the ${\bar l}$'s parameters and propagate those to the
scattering phase shifts, effective range parameters, etc$\dots$ by
means of a Monte Carlo simulation. Central values for the phase-shifts
have been computed using the central values of the ${\bar l}$'s
parameters. Dashed lines in the plots are the 68\% confidence limits.

As we see for both sets of constants, the simple approach presented here
describes the isovector and isotensor channels up to energies above 1
GeV, whereas the isoscalar channel is well reproduced up to 0.8--0.9
GeV. In the latter case, and for these high energies, one should also
include the mixing with the $K\bar K$ channel as pointed out recently
in Refs.~\cite{O97}--\cite{O98}. Regarding the deduced threshold
parameters we find agreement with the measured values when both
theoretical and experimental uncertainties are taken into
account. Furthermore, both
sets of parameters predict the existence of the $\rho$
resonance\footnote{We determine the position of the resonance, by 
assuming zero background and thus demanding the phase-shift, 
$\delta_{11}$, to be $\pi/2$. Furthermore, we obtain the width of the 
resonance from 
\begin{equation}
\frac{1}{\Gamma_\rho} = \frac{m_\rho}{(m_\rho^2-s)\tan \delta_{11}(s)}
= m_\rho \frac{d\delta_{11}(s)}{d s}\Big|_{s=m_\rho} = \frac{16\pi
m_\rho^3 }{\lambda^\frac12(m^2_\rho,m^2,m^2)}\frac{d {\rm
Re}\,T_{11}^{-1}(s)}{d s}\Big|_{s=m_\rho}\label{eq:width}
\end{equation}} in good agreement with the experimental data,
\begin{eqnarray}
{\rm set\,\, {\bf A}:\,\,\,} && m_\rho = 770\er{90}{60} \,\,{\rm
[MeV]}, \,\,\,  \Gamma_\rho = 180\er{80}{50} \,\,{\rm
[MeV]}\nonumber\\
{\rm set\,\, {\bf B}:\,\,\,} && m_\rho = 715\er{70}{50} \,\,{\rm
[MeV]}, \,\,\,  \Gamma_\rho = 130\er{60}{30} \,\,{\rm
[MeV]}\nonumber\\
\label{eq:mrho} 
\end{eqnarray}
In this way the ``existence'' of the $\rho$ resonance can be regarded
as a prediction of the BSE with ChPT and the parameters obtained from
some low energy data. Favoring one of the considered sets of
parameters, implies an assumption on the size of the ${\cal O}(p^6)$
contributions not included at this level of approximation. Finally, 
it is worth  mentioning that the predicted parameters $I^R_n(4m^2)$, 
agree reasonably well with those fitted to $\pi\pi$ scattering 
data in Ref.~\cite{ej99}.
\begin{table}[t]
\begin{center}
\begin{tabular}{c|cc|cc|cc}
& \multicolumn{2}{c|}{$J=I=0$} &  \multicolumn{2}{c|}{$J=I=1$}&
\multicolumn{2}{c}{$J=0, I=2$} \\
& set {\bf A}  & set {\bf B}    & set {\bf A}  & set {\bf B} 
& set {\bf A}  & set {\bf B}  \\\hline\tstrut
$-10^2 I^R_0(4m^2)$ & $3.1 (4) $ & $2.6 (5) $ 
                    & $9.8 (14)$ & $10.9 (14)$
                    & $6.6 (6) $ & $6.1 (6)$  \\\tstrut
$-10^3 I^R_2(4m^2)$ & $1.7 (11)$ & $3.0 (12)$
                    & $-77 (7)  $ & $-83 (8)  $
                    & $7.5 (12) $ & $8.4 (12) $  \\\tstrut
$-10^3 I^R_4(4m^2)$ & $ 3.0 (12)$ & $ 2.7 (12)$ 
& \multicolumn{2}{c|}{} 
                    & $2.1 (5) $  & $2.1 (5) $  \\\hline\tstrut 
$10^3 m^{2J+1} a_{IJ}$ & $225 (7) $ & $218 (7) $ 
                       & $39.5 (11)$ & $39.5 (12)$ 
                       & $-41.0 (12) $ & $-42.1 (13) $ \\\tstrut
$10^3 m^{2J+3} b_{IJ}$ & $308 (20) $ & $286 (20) $ 
                       & $7.3 (14) $ & $8.6 (16) $ 
                       & $-72.5 (20) $ & $-74.4 (23) $ \\
\end{tabular}
\end{center}
\caption[pepe]{\footnotesize Off-shell BSE approach parameters
($I^R_n(4m^2)$) obtained from both sets of ${\bar l}$'s parameters
given in Eq.~(\protect\ref{eq:elesab}).  $I^R_n(4m^2)$ are given in
units of $(2m)^n$.  We also give the threshold parameters $a_{IJ}$ and
$b_{IJ}$ obtained from an expansion of the scattering
amplitude~\protect\cite{dgh92}, Re$T_{IJ} = -16\pi m (s/4 -m^2)^J [
a_{IJ} + b_{IJ} (s/4 -m^2) + \cdots ]$ close to threshold.  Errors
have been propagated by means of a Monte Carlo simulation, they are given
in brackets and affect to the last digit of the quoted quantities.}
\label{tab:off-res}
\end{table}

\section{ On-shell BSE Scheme}
\label{sec:on}

As already recognized in previous work~\cite{ej99}, it is very hard in
practice to pursue the calculation of higher orders, within the
off-shell scheme presented in the previous section. The difficulty
lies on the non-locality of the four point Green-functions involved
in the BSE in momentum space, and the subsequent and unavoidable 
renormalization. More specifically, to keep the full off-shell
dependence of $V$ and $T$ in the BSE is the origin of most of the
difficulties.

However, if we look at the amplitude in each isospin channel
separately (Eqs.~(\ref{eq:i_0}),(\ref{eq:i_1}) and ~(\ref{eq:i_2})) it
seems that the off-shellness can be ignored by simply renormalizing
the parameters of the lowest order amplitude ($f$ and $m$). The
authors of Refs.~\cite{O97} take advantage of this observation.
Though, it is important to stress here, that this renormalization is
different for each $IJ-$channel and therefore this procedure does not
provide a satisfactory renormalization scheme.

In this section, we come up with a consistent and computationally
feasible renormalization program where all off-shellness effects can
be incorporated into a legitimate redefinition  of the two particle
irreducible amplitude, $V$, at all orders of the chiral expansion.
  
This new scheme allows us not only to describe the data but also to
make predictions for some of the ChPT--low energy constants.

\subsection{Renormalization of power divergences}
\label{sec:power}

 Let us consider the iteration in the BSE of a renormalized potential,
instead of the ``bare'' potential . Obviously, to define such a
quantity requires a renormalization prescription\footnote{In the
following we will assume a mass independent regularization scheme,
such as e.g. dimensional regularization. The reason for doing this is
that it preserves chiral symmetry.}. This new viewpoint can only be
taken, as we will see, because the divergence
structure of the non-linear sigma model is such that, all
non-logarithmic divergences can be absorbed by a suitable
renormalization of the parameters of the Lagrangian. For completeness,
we repeat here the argument given in Ref.~\cite{AB81} in a way that it
can be easily applied to our case of interest. It is easier to
consider first the chiral limit $m=0$. The Lagrangian at lowest order
is given by
\begin{equation}
{\cal L} = {f^2 \over 4} {\rm tr} \Big( \partial^\mu U^\dagger
\partial_\mu U \Big)\label{eq:kin-la}
\end{equation}
with $U$ a dimensionless unitarity matrix, involving the Goldstone
fields, which transforms linearly under the chiral group.  Thus, 
the necessary counter-terms  to cancel $L$ loops are
suppressed by the power $f^{2(1-L)}$. Let $\Lambda$ be a chiral
invariant regulator with dimension of energy and $D$ the dimension of
a counter-term which appears at $L$ loops. Since the $U$ field is
dimensionless, $D$ simply counts the number of derivatives. The
operator appearing in the counter-term Lagrangian should have dimension
4, thus we have
\begin{equation}
D = 2+ 2L-r
\end{equation}
with $r$ the number of powers of the regulator $\Lambda$ which
accompany this counter-term in the Lagrangian 
(degree of divergence). If we denote the
counter-terms in the schematic way
\begin{eqnarray}
{\cal L}_{\rm ct} = \sum_{L=0}^\infty  f^{2(1-L)} \sum_i c_{i,r}^L
\sum_{r=0}^{2L}
\Lambda^r \langle \partial^{2+2L-r} \rangle_i
\end{eqnarray}
where $ \langle \partial^D \rangle_i $ denotes a set of chiral
invariant linear independent operators made out of the matrix $U$ and
comprising $D$ derivatives and $c_{i,r}^L$ are suitable dimensionless
coefficients. We see that at $L$ loops the only new structures are
those corresponding to $\Lambda^0 $ ($r=0$) which in actual
calculations corresponds to logarithmic divergences. Higher powers in
$\Lambda$ generate structures which were already present at $L-1$
loops. For instance, let us consider the one-loop correction, by
construction it is down respect to the leading order by two powers of
$f$, i.e., the corresponding counter-terms in the Lagrangian will be
made out of chiral invariant operators constructed by means of
derivatives of the dimensionless $U$ matrix, defined above, and any of
them will be multiplied by an overall factor $f^0$. In addition, the
total energy dimension of the counter-terms has to be four. Hence, only
terms with two and four derivatives\footnote{Note that an odd-number
of derivatives are forbidden by parity conservation and the only
chirally invariant operator with zero derivatives is ${\rm tr}(
UU^\dagger) = 2$, an irrelevant constant.} might appear. To get the
proper dimensions, these terms will have to be multiplied by
$\Lambda^2$ and $\Lambda^0\ln \Lambda$ respectively. Thus, the
quadratic divergence, first term, is a renormalization of the leading
kinetic energy piece (Eq.~(\ref{eq:kin-la})). Besides, to renormalize
the logarithmic divergence, one has to add new counter-terms, not
present in the original Lagrangian (Eq.~(\ref{eq:kin-la})). For higher
loops, the argument can be easily generalized, see for more details
Ref.~\cite{AB81}.

This result is very important because it means that all but the
logarithmic divergences can effectively be ignored. An efficient
scheme which accomplishes this property is dimensional regularization,
since
\begin{equation}
\int \frac{d^4 q}{(2\pi)^4}(q^2)^n = 0 \qquad n=-1,0,1,
\dots
\end{equation}
This argument can be extended away from the chiral limit, although
in this case new terms appear which vanish in the limit $m^2 \to 0$.
The conclusion is again the same. If every possible counter-term
compatible with chiral symmetry is written down, all but the logarithmic
divergent pieces can be ignored. 

The above discussion means that when renormalizing the BSE, we may set
to zero (at a given renormalization point) all power divergent
integrals which appeared in Sect.~\ref{sec:off}. This is because they
only amount to a renormalization of the undetermined parameters of the
higher order terms of the Lagrangian. The power of this result is a
direct consequence of the symmetry, the derivative coupling of the
pion interactions and the fact that in ChPT there is an infinite tower
of operators. In a sense this is only true for the exact theory with
an infinite set of counter-terms, it is to say when we iterate by means
of the BSE a ``renormalized'' two particle irreducible 
amplitude,$V$, with an infinite number of terms.

To keep all power divergences and not only the logarithmic ones and
simultaneously iterate the most general two particle irreducible 
amplitude leads to redundant
combinations of undetermined parameters. For instance, to include in
the potential the tree ${\cal O}(p^4)$ Lagrangian (${\bar l}$'s) and
keep the power (non logarithmic) divergences produced by the loops
made out of the ${\cal O}(p^2)$ pieces of the Lagrangian produces
redundant contributions at ${\cal O}(p^4)$ (Eq.~(\ref{eq:li})
illustrates clearly the point). These redundancies also appear at
higher orders if higher order tree Lagrangian terms are also included
in the potential, $V$.

\subsection{ Off-shell versus on-shell}

Let us consider the BSE in the case of $\pi \pi$ scattering. The isospin
amplitudes satisfy the symmetry properties given in Eqs.~(\ref{eq:iden0})
and~(\ref{eq:cross0}).
Thus, they admit the following expansion
\begin{equation}
T_P^I (p,k) = \sum_{N_1} \sum_{N_2} \sum_{\mu_1 \mu_2 \dots}
T^I_{\mu_1 \dots \mu_{N_1} ; \nu_1 \dots \nu_{N_2}} [P]
k^{\mu_1} \cdots k^{\mu_{N_1}} p^{\nu_1} \cdots p^{\nu_{N_2}}
\end{equation}
where $ T^I_{\mu_1 \dots \mu_{N_1} ; \nu_1 \dots \nu_{N_2}} [P] =
T_{\nu_1 \dots \nu_{N_2} ; \mu_1 \dots \mu_{N_1}} [P] $ and $N_1$ and
$N_2$  run only 
over even (odd) natural numbers for the $I=0,2$ $(I=1)$ isospin channel.
 In short hand notation
we may write
\begin{equation}
T_P^I (p,k) = \sum_{(\mu)(\nu)} T^I_{(\mu)(\nu)} 
[P] k^{(\mu)} p^{(\nu)}
\end{equation}
Inserting this ansatz in the BSE, for simplicity let us consider first
only those contributions where the free meson propagators
($\Delta^{(0)}$) are used, we get
\begin{eqnarray}
T_{(\mu)(\nu)}^I [P] &=& V_{(\mu)(\nu)}^I [P]+ {\rm i}
 \sum_{(\alpha) (\beta)} T_{(\mu) (\alpha)}^I [P] 
\int \frac{d^4 q}{(2\pi)^4} q^{(\alpha)} q^{(\beta)} \Delta^{(0)} (q_+)
\Delta^{(0)}(q_-) V_{(\beta)(\nu)}^I [P]
\end{eqnarray}
Due to parity the number of indices in $(\alpha)$ and in $(\beta)$ ought
to be either both even or both odd. Thus, we are led to consider the integral
\begin{equation}
I_{2k}^{\mu_1 \dots \mu_{2n} } [P] :=
{\rm i} \int \frac{d^4 q}{(2\pi)^4} q^{2k}q^{\mu_1} \cdots q^{\mu_{2n}}
\Delta^{(0)} (q_+) \Delta^{(0)}(q_-)
\end{equation}
If we contract the former expression with $ P^{\mu_1} $ we get, using that
\begin{equation} 
2 P \cdot q = \Big( \frac{1}{\Delta^{(0)}(q_+)}- 
\frac{1}{\Delta^{(0)}(q_-)} \Big), 
\end{equation}
the identity
\begin{equation}
P_{\mu_1} I_{2k}^{\mu_1 \dots \mu_{2n} } [P] = -{\rm i}
\int \frac{d^4 q}{(2\pi)^4} q^{2k} q^{\mu_2} \cdots q^{\mu_{2n}} 
\Delta^{(0)} (q_+)
\end{equation}
Shifting the integration variable
\begin{eqnarray}
P_{\mu_1} I_{2k}^{\mu_1 \dots \mu_N } [P] &=& -{\rm i}
\int \frac{d^4 q}{(2\pi)^4} \Delta^{(0)} (q) \left (
(q^2-m^2)+(\frac{s}{4}+m^2)-q\cdot P \right)^k \nonumber\\ 
&\times & (q-P/2)^{\mu_2} \cdots (q-P/2)^{\mu_{2n}}
\end{eqnarray}
which corresponds to quadratic and higher divergences. As we have
said before, only the logarithmic divergences should be taken into
account since higher order divergences can be absorbed as
redefinition of the renormalized parameters of the higher order terms
of the Lagrangian. Thus, we would get that, up to these
non-logarithmic divergences, the integral is transverse, i.e. when
contracted with some $P^\mu$ becomes zero. The transverse part can be
calculated to give
\begin{eqnarray}
I_{2k}^{\mu_1 \dots \mu_{2n} } [P] &:=& C_{nk} (s)
\left [ \Big( g^{\mu_1\mu_2} - {P^{\mu_1} P^{\mu_2} \over s} \Big) \cdots
\Big( g^{\mu_{2n-1} \mu_{2n}} - { P^{\mu_{2n-1}} P^{\mu_{2n}} \over s}
\Big)\right.\nonumber\\ 
&&\nonumber\\ &+& \left.
\phantom{\Big(} {\rm Permutations} \right ]+ {\rm P. D.}
\end{eqnarray}
where $C_{nk}(s)$ is a function\footnote{It can be shown that 
\begin{equation}
C_{nk}(s) = {1\over (2n+1) !!} I_{2n+2k}(s).
\end{equation}
} 
of $s$ and P. D. means power divergences. 
Let us study now the integral $ I_{2n}(s) $ and use that
\begin{equation}
 q^2 = (m^2 - \frac{s}{4}) + {1\over 2} \Big( \Delta^{(0)}(q_+)^{-1}+
\Delta^{(0)}(q_-)^{-1} \Big)  
\end{equation}
then we get
\begin{equation}
I_{2n} (s) = (m^2-\frac{s}{4}) I_{2n-2} (s) +
{\rm i} \int \frac{d^4 q}{(2\pi)^4} \Delta^{(0)}(q) q_+^{2n-2} 
\end{equation}
Again, the second term corresponds to power divergences. Applying this
formula $n$ times we get
\begin{equation}
I_{2n} (s)= (m^2-\frac{s}{4})^n I_0 (s) +
{\rm P. D.}
\end{equation}
All this discussion means that under the integral sign we may set
$ q^2 =m^2-s/4$ and  $ P \cdot q =0 $ up to non-logarithmic
divergences, i.e.
\begin{equation}
{\rm i} \int \frac{d^4 q}{(2\pi)^4} F(q^2 ; P \cdot q) \Delta^{(0)}
(q_+)
\Delta^{(0)} (q_-) = F(m^2-\frac{s}{4}; 0) I_0 (s) + {\rm
P. D.} \label{eq:onsell-bare}
\end{equation}
Up to now we have considered the bare (free) meson propagator $\Delta^{(0)}$.
If we have the full renormalized propagator
\begin{equation}
\Delta (q) = \left ( q^2 -m^2 -
\Pi(q^2) \right )^{-1}
\end{equation}
where $\Pi(q^2)$ is the meson self-energy with 
the on-shell renormalization conditions
\begin{equation}
\Pi (m^2)=0 \, , \quad \Pi'(m^2) =0
\end{equation}
then we have the Laurent expansion around the mass pole, $q^2 = m^2 $, 
\begin{equation}
\Delta (q) = {1\over q^2 -m^2 }  + {1\over 2} \Pi''(m^2) + \cdots
\end{equation}
The non-pole terms of the expansion generate, when applied to the kernel
of the BSE, power divergences. Therefore, 
Eq.~(\ref{eq:onsell-bare}) remains valid
also when the full renormalized propagator $\Delta(q)$ is used.

In conclusion, the off-shellness 
of the BSE kernel leads to power divergences which have to be
renormalized by a suitable Lagrangian of counter-terms (for instance
the bare $l's-$Lagrangian pieces at order ${\cal O}(p^4)$), leaving the
resulting finite parts as undetermined parameters of the EFT . Thus, if we
iterate a renormalized potential we should ignore the power
divergences and therefore ignore the off-shell behavior within the
BSE and then solve
\begin{eqnarray}
T_P^I(p,k) &=& V_P^I(p,k) + {\rm i}\int\frac{d^4
q}{(2\pi)^4}T_P^I(\bar{q},k)\Delta(q_+) \Delta(q_-)
V_P^I(p,\bar{q})\label{eq:bs-on} 
\end{eqnarray}
with 
\begin{eqnarray}
\bar{q}^{\,\mu} \bar{q}^{\,\nu} = \frac{m^2-s/4}{q^2-(P\cdot
q)^2/s}\left ( q^\mu -P^\mu \frac{(P\cdot q)}{s} \right ) \left (
q^\nu -P^\nu \frac{(P\cdot q)}{s} \right );\qquad \bar{q}^2 = m^2-s/4
\label{eq:defqbar}
\end{eqnarray} 
Thus, given the exact two particle irreducible amplitude, potential
$V^I$, the solution of Eq.~(\ref{eq:bs}), in a given isospin channel,
and that of Eq.~(\ref{eq:bs-on}) are equivalent.  This important
result will allow us in the next subsection to find an exact solution
of the BSE given an exact two particle irreducible amplitude, $V$. It
also might justify the method used in Refs.~\cite{O97},~\cite{O98}
where the off-shell behavior of the Lippmann-Schwinger equation is
totally ignored.

\subsection{ Partial wave expansion }
\label{sec:pwd}

We now show how the BSE can be diagonalized within this on-shell
scheme for $s>0$. 
To this end we consider the following off-shell ``partial wave''
expansions for $T^I_P(p,k)$ and $V^I_P(p,k)$
\begin{equation}
T^I_P(p,k)= \sum_{J=0}^\infty (2J+1) 
T_{IJ} ( p^2 , P \cdot p \,; \,k^2 , P \cdot k )
 P_J ( {\rm cos} \theta_{k,p} ) \label{eq:wave-exp}
\end{equation}
and a similar one for $V^I_P(p,k)$.  The ``angle'' $\theta_{k,p}$ is given by
\begin{equation}
{\rm cos} \theta_{k,p} := { - p \cdot k + (P \cdot p) (P \cdot k) / s
\over \left [ (P \cdot p)^2/s- p^2 \right ] ^\frac12
       \left[ (P \cdot k)^2/s- k^2 \right ] ^\frac12 }
\end{equation}
and it reduces to the scattering angle in the CM
system for mesons on the mass shell. For $s>0$, 
the Legendre polynomials satisfy a
orthogonality relation of the form
\begin{eqnarray}
{\rm i} \int \frac{d^4 q}{(2\pi)^4} \Delta(q_+) \Delta(q_-)
P_J ( {\rm cos} \theta_{k,{\bar q}} )
P_{J'} ( {\rm cos} \theta_{{\bar q},p} )= \delta_{J J'} { P_J ( {\rm cos}
\theta_{k,p} ) \over 2J+1 }
 I_0 (s) \label{eq:ort}
\end{eqnarray}
which is the generalization of the usual one. Plugging the partial
wave expansions of Eq.~(\ref{eq:wave-exp}) 
for the amplitude and potential in Eq.~(\ref{eq:bs-on}) we 
get\footnote{ We preserve the ordering in multiplying the expressions 
in a way that the generalization to the coupled channel case becomes
evident.}   
\begin{eqnarray}
T_{IJ} ( p^2 , P \cdot p \,; \, k^2 , P \cdot k )&=&
V_{IJ} ( p^2 , P \cdot p \,; \, k^2 , P \cdot k )\nonumber\\&+&
T_{IJ} ( m^2-\frac{s}{4} , 0 \,; \, k^2 , P \cdot k ) 
I_0 (s) V_{IJ} ( p^2 , P \cdot p \,;\,  m^2-\frac{s}{4} , 0 )
\label{eq:full-off}
\end{eqnarray}
To solve this equation, we set the variable $p$ on-shell, and get for 
the half off-shell amplitude
\begin{eqnarray}
T_{IJ} ( m^2-\frac{s}{4} , 0 \,; \, k^2 , P \cdot k )^{-1}  
= -I_0(s) + \, V_{IJ}(m^2-\frac{s}{4} , 0 \,; \, k^2 , P \cdot k)^{-1}
\end{eqnarray}
Likewise, for the full on-shell amplitude we get
\begin{eqnarray}
T_{IJ} ( m^2-\frac{s}{4} , 0 \,; \, m^2-\frac{s}{4}, 0 )^{-1} := 
T_{IJ}(s)^{-1} = -I_0(s) + \,
V_{IJ} (s)^{-1}
\end{eqnarray}
where $V_{IJ}(s) = V_{IJ} ( m^2-s/4 , 0 \,; \, m^2-s/4, 0 )$. Choosing
the subtraction point $s_0$ we have a finite on-shell amplitude
\begin{eqnarray}
T_{IJ} (s)^{-1} - T_{IJ} (s_0)^{-1} = -( I_0 (s)-I_0 (s_0) )
+ V_{IJ} (s)^{-1} - V_{IJ} (s_0)^{-1} \label{eq:sol-bse-on}
\end{eqnarray}
A finite (renormalized) inverse half-off-shell can now be  obtained
from the  finite (renormalized) on-shell amplitude
\begin{eqnarray}
T_{IJ} ( m^2-\frac{s}{4} , 0 \,; \, k^2 , P \cdot k )^{-1} &=&
T_{IJ} (s)^{-1} + V_{IJ} ( m^2-\frac{s}{4} , 0 \,; \, k^2 , P \cdot k )^{-1} -
V_{IJ} (s)^{-1} \nonumber \\ 
&=& T_{IJ}
(s_0)^{-1}  -\left( I_0 (s)-I_0 (s_0) \right) \nonumber\\
&+& \left( V_{IJ} ( m^2-\frac{s}{4} , 0 \,;\, k^2 , P \cdot k )^{-1} -
V_{IJ} ( s_0 )^{-1} \right)
\end{eqnarray}
Note that once we have the half off shell amplitude we might, through
Eq.~(\ref{eq:full-off}), get the full off-shell one. The idea of
reconstructing the full off-shell amplitude from both the on- and half
off-shell amplitudes was first suggested in Ref.~\cite{Na68}, where a
dispersion relation inspired treatment of the BSE, with certain
approximations, was undertaken.

\subsection{ Off-shell unitarity}
\label{sec:uni}

The off-shell unitarity condition becomes particularly simple in terms
of partial wave amplitudes. Taking into account
Eq.~(\ref{eq:off-uni}), that in this equation and due to the on-shell
delta functions $q^\mu$ can be replaced by $\bar{q}^{\,\mu}$ in the
arguments of the $T-$matrices, and taking discontinuities in
Eq.~(\ref{eq:ort}) one can see that in terms of partial waves,
off-shell unitarity reads:
\begin{eqnarray}
T_{IJ} ( p^2 , P \cdot p \,; \, k^2 , P \cdot k )&-&
T_{IJ}^* ( k^2 , P \cdot k \,; \, p^2 , P \cdot p )\nonumber\\& =&
T_{IJ} (m^2-\frac{s}{4},0 \,;\, k^2 , P\cdot k)\,{\rm Disc}\, [I_0 (s)]
T_{IJ}^* ( m^2-\frac{s}{4},0\,;\, p^2 , P\cdot p)
\label{eq:off-uni-pw}
\end{eqnarray}
A solution of the form in Eq.~(\ref{eq:full-off}) automatically
satisfies\footnote{To prove this statement it is advantageous to note
that above the unitarity cut, where the potential $V$ is hermitian,
\begin{eqnarray}
T_{IJ}^* ( k^2 , P \cdot k  \,; \,p^2 , P \cdot p  )&=&
V_{IJ} ( p^2 , P \cdot p \,; \, k^2 , P \cdot k )\nonumber\\&+&
V_{IJ} ( m^2-\frac{s}{4} , 0 \,; \, k^2 , P \cdot k ) 
I_0^* (s) T_{IJ}^* (m^2-\frac{s}{4} , 0 \,;\,  p^2 , P \cdot p )
\end{eqnarray}
} the off-shell unitarity condition of Eq.~(\ref{eq:off-uni-pw}). This
property is maintained even if an approximation to the exact potential in
Eq.~(\ref{eq:full-off}) is made.

\subsection{ Approximations to the Potential}
\label{sec:app-pot}

Given the most general two particle irreducible renormalized
amplitude, $V_P(p,k)$, compatible with the chiral symmetry for
on-shell scattering, Eq.~(\ref{eq:sol-bse-on}) provides an exact
$T-$matrix for the scattering of two pions on the mass shell, for any
$IJ-$channel. Eq.~(\ref{eq:sol-bse-on}) can be rewritten in the
following form
\begin{eqnarray}
T_{IJ} (s)^{-1} +\bar{I}_0 (s) - V_{IJ} (s)^{-1}  &=& 
T_{IJ} (s_0)^{-1} +\bar{I}_0 (s_0) - V_{IJ} (s_0)^{-1}  = -C_{IJ}
\end{eqnarray}
where $C_{IJ}$ should be a constant, independent of $s$. Thus we 
have 
\begin{eqnarray}
T_{IJ} (s)^{-1}    &=& -\bar{I}_0 (s) -C_{IJ} + V_{IJ} (s)^{-1} \label{eq:defV}
\end{eqnarray}
or defining $W_{IJ} (s, C_{IJ})^{-1} =-C_{IJ} + V_{IJ} (s)^{-1}$ the
above equation can be written
\begin{eqnarray}
T_{IJ} (s)^{-1}    &=& -\bar{I}_0 (s) +  W_{IJ} (s)^{-1} \label{eq:defW}
\end{eqnarray}
The functions $\bar{I}_0 (s)$ and $W_{IJ} (s)$ or
$V_{IJ}(s)$  account for the  right and left hand cuts~\cite{Ma58}
respectively of the inverse amplitude $T_{IJ}(s)^{-1}$. The function $W_{IJ}
(s)$, or equivalently $V_{IJ} (s)$, 
is a meromorphic function in $\comp-\reales^{-}$, is real 
through the unitarity cut and contains the essential dynamics of the process. 
Thus, analyticity considerations (\cite{Ma58}) tell us that the scattering 
amplitude should be given by 
\begin{eqnarray}
T^{-1}_{IJ}(s)  = -{\bar I}_0(s) +
\frac{P^{IJ}(s)}{Q^{IJ}(s)}
\label{eq:prev-pade}
\end{eqnarray}
where $P^{IJ}(s)$ and ${Q^{IJ}(s)}$  are analytical functions of the
complex variable $s \in \comp-\reales^{-}$. A systematic approximation to
the above formula can be obtained by a {\it Pade}
approximant to the {\it inverse potential}, although not
exactly in the form proposed in Ref.~\cite{dht90},
\begin{eqnarray}
T^{-1}_{IJ}(s)  = -{\bar I}_0(s) +
\frac{P^{IJ}_{n}(s-4m^2)}{Q^{IJ}_{k}(s-4m^2)}
\label{eq:pade}
\end{eqnarray}
where $P^{IJ}_n(x)$ and $Q^{IJ}_k(x)$ are polynomials\footnote{Note
that the first non-vanishing coefficient in $Q_n$ corresponds to the
power $(s-4m^2)^J$.} of order $n$ and $k$ with real
coefficients. Crossing symmetry relates some of the coefficients of
the polynomials for different $IJ-$channels, but can not determine all
of them and most of these coefficients have to be fitted to the data
or, if possible, be understood in terms of the underlying QCD
dynamics. This type of {\it Pade} approach was already suggested in
Ref.~\cite{ej99}. Actually, the results of the off-shell scheme presented in
Sect.~\ref{sec:off} are just {\it
Pade} approximants of the type $[n,k]=[1,1]$ in Eq.~(\ref{eq:pade}).

A different approach would be to propose an effective range expansion 
for the function $W_{IJ} (s)$, namely
\begin{equation}
W_{IJ} (s) = \sum_{n=0}^\infty \alpha_{IJ}^{(n)} (s-4m^2)^{n+J}
\label{eq:reff}
\end{equation}
or  a chiral expansion of the type
\begin{equation}
W_{IJ} (s) = \sum_{n=1}^\infty \beta_{IJ}^{(n)}(s/m^2) \left (
\frac{m^2}{f^2} \right)^n
\label{eq:chiral}
\end{equation}
where the $\beta_{IJ}^{(n)}(s/m^2)$ functions are made of polynomials
and logarithms (chiral logs). As we explain below, the ability to
accommodate resonances reduces the applicability of these
approximations (Eqs.~(\ref{eq:reff}) and~(\ref{eq:chiral})) in
practice.

As discussed above, the function $W_{IJ}(s)$ is real in the elastic
region of scattering. Obviously, the zeros of $ T_{IJ}(s) $ and $
W_{IJ}(s) $ coincide in position and order\footnote{ In $\pi\pi$
scattering, the location of one zero for each $I$ and $J$ is known
approximately (see e.g.  Ref.\cite{We66}) in the limit of small $s$
and $m$ and are called Adler zeros.}. For real $s$ above the two
particle threshold, zeros could appear for a phase shift
$\delta_{IJ}(s)= n \pi, \, n \in \naturales$, in which case a
resonance has already appeared at a lower energy. For the present
discussion we can ignore these zeros. The first resonance would appear
at $\sqrt{s}=m_R $ for which $\delta_{IJ}(m_R)=\pi/2$ and hence $ {\rm
Re} T^{-1}_{IJ}(m_R)=0 $, i.e., $-{\rm Re} {\bar I}_0 (m_R) +
W^{-1}_{IJ}(m_R) =0 $, but $ {\rm Re} {\bar I}_0 (s) > 0 $, so a
necessary condition for the existence of the resonance is that
$W_{IJ}(m_R) > 0 $. On the other hand, the sign of $W_{IJ}(s) $ is
fixed between threshold and the zero at $\delta_{IJ}(s)=n \pi $. It
turns out that due to chiral symmetry the functions $W_{11}$ and
$W_{00}$ are always negative for $\pi\pi$ scattering so the existence
of the $\sigma$ and $\rho$ resonances can not be understood.

\begin{figure}


\begin{center}                                                                
\leavevmode
\epsfysize = 650pt
\makebox[0cm]{\epsfbox{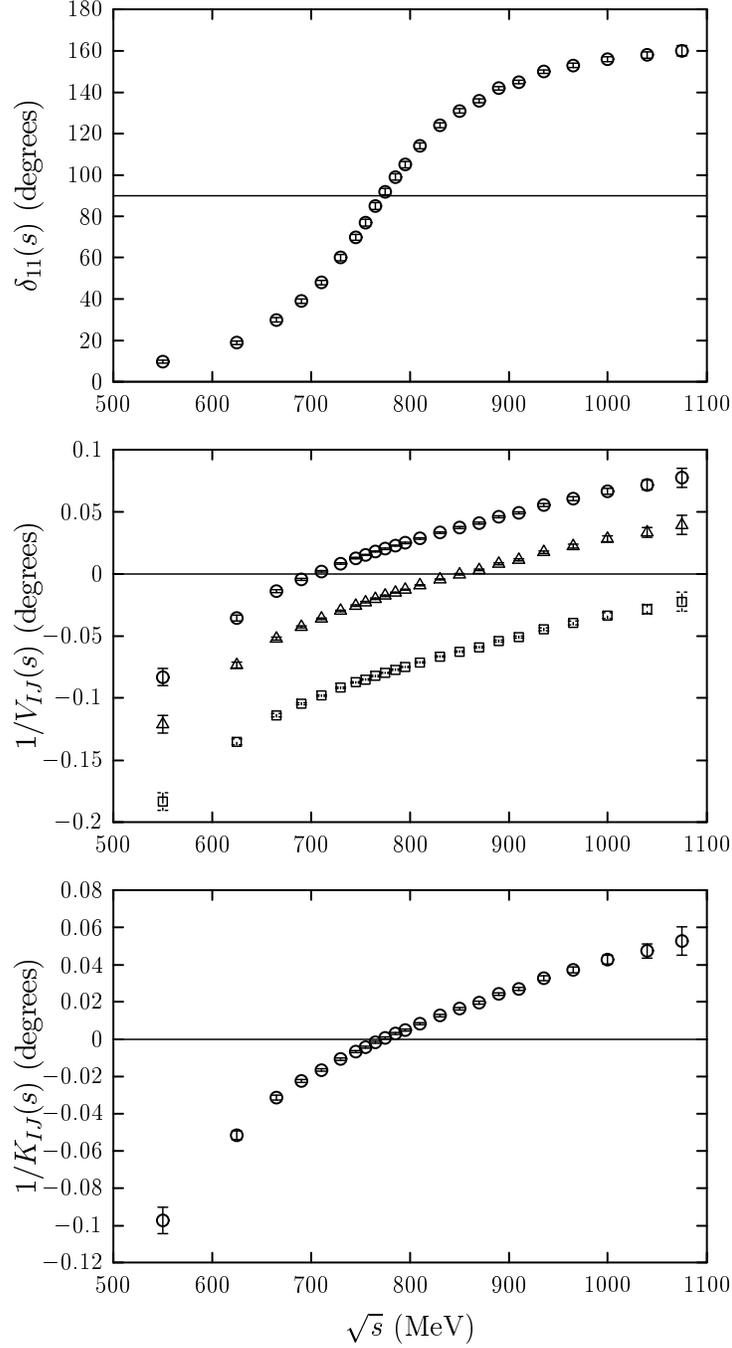}}
\end{center}
\vspace{-3.5cm}
\caption[pepe]{\footnotesize Experimental isovector $p-$wave phase
shifts (top panel), inverse of the functions $V_{11}(s)$ (middle
panel) and $K_{11}(s)$ (bottom panel) as a function of the total CM
energy $\protect\sqrt s$.  Phase shifts are taken from the
experimental analysis of Ref.~\protect\cite{pa73} and $V_{11}^{-1}(s)$
is determined through Eq.~(\ref{eq:defV}) from the data of the top
panel and using three values of the constant $C_{11}$: 0 ($V_{11}$ =
$W_{11}$), $-0.1$ (as suggested by the results of
Table~\protect\ref{tab:off-res}) and $-0.038$ (as suggested by the
formula $[ \log(m/\mu)-1]/8\pi^2$ with a scale $\mu$ of the order of 1
GeV, see Eq.~(\protect\ref{eq:cdim})), which are represented by
circles, squares and triangles respectively. Finally, $K_{11}$ is also
determined from the data of the top panel through
Eq.~(\protect\ref{eq:kmatrix}).}
\label{fig:pole}
\end{figure}

The above argument overlooks the fact that the change in sign of
$W_{IJ}$ may be also due to the appearance of a pole at $s=s_p$, which
does not produce a pole in $T_{IJ}$, since $T_{IJ}(s_p)= -1/{\bar
I}_0(s_p)$. The problem is that if that $W_{IJ}$ has to diverge before
we come to the resonance, how can an approximate expansion, of the
type in Eqs.~(\ref{eq:reff}) and~(\ref{eq:chiral}), for $W_{IJ}$
produce this pole, and if so how could the expansion be reliable at
this energy region ?  To get the proper perspective we show in
Fig.~\ref{fig:pole} the inverse of the function $W_{11}$ (circles in
the middle plot) for $\pi\pi$ scattering extracted from experiment
through Eq.~(\ref{eq:defW}). The presence of a pole (zero in the
inverse function) in $W_{11}$ is evident since the $\rho$ resonance
exists!

We suggest to use Eq.~(\ref{eq:defV}) rather than Eq.~(\ref{eq:defW})
with the inclusion of a further unknown parameter $C_{IJ}$, because in
contrast to the function $W_{IJ}$, the potential $V_{IJ}$ can be
approximated by expansions of the type given in Eqs.~(\ref{eq:reff})
and~(\ref{eq:chiral}).  It is clear that with the new constant
$C_{IJ}$,  ${\rm Re} T^{-1}_{IJ} $ can vanish
without requiring a change of sign in the new potential $
V_{IJ}(s)$. Thus, $ V_{IJ}(s)$ can be kept small as to make the use of
perturbation theory credible. The effect of this constant on the
potential extracted from the data can be seen in Fig.~\ref{fig:pole} 
for several particular values.

After this discussion, it seems reasonable to propose some kind of 
expansion (effective range, chiral,\dots ) for the potential 
$V_{IJ}(s)$ rather than for the  function  $W_{IJ}(s)$. The role played
by the renormalization constant $C_{IJ}$ will have to be
analyzed. Thus, in the next subsection we use a chiral expansion of the
type 
\begin{equation}
V_{IJ} (s) = \sum_{n=1}^{\infty} \phantom{p}V^{(2n)}_{IJ}(s/m^2) \left (
\frac{m^2}{f^2} \right)^n
\label{eq:vexp}
\end{equation}
where the $^{(2n)}V_{IJ}(s/m^2)$ functions are made of polynomials and
logarithms (chiral logs). In this way we will be able to determine the
higher order terms of the ChPT Lagrangian not only from the threshold
data, as it is usually done in the literature, but from a combined
study of both the threshold and the low-lying resonance region.

\subsection{ Chiral expansion of the on-shell potential}
\label{sec:expan-pot}
The chiral expansion of the $\pi\pi$ 
elastic scattering amplitude reads
\begin{eqnarray}
T_{IJ}(s)  = T_{IJ}^{(2)} (s) /f^2 + T_{IJ}^{(4)} (s) /f^4 +
T_{IJ}^{(6)} (s) /f^6 +
\dots \label{eq:tchpt}
\end{eqnarray}
where $T_{IJ}^{(2)} (s)$, $T_{IJ}^{(4)} (s)$ and $T_{IJ}^{(6)}(s)$ can
be obtained from Refs.~\cite{We66},~\cite{GL84} and~\cite{mksf95}--
\cite{bc97} respectively. For the sake of completeness we give in the
Appendix~\ref{sec:appeb} the leading and next-to-leading orders.

If we consider the similar expansion for the potential $V_{IJ}$ 
given in Eq.~(\ref{eq:vexp}) and expand in power series of $m^2/f^2$ the
amplitude given in Eq.~(\ref{eq:defV}) we get
\begin{eqnarray}
T_{IJ}  &=& \frac{m^2}{f^2} \phantom{p}V^{(2)}_{IJ}
            + \frac{m^4}{f^4} \left ( \phantom{p}V^{(4)}_{IJ} + ({\bar
I}_0 + C_{IJ}) \phantom{p}[V^{(2)}_{IJ}]^2 \right ) \nonumber  \\
&+& \frac{m^6}{f^6} \left ( \phantom{p}V^{(6)}_{IJ} + 2 ({\bar
I}_0 + C_{IJ}) \phantom{p}V^{(4)}_{IJ} \phantom{p}V^{(2)}_{IJ} + ({\bar
I}_0 + C_{IJ})^2 \phantom{p}[V^{(2)}_{IJ}]^3 \right ) \cdots \label{eq:tbse}
\end{eqnarray}
Matching the expansions of Eqs.~(\ref{eq:tchpt}) and~(\ref{eq:tbse})
we find
\begin{eqnarray}
m^2 \phantom{p}V^{(2)}_{IJ} & = & T_{IJ}^{(2)}  \nonumber\\
m^4 \phantom{p}V^{(4)}_{IJ} & = & T_{IJ}^{(4)}  - ({\bar
I}_0 + C_{IJ}) [T_{IJ}^{(2)}]^2 =  \tau_{IJ}^{(4)}-C_{IJ}
[T_{IJ}^{(2)}]^2 \nonumber\\
\dots \label{eq:approx_v}
\end{eqnarray}
with $\tau_{IJ}^{(4)}$ defined in Eq.~(\ref{eq:tau}). 
From the unitarity requirement 
\begin{equation} 
{\rm Im} T_{IJ}^{(4)}(s) =
  - {1\over 16\pi }  \sqrt{1-{4m^2 \over s}}
[T_{IJ}^{(2)}(s)]^2,\,\,\, s>4m^2
\end{equation}
 and thus, we see that  $\phantom{p}V^{(4)}_{IJ}$ is real above the
unitarity cut, as it should be.

Thus, in this expansion we get the following formula for the inverse
scattering amplitude
\begin{eqnarray}
T^{-1}_{IJ}(s)  &=& - {\bar
I}_0 (s) - C_{IJ} + {1\over
T_{IJ}^{(2)} (s)/f^2 + T_{IJ}^{(4)} (s)/f^4 - ({\bar
I}_0 (s) + C_{IJ}) [T_{IJ}^{(2)} (s)]^2 /f^4 + \cdots} \nonumber \\
&&\nonumber \\
&=&- {\bar
I}_0(s) - C_{IJ} + {1\over
T_{IJ}^{(2)} (s)/f^2 + \tau_{IJ}^{(4)} (s)/f^4 - 
C_{IJ} [T_{IJ}^{(2)} (s)]^2 /f^4 + \cdots}\label{eq:tinv_fin}
\end{eqnarray}

Notice that reproducing ChPT to some order means neglecting higher
order terms in $ s/f^2 $ and $ m^2 / f^2 $, which is not the same as
going to low energies. With the constant $C_{IJ}$ we may be able to improve the
low energy behavior of the amplitude. Obviously, the {\it exact}
amplitude $T_{IJ}(s)$ is independent on the value of the constant
$C_{IJ}$, but the smallness of $V_{IJ}(s)$ depends on $C_{IJ}$. Ideally,
with an appropriated choice of $C_{IJ}$,  $V_{IJ}(s)$ 
would be, in a determined region of energies, as small as
to make the use of perturbation theory credible and simultaneously fit
the data.

Any unitarization scheme which reproduces ChPT 
to some order, is necessarily generating all higher orders. For
instance, if we truncate the expansion at fourth order we would
``predict'' a sixth order 
\begin{equation}
{\bar T}_{IJ}^{(6)} =   2 ({\bar
I}_0 + C_{IJ}) \left(\tau_{IJ}^{(4)}-C_{IJ}[T_{IJ}^{(2)}]^2\right) 
T_{IJ}^{(2)} + ({\bar
I}_0 + C_{IJ})^2 [T_{IJ}^{(2)}]^3 \label{eq:next-order} 
\end{equation}
and so on. 

Finally, we would like to point out that within this on-shell scheme
crossing symmetry is restored more efficiently than within the
off-shell scheme exposed in Sect.~\ref{sec:off}. Thus, for instance,
if we truncate the expansion at fourth order and neglect terms of order
$1/f^6$, crossing symmetry is exactly restored at all orders in the
expansion $(s-4m^2)$, whereas in the off-shell scheme this is only
true if the $u-$ and $t-$ unitarity corrections are Taylor expanded
around $4m^2$ and only the leading terms in the expansions are kept.

\subsection{Form Factors}
\label{sec:form}

\begin{figure}[t]
\vspace{-9.5cm}
\hbox to\hsize{\hfill\epsfxsize=0.65\hsize
\epsffile[52 35 513 507]{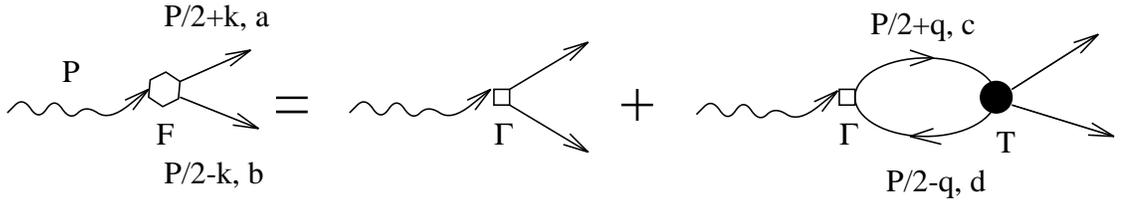}\hfill}
\vspace{1.5cm}
\caption[pepe]{\footnotesize Diagrammatic representation of the BSE
type equation used to compute vertex functions. It is also 
sketched the used kinematics and $a,b,c,d$ are isospin indices.}
\vspace{0.5cm}
\label{fig:formkin}
\end{figure}

The interest in determining the half-off shell amplitudes lies upon
their usefulness in computing vertex functions. Let $ \Gamma_P (p,k) $
be the irreducible three-point function, connecting the two meson state
to the corresponding current. The BSE for this vertex function,
Fig.~\ref{fig:formkin},  is then
\begin{eqnarray}
F^{ab}_P(k) &=& \Gamma^{ab}_P(k) + \frac12 \sum_{cd}{\rm i}\int\frac{d^4
q}{(2\pi)^4}T_P(q,k)_{cd;ab}\Delta(q_+) \Delta(q_-)
\Gamma^{cd}_P(q) \label{eq:vertex}
\end{eqnarray}
and using the BSE we get alternatively
\begin{eqnarray}
F^{ab}_P(k) &=& \Gamma^{ab}_P(k) + \frac12 \sum_{cd}{\rm i}\int\frac{d^4
q}{(2\pi)^4}V_P(q,k)_{cd;ab}\Delta(q_+) \Delta(q_-)
F^{cd}_P(q)
\end{eqnarray}

In operator language we have $ F = \Gamma + V G_0 F = \Gamma + T G_0
\Gamma $, with $G_0$ the two particle propagator. The discontinuity in
$F$ is then given by the discontinuities of the scattering
amplitude $T$ and the two particle propagator $G_0$. 

The off-shellness of the kernel of
Eq.~(\ref{eq:vertex}) leads to power divergences which have to be
renormalized by appropriate Lagrangian counter-terms. The
resulting finite parts are undetermined parameters of the EFT (for
instance ${\bar l}_6$, at order ${\cal O}(p^4)$, for the
vector vertex). The situation is similar to that discussed in
Subsect.~\ref{sec:power} for the scattering amplitude, and thus if we
iterate a renormalized irreducible three-point function, $\Gamma$, we
should ignore the power divergences and therefore ignore the
off-shell behavior of the kernel of Eq.~(\ref{eq:vertex}) and then
solve
\begin{eqnarray}
F^{ab}_P(k) &=& \Gamma^{ab}_P(k) + \frac12 \sum_{cd}{\rm i}\int\frac{d^4
q}{(2\pi)^4}T_P(\bar{q},k)_{cd;ab}\Delta(q_+) \Delta(q_-)
\Gamma^{cd}_P(\bar{q}) \label{eq:vertex-on}
\end{eqnarray}
with $\bar{q}$ given in Eq.~(\ref{eq:defqbar}). The above
equation involves the half-off or  on-shell amplitudes for off-shell
($k^2 \neq m^2$) or on-shell ($k^2 = m^2$) vertex functions respectively.

\subsubsection{Vector form factor}

The vector form factor, $F_V(s)$, is defined by
\begin{equation}
\left \langle \pi^a (P/2+k) \,\, \pi^b (P/2-k) \left | \frac12 \left
(\bar{u}\gamma_\mu u - \bar{d}\gamma_\mu d \right ) \right | 0 \right \rangle
= - 2 i F_V(s) \epsilon_{ab3} k_\mu
\end{equation}
where $s = P^2$, $a,b$ are Cartesian isospin indices, $u,d$ and
$\gamma_\mu$ are Dirac fields, with flavor ``up'' and ``down'', and
matrices respectively.  For on-shell mesons, $k \cdot P = 0$ and then the
vector current $\frac12 \left (\bar{u}\gamma_\mu u - \bar{d}\gamma_\mu
d \right )$ is conserved as demanded by gauge invariance.  The vector
form factor is an isovector, and we calculate here its $I_z=0$
component, thus the sum over isospin in Eq.~(\ref{eq:vertex-on})
selects the $I=1$ channel. Then, we have
\begin{eqnarray}
F_V(s) k^\mu &=& \Gamma_V (s) k^\mu + {\rm i} \Gamma_V(s) \int\frac{d^4
q}{(2\pi)^4}T^1_P(\bar{q},k)\Delta(q_+) \Delta(q_-) \bar{q}^\mu \label{eq:elec}
\end{eqnarray}
with $\Gamma_V (s)$ related with the irreducible vector three-point
vertex by means of
\begin{equation}
\Gamma^{ab}_P(k) =  - 2 i \Gamma_V(s) \epsilon_{ab3} k_\mu
\end{equation}
Using a partial-wave
expansion for $T^1_P(\bar{q},k)$ (Eq.~(\ref{eq:wave-exp})) 
and the orthogonality relation given 
in Eq.~(\ref{eq:ort}) the integral in Eq.~(\ref{eq:elec}) 
can be performed and  for on-shell pions we obtain
\begin{equation}
F_V(s) = \Gamma_V(s) + T_{11}(s) I_0(s) \Gamma_V(s) 
\end{equation}
The above expression presents a logarithmic divergence which
 needs one subtraction to be renormalized, choosing the subtraction
point $s_0$ we have a finite form-factor
\begin{eqnarray}
F_V(s) &=& \Gamma_V(s) + T_{11}(s) \left( {\bar I}_0(s) + C_V \right )
\Gamma_V(s)  \\
C_V &=& - {\bar I}_0(s_0) + \frac{F_V-\Gamma_V}{T_{11}\Gamma_V}\Big|_{s=s_0}
\label{eq:defcv}
\end{eqnarray}
Replacing ${\bar I}_0(s) = - T_{11}^{-1}(s)-C_{11}+V_{11}^{-1}(s)$ we
get 
\begin{eqnarray}
F_V(s) = \frac{T_{11}(s)}{V_{11}(s)}\left 
(1 + (C_V-C_{11})V_{11}(s) \right)\Gamma_V(s)
\end{eqnarray}
Notice that Watson's theorem~\cite{Wa55} reads
\begin{eqnarray}
{ F_V(s+ {\rm i} \epsilon ) \over F_V(s- {\rm i} \epsilon ) } =
{ T_{11} (s+ {\rm i} \epsilon ) \over T_{11} (s- {\rm i} \epsilon )} =
e^{ 2 {\rm i} \delta_{11} (s) },\,\,\, s > 4m^2 \label{eq:wat}
\end{eqnarray}
and it is automatically satisfied. That is a very
reassuring aspect of the BSE approach. Indeed, in the literature it is
usual~\cite{GM91},~\cite{bt94}--~\cite{GO99} to use Watson's theorem
as an input, and employ it to write a dispersion relation for the
form-factor. This procedure leads to so called
Omn\`es--Muskhelishvili~\cite{Om58} representation of the form-factor,
which requires the introduction of a polynomial of arbitrary
degree. In our case, not only the phase of $F_V (s)$ is fixed, in
harmony with Watson's theorem, but also the modulus, and hence the
polynomial, is fixed at the order under consideration.

The normalization of the form factor requires $ F_V(0)=1 $, which allows
to express the renormalization constant $(C_V-C_{11})$ in terms of 
$T_{11},\Gamma_V$ and $V_{11}$ at $s=0$, and thus we get
\begin{eqnarray}
F_V(s) = T_{11}(s)\,\Gamma_V(s)\left\{\frac{1}{V_{11}(s)} - 
\frac{1}{V_{11}(0)}+ \frac{1}{T_{11}(0)\Gamma_V(0)} \right\} \label{eq:fv}
\end{eqnarray}
From our formula it is clear that for $s > 4m^2$, where both
$\Gamma_V$ and $V_{11}$ are real,  ${\rm Re} F (s)^{-1} = 0 $
when ${\rm Re} T_{11} (s)^{-1} = 0$, in agreement with the vector
meson dominance hypothesis.  In Eq.~(\ref{eq:fv}), $T_{11}$ is 
obtained from the solution of the BSE for a given ``potential'',
$V_{11}$, which admits a chiral expansion, as discussed in
Subsect.~\ref{sec:expan-pot}. The irreducible three-point function,
$\Gamma_V(s)$ admits also a chiral expansion of the type
\begin{equation}
\Gamma_V(s) = 1 + \frac{\Gamma_2^V(s)}{f^2} + \cdots \label{eq:ga_chi}
\end{equation}
Note that at lowest order, ${\cal O}(p^0)$, $\Gamma_V(0) = F_V(0) =
1$. The next-to-leading function $\Gamma_2^V(s)$ can be obtained 
if one expands Eq.~(\ref{eq:fv}) in powers of $1/f^2 $ 
and compare it to the order ${\cal O}(p^2)$ deduced by Gasser-Leutwyler
in Ref.~\cite{GL84}. Thus, we get

\begin{equation}
\Gamma_V(s) = 1 + \left \{ (1-\frac{s}{4m^2})\,\Gamma_2^V(0) + 
\frac{s}{96\pi^2}({\bar l}_6 - \frac13 )\right\}/ f^2 + \cdots
\label{eq:gamma} 
\end{equation}
For $s >> 4m^2$, $V_{11}(s)$ and $\Gamma_V(s)$ might increase as a
power of $s$, whereas elastic unitarity ensures that $T_{11}(s)$
cannot grow faster than a constant. Indeed, if $V_{11}$ actually
diverges or remains at least constant in this limit, then $T_{11}(s)$
behaves like $1/{\bar I}_0(s)$, it is to say, it decreases
logarithmically.  To guaranty that the elastic form factor goes to
zero\footnote{This behavior is in agreement with the expected once
subtracted dispersion relation for the form factor~\cite{BD65}.} for
$s \to \infty$, we impose in Eq.~(\ref{eq:fv}) the constraint
\begin{equation}
\Gamma_V(0) = V_{11}(0) / T_{11}(0)\label{eq:gamma2_0} 
\end{equation}
and thus, we finally obtain 
\begin{eqnarray}
F_V(s) &=& \frac{T_{11}(s)\,\Gamma_V(s)}{V_{11}(s)} \label{eq:fv-fin} 
\end{eqnarray}
with $\Gamma_V(s)$ given in Eq.~(\ref{eq:gamma}) and $\Gamma_2^V(0)$
determined by the relation of Eq.~(\ref{eq:gamma2_0}). Thus, the 
vector form factor, at this order,  is determined by the
vector-isovector $\pi\pi$
scattering plus a new low-energy parameter: ${\bar l}_6$.
\subsubsection{Scalar form factor}

The scalar form factor, $F_S(s)$, is defined by
\begin{equation}
\left \langle \pi^a (P/2+k) \,\, \pi^b (P/2-k) \left |  \left
(\bar{u} u + \bar{d} d \right ) \right | 0 \right \rangle
= \delta^{ab} F_S(s) 
\end{equation}
The scalar form factor is defined through the isospin-zero scalar
source, thus the sum over isospin in Eq.~(\ref{eq:vertex-on}) selects
the $I=0$ channel. Following the same steps as in the case of the
vector form factor we get, after having renormalized, 
\begin{eqnarray}
F_S(s) &=& \frac{T_{00}(s)}{V_{00}(s)}\left 
(1 + d_S  V_{00}(s) \right)\Gamma_S(s) \nonumber\\
d_S &=& C_S-C_{00}\label{eq:fs}
\end{eqnarray}
where $C_S$ is given by Eq.~(\ref{eq:defcv}), with the obvious
replacements $V \to S$ and $11 \to 00$. Watson's theorem is here again
automatically satisfied by Eq.~(\ref{eq:fs}). A similar discussion as
in the case of the vector form-factor leads to set the $d_S$ parameter
to zero, since the form factor goes to zero for $s\to
\infty$~\cite{bct98}. Here again we propose a chiral expansion of the
type of Eq.~(\ref{eq:ga_chi}) for the irreducible three-point function,
$\Gamma_S(s)$ 
\begin{eqnarray}
\Gamma_S(s) = \Gamma_0^S(s) + \frac{\Gamma_2^S(s)}{f^2} + \cdots 
\end{eqnarray}
and expanding the form factor in powers of $1/f^2 $ and
comparing it to the order ${\cal O}(p^2)$ deduced by Gasser-Leutwyler
in Ref.~\cite{GL84}, we finally get
\begin{eqnarray}
\frac{F_S(s)}{F_S(0)} &=& \frac{\Gamma_S(s)}{\Gamma_S(0)}
\frac{T_{00}(s)}{T_{00}(0)} \frac{V_{00}(0)}{V_{00}(s)} \nonumber\\
&&\nonumber\\
\frac{\Gamma_S(s)}{\Gamma_S(0)} &=& \frac{1}{1+
\frac{1}{f^2}\frac{\Gamma_2^S(0)}{\Gamma_0}} \left \{ 
1+\frac{1}{f^2}\left [ \frac{\Gamma_2^S(0)}{\Gamma_0}+
\frac{s}{16\pi^2} ({\bar l}_4 + 1 + 16\pi^2 C_{00})\right ] \right
\}\nonumber \\
&&\nonumber\\
\frac{\Gamma_2^S(0)}{\Gamma_0} & = & -\left (
\frac{m^2}{16\pi^2}({\bar l}_3 + \frac12) + \frac{m^2\,C_{00}}{2}\right ) 
\nonumber \\
&&\nonumber\\
\Gamma_0^S(s) &=& \Gamma_0 = 2B \nonumber \\
F_S(0) &=& 2B\left\{ 1-\frac{m^2}{16\pi^2f^2}({\bar l}_3 -\frac12)\right\}
\label{eq:fm_esc}
\end{eqnarray}
where $B$ is a low energy constant which measures the vacuum
expectation value of the scalar densities in the chiral
limit~\cite{GL84}. Note that in contrast to the vector case the
normalization at zero momentum transfer of the scalar form-factor is
unknown.

\subsection{ Numerical results: $\pi\pi$-phase shifts and form-factors.}
\label{sec:num-res-on}

The lowest order of our approach is obtained by approximating the potential, 
\begin{equation}
V_{IJ} \approx
\frac{m^2}{f^2}\phantom{p}V^{(2)}_{IJ}\label{eq:leading}
\end{equation}
with $\phantom{p}V^{(2)}_{IJ}$ given in Eq.~(\ref{eq:approx_v}). Thus,
at this order we have three undetermined parameters
$C_{00},C_{11},C_{20}$. At this level of approximation the $d-$wave
phase-shifts are zero, which is not completely unreasonable given
their small size, compatible within experimental uncertainties with
zero in a region up to $500$ MeV. Note that there is a clear
parallelism between the $C's$ parameters here and the $I^R_0(4m^2)$
low energy parameters introduced in Sect.~\ref{sec:off}. Thus, the
$C's$ parameters will be given in terms of the ${\bar l}_{i}$
parameters, as we found in Subsect.~\ref{sec:num-res-off}, although
some constraints on the ${\bar l}'s$ would be imposed since $I_2^R =
I_4^R=0$. However, one should expect more realistic predictions for
the phase-shifts in the off-shell case, since there were more freedom
to describe the data (four versus three parameters). For the sake of
shortness, we do not give here any numerical results for this lowest
order of the proposed approximation and also because they do not
differ much from those already presented in
Subsect.~\ref{sec:num-res-off}. The interesting aspect is that already
this lowest order approximation is able to describe successfully the
experimental phase shifts for energies above the certified validity
domain of ChPT. We come back to this point in
Sect.~\ref{sec:c_non}. In Subsect.~\ref{sec:compari} we also investigate
the convergence of the expansion for the potential, $V$, proposed in
this work. Thus we will compare results  at
leading (discussed above) and next-to-leading (discussed below) orders
for the function $W_{IJ}$ defined in Eq.~(\ref{eq:defW}).

Further improvement can be gained by  considering  the  next-to-leading 
order correction to the potential, which  is determined by the approximation:
\begin{equation}
V_{IJ} \approx
\frac{m^2}{f^2}\phantom{p}V^{(2)}_{IJ} + 
\frac{m^4}{f^4}\phantom{p}V^{(4)}_{IJ} \label{eq:nl}
\end{equation}
with $\phantom{p}V^{(2,4)}_{IJ}$ given in Eq.~(\ref{eq:approx_v}). At
next-to-leading, we have nine free parameters ($C_{IJ},\,$ with
$IJ=00,11,20,02,$ $22$ and ${\bar l}_i,\, i =1,\cdots 4$ ) for
$\pi\pi$ scattering in all isospin channels and $J\le 2$. Besides we
have two additional ones,
($B$ and ${\bar l}_6$) to describe the scalar and vector form
factors. As we discussed in Eq.~(\ref{eq:next-order}), in this context
the $C$'s parameters take into account partially the two loop
contribution, and thus they could be calculated in terms of the two
loop ChPT low energy parameters. That is similar to what we did for
the $I^R_n(4m^2)$ parameters in Sect.~\ref{sec:off} or what we could
have done above with the $C$'s parameters at the lowest order of our
approach\footnote{In both cases the unknown parameters could be
calculated in terms, of the one loop ChPT low energy parameters,
${\bar l}$'s.}. However, we renounce to take to practice this program,
because of the great theoretical uncertainties in the determination of
the needed two loop ChPT contributions: the two loop contribution have
been computed in two different frameworks: ChPT~\cite{bc97} and
generalized ChPT~\cite{mksf95}, and only in the former one a complete
quantitative estimate of the low energy parameters is
given. Furthermore, this latter study lacks a proper error analysis
which has been carried out in Ref.~\cite{ej99_2}.  On top of that a
resonance saturation assumption~\cite{E89} has been relied upon. Therefore, we
are led to extract, at least, the $C$'s parameters from experimental
data. For the ${\bar l}'\,s$ parameters, we could either fit all or
some of them to data and fix the remainder to some reasonable values of the
literature, as we did in Subsect.~\ref{sec:num-res-off}. Here, we will
follow a hybrid procedure. The justification will be provided {\it a
posteriori}, since the error bars in the $\bar l$'s are reduced in
some cases.

\subsubsection{ Electromagnetic pion form factor.} 

\begin{figure}


\begin{center}                                                                
\leavevmode
\epsfysize = 600pt
\epsfbox{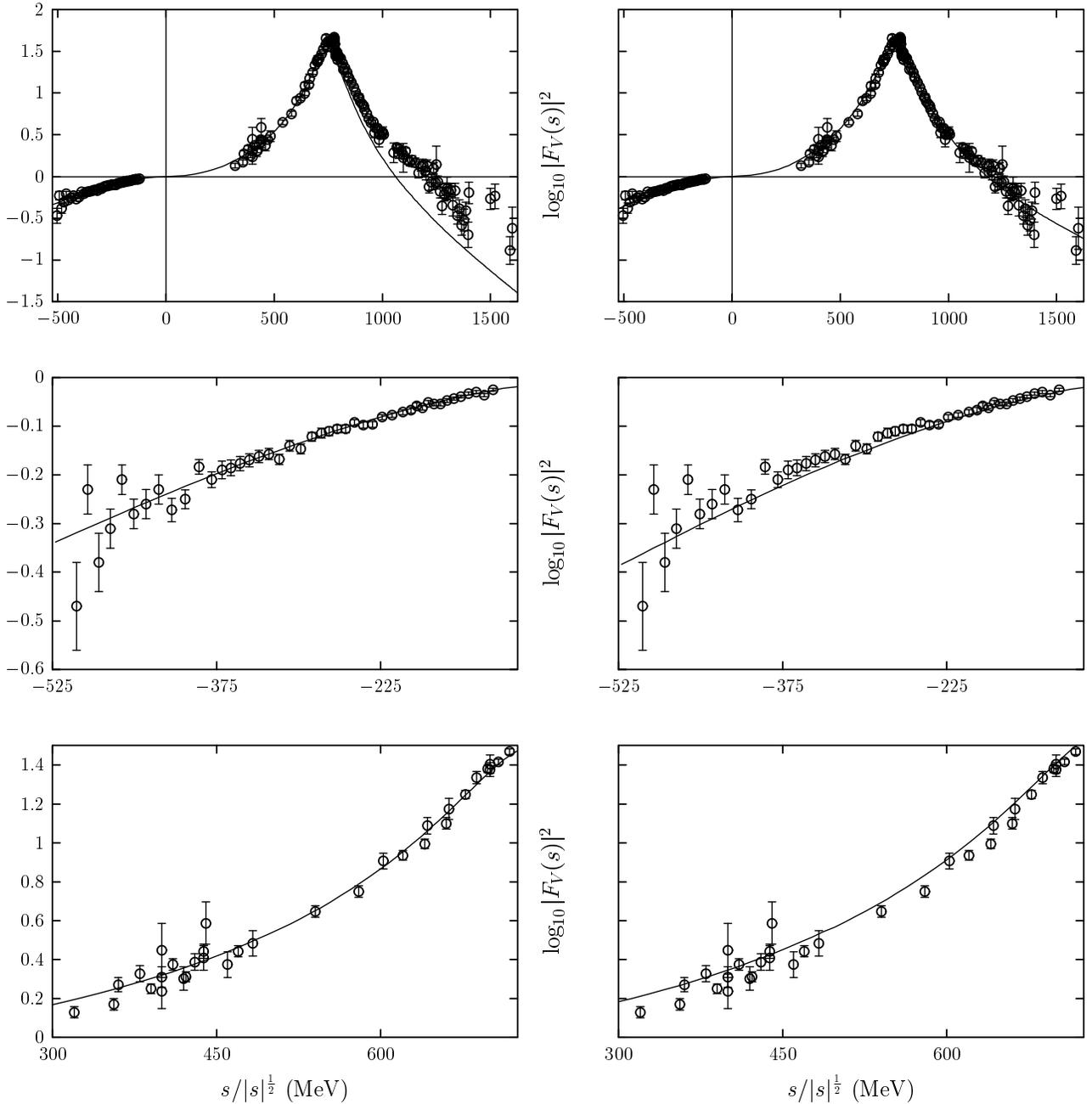}\phantom{tonto}
\end{center}
\vspace{-2.5cm}
\caption[pepe]{\footnotesize Best fits of the next-to-leading approach
(Eqs.~(\protect\ref{eq:defV}) and~(\protect\ref{eq:fv-fin})) to the
electromagnetic pion form factor. Data are taken from
Refs.~\protect\cite{na7} (space--like) and ~\protect\cite{bar85}
(time--like).  Results obtained with the parameter set determined by
the entry $\protect\sqrt s \le 700 $ ($\protect\sqrt s \le 1600 $) of
Table~\protect\ref{tab:elec} are displayed in the left (right)
panels.}
\label{fig:elec}
\end{figure}

Data on $\pi\pi$ scattering are
scarce and in most of cases the experimental analysis relies on some
theoretical assumptions because of the absence of direct $\pi\pi$
scattering experiments~\cite{mp95}. However, there are direct and
accurate measurements on the electromagnetic pion form factor in both
the space~\cite{na7}-- and time--like regions~\cite{bar85}. At
next-to-leading order, the vector form-factor depends on $C_{11}$, ${\bar
l}_1-{\bar l}_2$, ${\bar l}_4$ and ${\bar l}_6$ (see
Eq.~(\ref{eq:fv-fin})). As we did for both sets of parameters in
Eq.~(\ref{eq:elesab}) we take 
\begin{equation}
{\bar l}_4 = 4.4 \pm 0.3 \label{eq:l4}
\end{equation}
as determined
by the the scalar radius~\cite{bct98} and fit the another three
parameters to the data of Refs.~\cite{na7} and ~\cite{bar85}. Results
can be seen both in Fig.~\ref{fig:elec} and Table~\ref{tab:elec}, for 
different energy cuts. 
\begin{table}[t]
\begin{center}
\begin{tabular}{c|ccccc}
${\bar l}_4 = 4.4$ & $\sqrt s \le 483 $  & $\sqrt s \le 700 $  
                   & $\sqrt s \le 808 $  & $ \sqrt s \le 915 $ 
                   & $ \sqrt s \le 1600 $ \\\hline\tstrut
${\bar l}_1-{\bar l}_2$ & $-6 \er{1}{2}$ & $-6.1 \er{0.1}{0.3}$ 
& $-5.92 \er{0.03}{0.04}$ & $-5.93 \pm 0.03$ & $-5.861 \pm 0.010$ \\
$C_{11}$                & $-0.15 \er{0.04}{0.05}$  & $-0.112 \pm 0.010$ 
& $-0.1099 \pm 0.0018$  & $-0.1098 \pm 0.0018$ &$-0.1195
\er{0.0016}{0.0015}$ \\ 
${\bar l}_6$            & $19.0 \pm 0.3$ & $19.14 \pm 0.19$ 
& $20.84 \pm 0.06$ & $20.82 \pm 0.06$ & $21.09 \pm 0.06$ \\
$\chi^2/ dof $          & $1.2$ & $1.2$ & $3.9$  & $3.5$  & $3.3$  \\ 
num. data               & $63$  & $76$ & $112$ & $126$ & $184$ 
\end{tabular}
\end{center}
\caption[pepe]{\footnotesize Results of best fits of the
next-to-leading approach (Eqs.~(\protect\ref{eq:defV}),
~(\protect\ref{eq:fv-fin}) and
~(\protect\ref{eq:nl})) to the electromagnetic pion form factor. Data
are taken from Refs.~\protect\cite{na7} (space--like, $-(503 \,{\rm
MeV})^2 \, \le s \le -(122.47 \,{\rm MeV})^2$ ) and
~\protect\cite{bar85} (time--like, $320 \, {\rm MeV} \le \protect\sqrt
s \le 1600 \, {\rm MeV}$). Errors in the fitted parameters are
statistical and have been obtained by increasing the value of $\chi^2$
by one unit.  We fix ${\bar l}_4 = 4.4$ and show the variation of the
fitted parameters, their statistical errors and $\chi^2/ dof $ with
the used energy cut (given in MeV) for the time--like region. We also
give, in the last row the number of piece of data of each of the best fits.}
\label{tab:elec}
\end{table}
\begin{table}[t]
\begin{center}
\begin{tabular}{c|ccc}
${\bar l}_4$   &  ${\bar l}_1-{\bar l}_2$ & $C_{11}$ & ${\bar l}_6$
\\\hline\tstrut
4.1 &  $-6.2 \er{0.1}{0.3}$& $-0.112 ~\er{~0.011}{~0.010}$ &$19.13 \pm 0.19$ \\ 
4.4 &  $-6.1 \er{0.1}{0.3}$  & $-0.112 \pm 0.010$  & $19.14 \pm 0.19$ \\ 
4.7 &  $-6.1 \er{0.1}{0.3}$ & $-0.111 \pm 0.010$& $19.15~ \er{~0.19}{~0.18}$ \\ 
\end{tabular}
\end{center}
\caption[pepe]{\footnotesize Results of best fits of the
next-to-leading approach (Eqs.~(\protect\ref{eq:defV}),
~(\protect\ref{eq:fv-fin}) and
~(\protect\ref{eq:nl})) to the electromagnetic pion form factor for
different values of ${\bar l}_4$. Data are taken from
Refs.~\protect\cite{na7} and ~\protect\cite{bar85}. In all cases, the
fit range is $-503\,\, {\rm MeV} \le s/|s|^\frac12 \le 700 \,\, {\rm
MeV}$. Errors in the fitted parameters are statistical and have been
obtained by increasing the value of $\chi^2$ by one unit.}
\label{tab:elec-l4}
\end{table}
As can be seen both in the last column of Table~\ref{tab:elec} and in
 the right plots of Fig.~\ref{fig:elec}, a fairly good description of
 the data from $-500\,\, {\rm MeV} \le s/|s|^\frac12 \le 1600 \,\,
 {\rm MeV}$ can be achieved. However, our aim is not only to describe
 the data but also to determine the low energy parameters which
 determine the chiral expansion of both the electromagnetic form
 factor and the $\pi\pi$-amplitude. If one look at the values of
 $\chi^2/dof$ quoted in Table~\ref{tab:elec}, one realizes that there
 is a significant change (1.2 versus 3.9) when data above $700$ MeV
 are considered. Besides, the fitted parameters turn out to be not
 statistically compatible below and above this cut in
 energies. Although with a small variation on ${\bar l}_1 - {\bar
 l}_2$, ${\bar l_6}$ and $C_{11}$ one still gets a reasonable
 description of the form-factor in the whole range of $s$, one can
 appreciate, in the two last rows of plots of Fig.~\ref{fig:elec},
 that the higher the energy cut, the worse the description of the low
 energy region ($-500\,\, {\rm MeV} \le s/|s|^\frac12 \le 600 \,\,
 {\rm MeV}$). Indeed, this is the region where the chiral expansion is
 supposedly more trustable. Neither higher orders in the chiral
 expansion proposed in this work, nor the effect of non-elastic
 channels (like $4\pi$ and $K \bar K$ production) have been included
 at this level of approximation, and both might account for the
 discrepancies appreciated in the top left panel of
 Fig.~\ref{fig:elec} at high energies.  The small changes in the
 fitted parameters above and below the energy cut of 700 MeV
 effectively incorporate these new effects into the model. Thus, our
 best determination of the one-loop (${\bar l}_1 - {\bar l}_2$ and
 ${\bar l_6}$) and two-loop $C_{11}$ low energy parameters come from
 the best fit with an energy cut of 700 MeV in the time-like
 region\footnote{Lower cuts lead to similar $\chi^2/dof$ and
 statistically compatible fitted parameters, but with appreciably
 higher errors, as expected from the reduction on the amount of
 experimental data points. Obviously, we take the highest cut-off, 700
 MeV, which keeps the $\chi^2/dof$ around
 one, maximizes the number of data and hence minimizes the
 errors on the fitted parameters.}. Besides the statistical errors quoted in
 Table~\ref{tab:elec}, we should incorporate some systematics due to
 the uncertainties in ${\bar l}_4$. From the results reported in
 Table~\ref{tab:elec-l4} we deduce that such systematic errors are
 much smaller than the statistical ones quoted in Table~\ref{tab:elec}
 and can be safely ignored\footnote{Vector form factor data are almost
 insensitive to ${\bar l}_4$ which has prevented us to determine it
 from a four parameter fit, but rather to fix it to the value obtained
 in Ref.~\protect\cite{bct98}.}. Thus, we get
\begin{equation}
{\bar l}_1 - {\bar l}_2 = -6.1\er{0.1}{0.3}, \,\,\,\, {\bar l}_6 = 19.14
\pm 0.19, \,\,\,\, C_{11} = -0.112 \pm 0.010 \label{eq:form-res}
\end{equation}
which provides us with an extraordinarily precise determination of the
difference ${\bar l}_1 - {\bar l}_2$ and of ${\bar l}_6$. The authors of 
Ref.~\cite{bct98} have completed a two loop calculation for the scalar and 
vector form factors, finding  ${\bar l}_6 = 16.0 \pm 0.5 \pm 0.7$ which
differs by as much as two standard deviations from our result. 

\subsubsection{ Elastic $\pi\pi-$scattering in the $\rho-$channel.}

\begin{figure}


\begin{center}                                                                
\leavevmode
\epsfysize = 650pt
\makebox[0cm]{\epsfbox{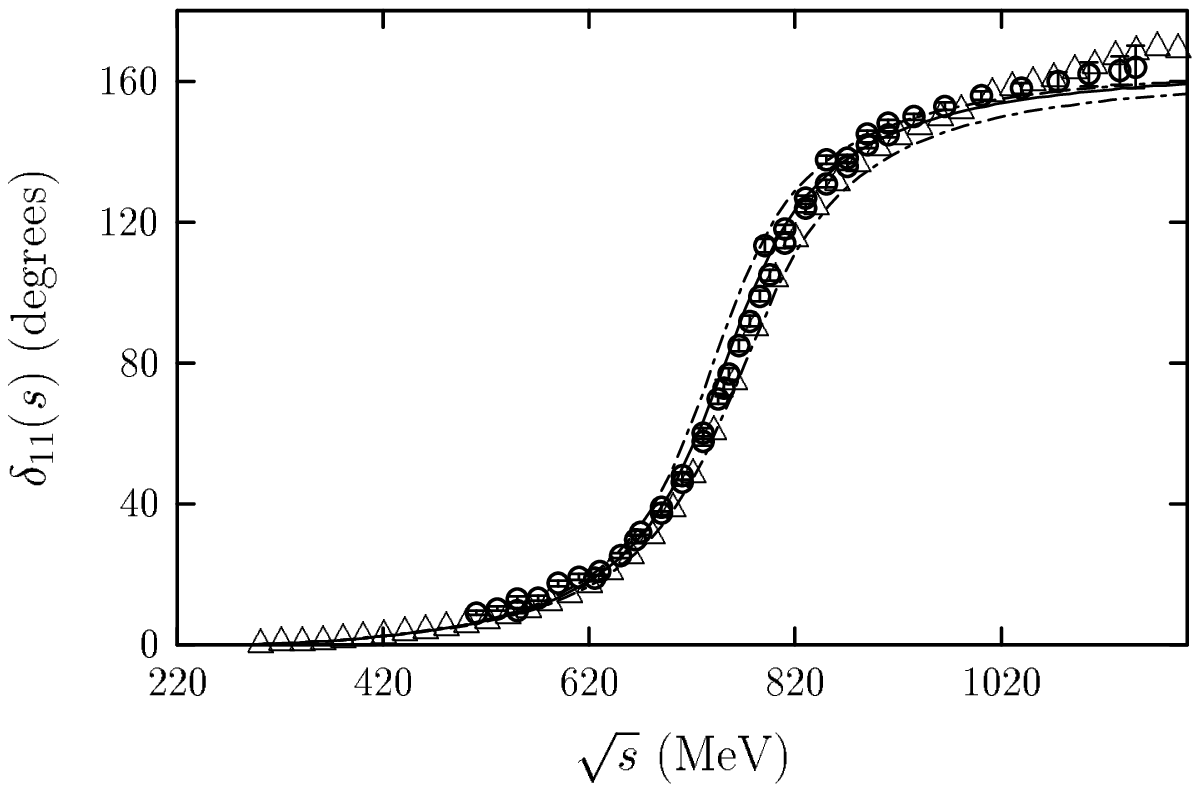}}
\end{center}
\vspace{-14.5cm}
\caption[pepe]{\footnotesize On-shell next-to-leading order BSE
(Eqs.~(\protect\ref{eq:defV}) and~(\protect\ref{eq:nl}))
vector-isovector phase shifts as a function of the total CM energy
$\protect\sqrt s$. The solid line has been obtained using the central
values quoted in Eqs.~(\protect\ref{eq:l4})
and~(\protect\ref{eq:form-res}).  Dashed lines are the 68\% confidence
limits assuming Gaussian and uncorrelated error distributions for the
parameters. For the difference ${\bar l}_1 - {\bar l}_2$ we have
assumed a symmetric error of magnitude 0.3.  Circles stand for the
experimental analysis of Refs.~\protect\cite{pa73} and
~\protect\cite{em74} and the triangles stand for the Frogatt and
Petersen phase-shifts (Ref.~\protect\cite{fp77}) with no errors due to
the lack of error estimates in the original analysis. }
\label{fig:pwave}
\end{figure}

The parameters of Eqs.~(\ref{eq:l4}) and~(\ref{eq:form-res}) uniquely
determine, at next-to-leading order in our expansion
(Eq.~(\protect\ref{eq:nl})), the vector--isovector $\pi\pi$ scattering
phase shifts. Results are shown in Fig.~\ref{fig:pwave}. The solid
line has been obtained with the central values quoted in
Eqs.~(\ref{eq:l4}) and~(\ref{eq:form-res}). There also we show the
68\% confidence limits (dashed lines), obtained by assuming
uncorrelated Gaussian distributed errors in the parameters quoted in
Eqs.~(\ref{eq:l4}) and~(\ref{eq:form-res}) and propagate those to the
scattering phase shifts by means of a Monte Carlo simulation. For the
difference ${\bar l}_1 - {\bar l}_2$ we have assumed a symmetric error
of magnitude 0.3. 
Besides we also find\footnote{We use
the procedure described in Subsect.~\ref{sec:num-res-off}.}
\begin{equation}
m_\rho = 764\er{21}{12} \,\,{\rm
[MeV]}, \,\,\,  \Gamma_\rho = 149\er{18}{7} \,\,{\rm
[MeV]}
\end{equation}
in fairly good agreement with experiment and we also obtain the
threshold parameters reported in Table~\ref{tab:th-par}.

\subsubsection{ Elastic $\pi\pi-$scattering for $s-$ and $d-$waves.}

In the original work of Gasser-Leutwyler~\cite{GL84} and in the
subsequent analysis of Ref.~\cite{bc97} (set {\bf II} in this latter
reference) the parameters ${\bar l}_{1,2}$
are measured through the $d-$wave scattering lengths. Both scattering
lengths and specially the isotensor one, suffer from large
experimental uncertainties~\cite{Pe70} which lead to quite big errors
on ${\bar l}_{1,2}$. Here, we have in Eq.~(\ref{eq:form-res})  a
precise determination of the difference ${\bar l}_1 - {\bar l}_2$,
thus we can combine it with the experimental measurement of $a_{02}$
to determine both parameters ${\bar l}_{1,2}$. In this way we avoid to
use $a_{22}$, whose relative error is about 13 times bigger than that
of $a_{02}$. We get:
\begin{equation}
{\bar l}_1 = -0.5 \pm 0.5,  \,\,\,\, {\bar l}_2 = 5.6 \pm 0.5 \label{eq:l12}
\end{equation} 
where we have used $m^5\,a_{02} = (17 \pm
3)\cdot10^{-4}$~\cite{Pe70}. The error on the scattering length
dominates by far the errors on ${\bar l}_{1,2}$. Fixing ${\bar l}_3$
from the $SU(3)$ mass formulae~\cite{GL84},
\begin{equation}
{\bar l}_3 = 2.9 \pm 2.4 \label{eq:l3}
\end{equation} 
to complete the analysis of the $\pi\pi-$scattering at next-to-leading
order in our expansion (Eq.~(\ref{eq:nl})), we still have four
undetermined parameters ( $C_{IJ},\,$ with $IJ=00,20,02,$ and $22$).
We determine these four parameters from four independent best fits to
the data, fixing the ${\bar l}_i$ parameters to the central values
given in Eqs.~(\ref{eq:l4}),(\ref{eq:l12}) and~(\ref{eq:l3}). Results
are reported in Table~\ref{tab:on-fit}. There, we give $C_{IJ}$ and
their statistical errors which have been obtained by increasing the
value of the corresponding $\chi^2$ by one unit. In addition to these
statistical errors, we have some systematic errors due to
uncertainties in the one loop parameters ${\bar l}_i$. One way of
taking into account both types of errors would be to perform a
simultaneous eight-parameter ($C_{IJ},\,$ with $IJ=00,20,02,$ and $22$
and ${\bar l}_{1,2,3,4}$) fit to data. However and due to the limited
quality of the available experimental data and the multidimensional
character of the fit, it is very difficult to single out a stable
minimum advising against carrying out this procedure directly. Thus,
we have designed an alternative strategy to estimate the systematic
uncertainties in the $C's$ parameters induced by the errors in the one
loop parameters. We fix central and statistical errors of the
parameters ${\bar l}_{1,2,3,4}$ to the values quoted in
Eqs.~(\ref{eq:l4}),(\ref{eq:l12}) and~(\ref{eq:l3}), and generate a
sufficiently large sample of low energy parameters ${\bar
l}_{1,2,3,4}$ randomly distributed according to a Gaussian. For each
set of four ${\bar l}$'s parameters we fit, in each angular momentum
and isospin channel, the $C_{IJ}$ parameters to the experimental phase
shifts. In this way we generate distributions for each of the $C$'s
parameters. For the central values of the $C$'s parameters we take the
results of the fits obtained with the central values of
Eqs.~(\ref{eq:l4}),(\ref{eq:l12}) and~(\ref{eq:l3})\footnote{This is
not the same as computing the mean of the generated
distribution. Indeed, the mean values in general can not be obtained
from any specific choice of the input parameters, in this case the
${\bar l}$'s. This is the reason to quote central values from results
obtained from central values of the input parameters. Furthermore, the
statistical errors obtained before by changing the $\chi^2$ by one
unit, are referred to the values resulting from best fits to the data
with input parameters fixed to their central values.  Both choices do
not differ much as long as we are dealing with relatively small
asymmetries in the distributions. We should stress that, what it is
statistically significant is the 68\% error band rather than the
choice of the central value.}. The 68\% confidence limits of each of
the distributions give an estimate of the systematic errors in the
$C$'s parameters. Since the out-coming $C$'s parameter distributions
are, in general, not Gaussian we end up with asymmetric errors.
Results of this procedure are also reported in Table~\ref{tab:on-fit}.

A final detail concerns the choice 610 MeV as the energy cut in the
scalar-isoscalar channel. This is justified to avoid any possible
contamination from the $K\bar K$ channel in the sub-threshold region
or from higher chiral orders, not included in neither case, at this
level of approximation.

\begin{figure}


\begin{center}                                                                
\leavevmode
\epsfysize = 650pt
\makebox[0cm]{\epsfbox{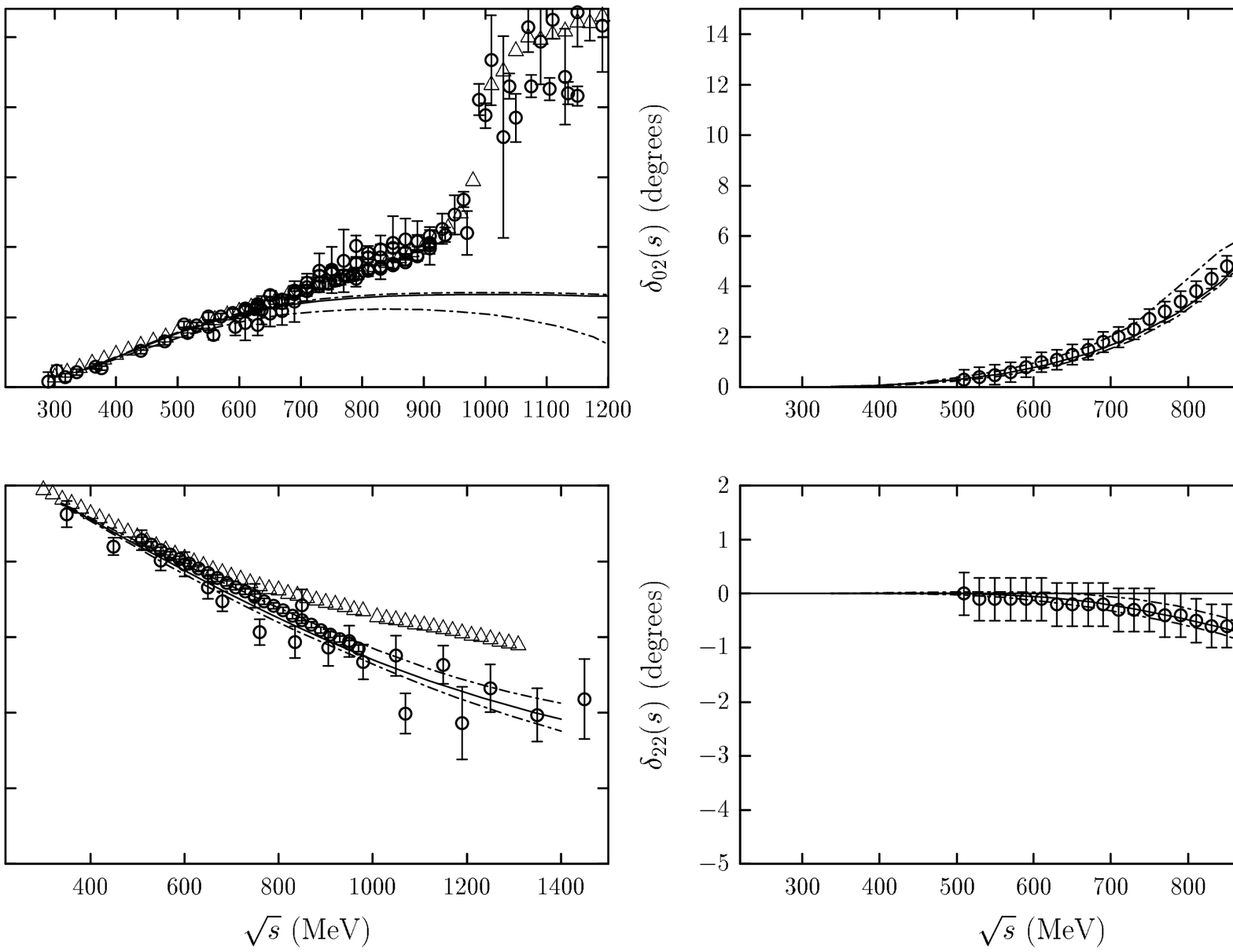}}
\end{center}
\vspace{-8.5cm}
\caption[pepe]{\footnotesize Several $\pi\pi$ phase shifts as a function
of the total CM energy $\protect\sqrt s$ obtained within our
next-to-leading approximation (see Eq.~(\protect\ref{eq:nl})). The top
(bottom) panels correspond to $I=0 \,(2)$ whereas the left (right)
panels correspond to $J=0\, (2)$.  Experimental data are taken from
Refs.~\protect\cite{pa73} --\protect\cite{klr97} ($I=J=0$) from
Refs.~\protect\cite{ho77} and~\protect\cite{lo74} ($I=2,\,J=0$) and
finally from Ref.~\protect\cite{em74} for the $d-$wave
channels. Triangles stand for the Frogatt and Petersen phase-shifts
(Ref.~\protect\cite{fp77}) with no errors due to the lack of error
estimates in the original analysis. Solid lines have been obtained
with the central values quoted in Table~\protect\ref{tab:on-fit} and
in Eqs.~(\ref{eq:l4}),(\ref{eq:l12}) and~(\ref{eq:l3}).  Dashed lines
show our total uncertainties, statistical and systematic, added in
quadratures, see text for more details.}
\label{fig:on-fit}
\end{figure}

Solid lines in Fig.~\ref{fig:on-fit} are the predictions for the phase
shifts in each angular momentum-isospin channel. At each fixed CM
energy, our prediction for the phase shift will suffer from
uncertainties due to both the statistical errors on ${\bar l}_i$ and
$C_{IJ}$ parameters and also the systematic errors on the latter ones
induced by the statistical fluctuations of the former ones, as we
discussed above. We use a Monte Carlo simulation to propagate
independently both
sources of errors and finally we add in quadratures, respecting the
possible asymmetries, the statistical and systematic phase shift
errors. The dashed lines in Fig.~\ref{fig:on-fit} join the central
value for the phase-shift plus its upper total error or minus its
lower total error for every CM energy. We see that, in general, the
agreement with experiment is good except for the scalar--isoscalar
channel, where one clearly sees that there is room for contributions
due to the $K\bar K$ inelastic channel above the energy cut (610 MeV)
used for the best fit.

\begin{table}[t]
\begin{center}
\begin{tabular}{c|cccc}
$IJ$ & 00 & 20 & 02  & 22  \\\hline\tstrut $C_{IJ}$ & $-0.022\er{0.001}{0.001}
\er{0.005}{0.003} $ & $-0.058\pm 0.002\er{0.025}{0.002}$ & $-0.203\pm
0.003\er{0.080}{0.300}$& $-0.5\pm 0.2\er{0.5}{0.3}$\\ $\chi^2 /{\rm
num.\, data}$ & 34.7 / 20 & 20.5 / 21 & 11.4 / 24 & 0.4 / 24
\end{tabular}
\end{center}
\caption[pepe]{\footnotesize Next-to-leading (Eq.~(\protect\ref{eq:nl}))
parameters fitted to the experimental data of
Refs.~\protect\cite{pa73} --\protect\cite{klr97} (with an energy cut of
$610$ MeV) for $I=J=0$, of Refs.~\protect\cite{ho77}
and~\protect\cite{lo74} for $I=2,\,J=0$ and finally of
Ref.~\protect\cite{em74} (with an energy cut of $970$ MeV) for the
$d-$wave channels.  In the latter case and due to the lack of error
estimates in the original references, we have assumed an error of 0.4
in all phase shifts, this assumes that the errors of the $d-$wave phase-shifts
given in Ref.~\protect\cite{em74} , affect only to the last digit. The
central values and statistical errors of the parameters ${\bar l}_{1,2,3,4}$
have been fixed to the values quoted in
Eqs.~(\ref{eq:l4}),(\ref{eq:l12}) and~(\ref{eq:l3}).  We give two sets
of errors in the fitted parameters. The first set corresponds to statistical 
errors and have been obtained by increasing the value of the corresponding
$\chi^2$ by one unit. The second set are systematic, induced by the
uncertainties in ${\bar l}_{1,2,3,4}$, and have been estimated by means
of a Monte Carlo simulation, see text for details.}
\label{tab:on-fit}
\end{table}

\begin{table}[t]
\begin{center}
\begin{tabular}{c|ccccc}
 $IJ$  & $00$ & $11$ & $20$ & $02$ & $22$
\\\hline\tstrut $m^{2J+1} a_{IJ}$ & $0.216 \er{0.004}{0.006} $ & $0.0361 \pm
0.0003$ & $-0.0418 \pm 0.0013 $ & input & $0.00028 \pm
0.00012 $ \\\tstrut 
(exp) & $0.26 \pm 0.05$ &  $0.038 \pm 0.002$ &  $-0.028 \pm 0.012$  
&  $0.0017 \pm 0.0003$ & $0.00013 \pm 0.00030$ \tstrut\\\hline
$m^{2J+3} b_{IJ}$ & $0.284 \er{0.009}{0.014} $ & $0.0063
\pm 0.0005$ & $-0.075 \er{0.002}{0.003} $ 
&$-\frac{481}{201600}\frac{m^4}{\pi^3f^4}$ 
& $-\frac{277}{201600}\frac{m^4}{\pi^3f^4}$ \\\tstrut
(exp) & $0.25 \pm 0.03$ &   &  $-0.082 \pm 0.008$  
&  & \\
\end{tabular}
\end{center}
\caption[pepe]{\footnotesize Threshold parameters $a_{IJ}$ and
$b_{IJ}$ (obtained from an expansion of the scattering
amplitude~\protect\cite{dgh92}, Re$T_{IJ} = -16\pi m (s/4 -m^2)^J [
a_{IJ} + b_{IJ} (s/4 -m^2) + \cdots ]$ close to threshold) deduced
from the ${\cal O}(p^4)$ results reported in
Table.~\protect\ref{tab:on-fit} and
Eq.~(\protect\ref{eq:form-res}). For all channels, except for the
$I=J=1$ one, the errors have been obtained by adding in quadratures
those induced by the uncertainties in the ${\bar l}$'s parameters and
those induced by the experimental errors in the $\pi\pi$
phase-shifts. For the vector-isovector channel, the error treatment is
simpler thanks to the simultaneous determination of ${\bar l}_1 -
{\bar l}_2$ and $C_{11}$ from a best fit to data and to the negligible
effect of the error bars of ${\bar l}_4$ on the latter fit. Thus in
this channel, uncertainties have been obtained by assuming
uncorrelated Gaussian distributed errors in the parameters quoted in
Eqs.~(\protect\ref{eq:l4}) and~(\protect\ref{eq:form-res}). We also
give the known experimental values compiled in
Ref.~\protect\cite{Du83}. }
\label{tab:th-par}
\end{table}

In Table~\ref{tab:th-par} we also give the deduced threshold
parameters.  We propagate both types of errors discussed above to
these observables, by means of a Monte Carlo simulation, and finally
we add both errors in quadratures. In general, for those parameters which
have been measured, we find agreement within experimental
uncertainties, though our final errors are always significantly
smaller than the experimental ones compiled in Ref.~\cite{Du83}.

\subsubsection{ Numerical comparison with two loop ChPT.}
\begin{table}[t]
\begin{center}
\begin{tabular}{cc|ccc|c|c}
& Set & ${\rm (tree)}$ &  ${\rm+  (1 loop)}$&  ${\rm+ (2loop)}$
&  ${\rm total }$&  ${\rm experiment }$
\\\hline\tstrut
 &{\bf I} & $   $ & $ 0.044 \pm 0.005$ & $ 0.016 \pm 0.003 $
               & $ 0.216 \pm 0.009$ &  \\
$a_{00}$ &{\bf II} &$ 0.156  $ & $ 0.039 \pm 0.008$ & $ 0.013 \pm 0.003$
               & $ 0.208 \pm 0.011$ & $ 0.26 \pm 0.05 $ \\
        & BSE &  & $ 0.045 \pm 0.004$ & $ ~~0.011 ~\er{~0.001}{~0.002} $ & $ 0.216 \er{0.004}{0.006}$ &  \\
\hline\tstrut
 &{\bf I} & $   $ & $ 0.069 \pm 0.010$ & $ 0.027 \pm 0.007 $
               & $ 0.275 \pm 0.016$ &  \\
$b_{00}$ &{\bf II} & $ 0.179  $ & $ 0.059 \pm 0.024$ & $ 0.019 \pm 0.011 $
               & $ 0.256 \pm 0.034$ & $ 0.25 \pm 0.03 $ \\
& BSE &  & $ 0.072 \pm 0.009$ & $ 0.006 \er{0.005}{0.009} $
               & $ 0.284 \er{0.009}{0.014}$ &  \\
\hline\tstrut

 &{\bf I} &  & $ 0.073 \pm 0.010$ & $ 0.025 \pm 0.006$
               & $ 0.395 \pm 0.014$ &  \\
$10 \cdot  a_{11}$ 
&{\bf II} & $ 0.297  $ & $ 0.058 \pm 0.033$ & $ 0.018 \pm 0.005$
               & $ 0.374 \pm 0.034$ & $ 0.38 \pm 0.02$ \\
& BSE &  & $ 0.064 \pm 0.003$ & $ 0 $
               & $ 0.361 \pm 0.003$ &  \\
\hline\tstrut
   &{\bf I} & $     $ & $ 0.048 \pm 0.006$ & $
0.031~\er{~0.005}{~0.007}$
               & $ 0.080 ~\er{~0.007}{~0.009}$ &  \\
$10 \cdot
 b_{11}$ & {\bf II}& $  0    $ & $ 0.034 \pm 0.033$ & 
$ 0.020 ~\er{~0.005}{~0.008}$
               & $ 0.054 \pm 0.029$ & $  -   $ \\
& BSE &  & $ 0.0389 \pm 0.0017$ & 
$ 0.021 \pm 0.003 $
               & $ 0.063 \pm 0.005$ & \\
\hline\tstrut

  &{\bf I} &  & $ 0.028 \pm 0.018 $ & $ 0.004 \pm 0.002$
               & $-0.414 \pm 0.020 $ &  \\
$10 \cdot
 a_{20}$ &{\bf II}& $ -0.446  $ & $ 0.008 \pm 0.031 $ & 
$ 0.000~\er{~0.002}{~0.003}$
               & $-0.438 \pm 0.032 $ & $-0.28 \pm 0.12$ \\
&BSE& & $ 0.028 \pm 0.016 $ & 
$ 0.000~\er{~0.003}{~0.004}$
               & $-0.418 \pm 0.013 $ & \\
\hline\tstrut
 &{\bf I} & & $ 0.17 \pm 0.04$ & 
$ 0.01\pm 0.01 $
               & $ -0.72 \pm 0.04$ &  \\
$10 \cdot
 b_{20}$ &{\bf II}& $-0.89  $ & $ 0.10 \pm 0.05$ & $ 0.00 \pm 0.01 $
               & $ -0.79\pm 0.05 $& $ -0.82 \pm 0.08 $ \\
&BSE& & $ 0.16 \pm 0.04$ & $ 0.01 \er{0.01}{0.02} $
               & $ -0.75\er{0.02}{0.03} $&  \\
\hline\tstrut
&{\bf I}   & & $  0.181 \pm 0.025 $ & $ 0.079 \pm 0.016$
               & $  0.260 \pm 0.036  $ &  \\
$10^2 \cdot
 a_{02}$ &{\bf II}  & $  0  $ & $  0.117 \pm 0.026 $ & $ 0.053 \pm 0.018$
               & $  0.170 \pm 0.030  $ & $ 0.17 \pm 0.03 $ \\
&BSE  & & $  0.170 \pm 0.030 $ & $ 0 $
               & input &  \\
\hline\tstrut
 &{\bf I}& & $  0.21 \pm 0.13 $ & $-0.01~\er{~0.06}{~0.04} $
               & $  0.20 \pm 0.10 $ & \\
$10^3 \cdot
 a_{22}$ &{\bf II}& $0$& $  0.12\pm 0.44 $ & $ 0.01~\er{~0.17}{~0.12} $
               & $  0.13\pm 0.30 $ & $ 0.13 \pm 0.30$ \\
&BSE& & $  0.28\pm 0.12 $ & $ 0$
               & $  0.28\pm 0.12 $ &  \\
\end{tabular}
\end{center}
\caption[pepe]{\footnotesize Separation in powers of $1/f^2$ of some
of the threshold parameters reported in
Table~\protect\ref{tab:th-par}. For comparison we have also compiled
the results reported in Table 1 of Ref.~\protect\cite{ej99_2} which
are obtained from the two-loop analysis of Ref.~\protect\cite{bc97}
supplemented with proper error estimates. The Sets {\bf I } and {\bf
II } refer to those define in Ref.~\protect\cite{bc97}.  The authors
of Ref.~\protect\cite{bc97} assume some resonance saturation to give
numerical values to the ${\cal O}(p^6)$ parameters at certain scale
around $750$ MeV (see that reference for details).  In addition, Set
{\bf I} uses: $ \bar l_1 = -1.7 \pm 1.0 \, , \bar l_2 = 6.1 \pm 0.5 \,
, \bar l_3 = 2.9 \pm 2.4 \, , \bar l_4 = 4.3 \pm 0.9 $ and Set {\bf
II} uses: $ \bar l_1 = -0.8 \pm 4.8 \, , \bar l_2 = 4.45 \pm 1.1 \, ,
\bar l_3 = 2.9 \pm 2.4 \, , \bar l_4 = 4.3 \pm 0.9 $. On the other
hand, BSE stands for the results of the present work at
next-to-leading order, with parameters given in
Table~\protect\ref{tab:on-fit} and
Eqs.~(\ref{eq:l4}),~(\ref{eq:form-res}), (\ref{eq:l12}),
and~(\ref{eq:l3}).  Tree, one-loop, two-loops and total labels stand
for the threshold parameters calculated at ${\cal O}(p^2)$, ${\cal
O}(p^4)$, ${\cal O}(p^6)$ and all orders, respectively. We also give
the known experimental values compiled in
Ref.~\protect\cite{Du83}. All threshold parameters are given in pion
mass units. Note, that due to correlations between the several orders,
in the $1/f^2$ expansion, contributing to the threshold observables,
the errors of the total quantities can not be simply added in
quadratures; we use a Monte Carlo simulation to propagate errors.}
\label{tab:two-loop}
\end{table}
A very interesting feature of the present treatment at next-to-leading
order is that having fitted the $C$'s to the data, we can use those
parameters to learn about the two and even higher loop contributions
(see Eq.~(\ref{eq:next-order})). The separation of the amplitudes near
threshold, in powers of $1/f^2$, is presented in
Table~\ref{tab:two-loop}, and compared to recent two loop ChPT
calculations~\cite{bc97} supplemented with proper error
estimates~\cite{ej99_2}. Despite of the resonance saturation
hypothesis\footnote{Such an approximation reduces a priori the errors
on the two loop contributions with respect to the present framework
where these higher order corrections have to be fitted directly to
experimental data.} assumed in Refs.~\cite{bc97,ej99_2}, in general we
see that the BSE predictions are not less accurate than those of these
references. Both sets of results, BSE and ChPT, suffer from systematic
errors induced by the higher order contributions. Those are not
included in either case.

\subsubsection{ Scalar pion form factor.} 

\begin{figure}
\begin{center}                                                                
\leavevmode
\epsfysize = 650pt
\makebox[0cm]{\epsfbox{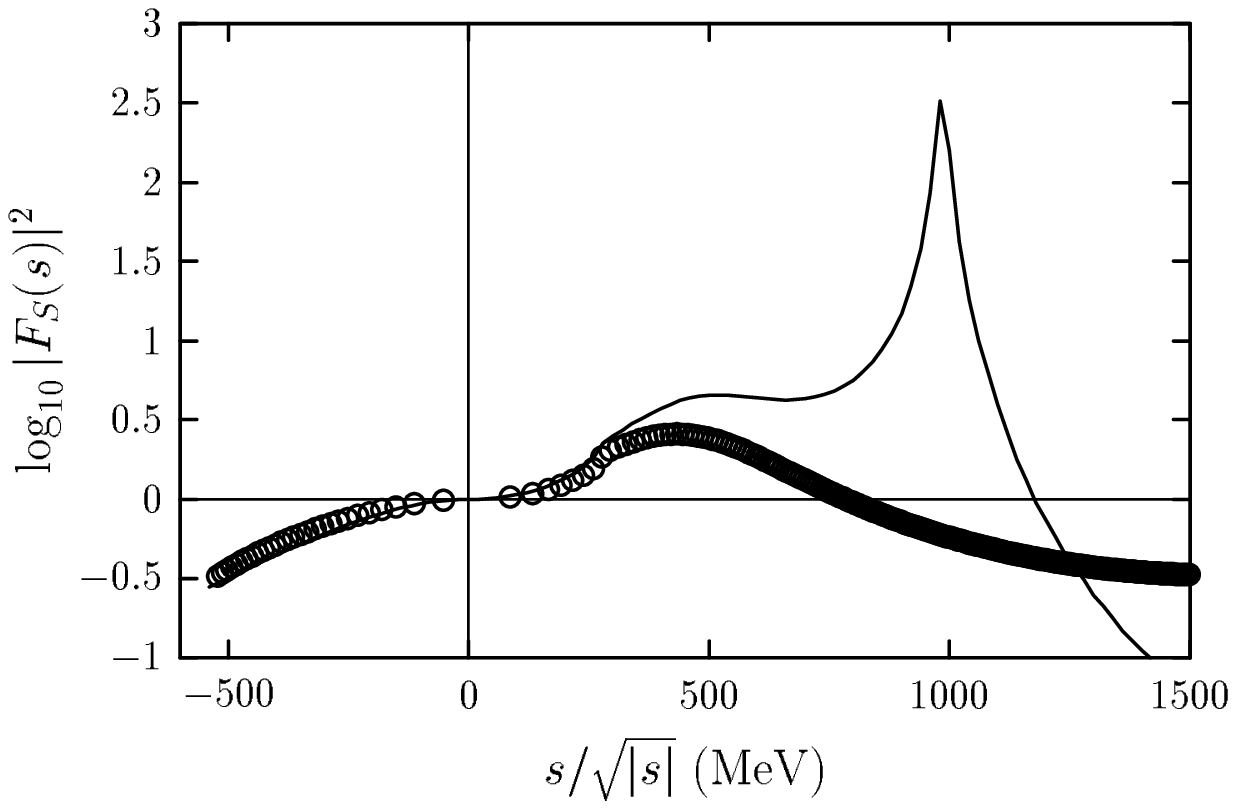}}
\end{center}
\vspace{-14.5cm}
\caption[pepe]{\footnotesize On-shell next-to-leading order BSE
(Eqs.~(\protect\ref{eq:fm_esc}) and~(\protect\ref{eq:nl})) scalar
form factor (circles) as a function of $s$. The used parameters are
those given in Eqs.~(\protect\ref{eq:l4}),(\protect\ref{eq:l12}) 
and~(\protect\ref{eq:l3}) and
Table~\ref{tab:on-fit}.  For comparison we also show the 
results of Refs.~\protect\cite{GO99} (solid line). }
\label{fig:form_esc}
\end{figure}

To end this subsection, in Fig.~\ref{fig:form_esc}, we present results
for the scalar form-factor (Eq.~(\ref{eq:fm_esc})), normalized to one
at $s=0$ and computed at next-to-leading order accuracy, within our
approach. It is to say, it uses, $V_{00}(s)$ given in
Eq.~(\ref{eq:nl}) with the parameters presented in
Eqs.~(\ref{eq:l4}),(\ref{eq:l12}) and~(\ref{eq:l3}) and
Table~\ref{tab:on-fit}. We also compare our results with those obtained
in Ref.~\cite{GO99}. Our results disagree with those
of Ref.~\cite{GO99} above 500 MeV, presumably due to the role played
by the sub-threshold $K\bar K$ effects.

\subsection{Comparison of leading and next-to-leading on--shell 
BSE predictions.}
\label{sec:compari}
\begin{figure}
\begin{center}                                                                
\leavevmode
\epsfysize = 650pt
\makebox[0cm]{\epsfbox{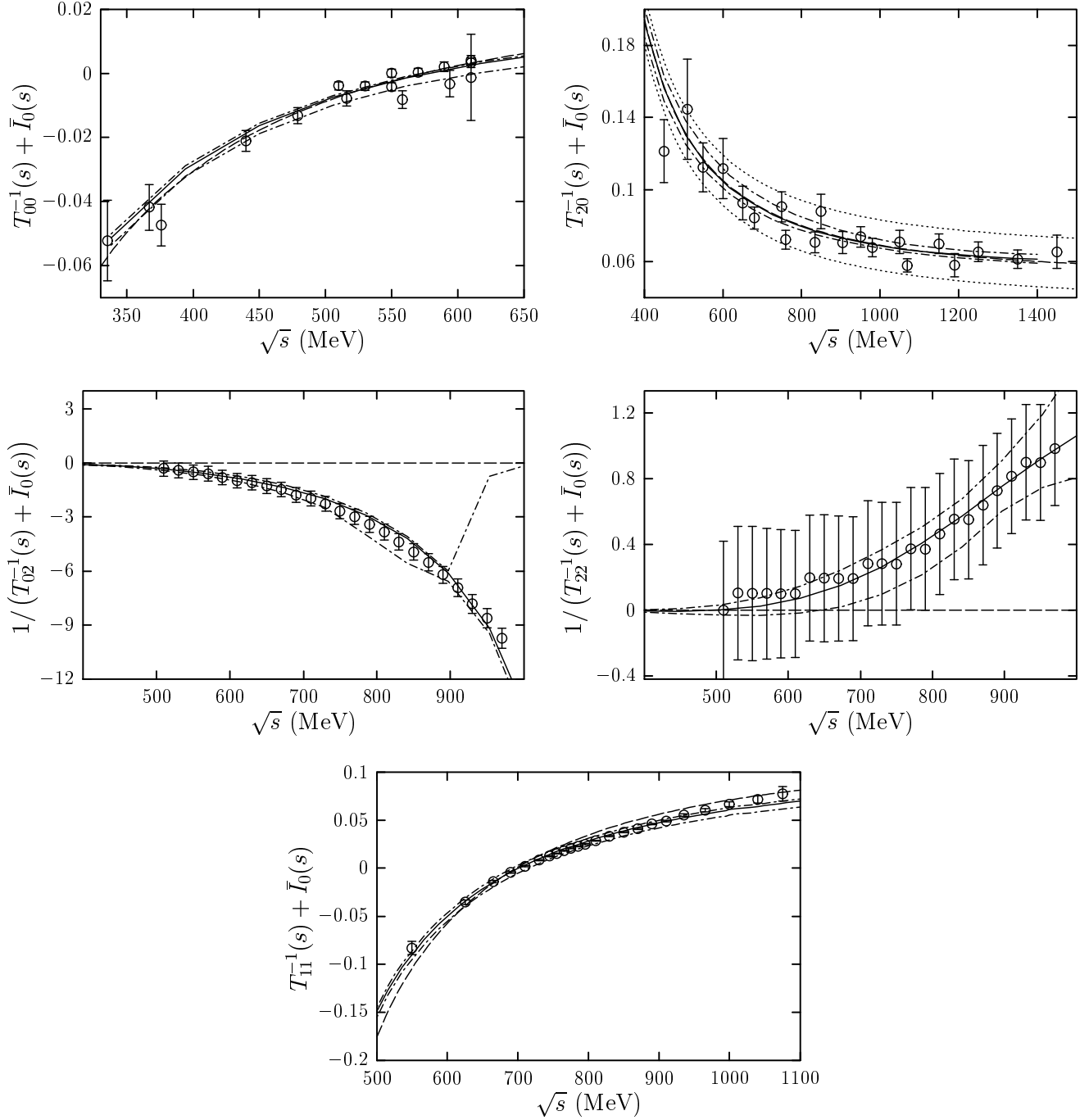}}
\end{center}
\vspace{-2.5cm}
\caption[pepe]{\footnotesize Experimental (circles), BSE--leading
(dashed lines) and --next-to-leading (solid lines) results, together with
both experimental and theoretical (dash-dotted and dotted lines for leading
and next-to-leading predictions respectively) error estimates, for
$W_{00}^{-1}, W_{20}^{-1}, W_{02},W_{22}$ and $W_{11}^{-1}$ as a
function of the CM energy. In some cases the theoretical errors can not
be seen due to their smallness. See Subsect.~\protect\ref{sec:compari}
for more details.}
\label{fig:comp}
\end{figure}

 In this subsection we compare results for elastic $\pi\pi-$scattering
 at leading and next-to-leading accuracy within the on-shell BSE
 scheme. This kind of comparison is not generally undertaken in the
 literature regarding unitarization methods, where the main goal is
 just to fit the data, rather than study also the convergence of the
 expansion. Since in our framework, the potential plays the central
 role, such a comparison is naturally done in terms of it or in terms
 of the function $W_{IJ}(s) = T^{-1}_{IJ }(s) + {\bar I}_0(s)$ defined
 in Eq.~(\ref{eq:defW}). These $W_{IJ}$ functions can be directly
 extracted from data\footnote{Note, the functions $V_{IJ}$ can not
 directly be obtained from experiment, because of the unknown
 constants $C_{IJ}$.}, in order words, they can be derived from
 experimental phase-shifts.  In Fig.~\ref{fig:comp} we present
 experimental, BSE--leading and --next-to-leading
 results\footnote{Note, that because the presence of resonances in the
 $\sigma-$ and $\rho-$channels, the corresponding $W-$functions should
 have a pole (see discussion related to Fig.~\ref{fig:pole}), so we
 present the inverse for these channels and also, for a better
 comparison, for the isotensor-scalar one. On the other hand for $d-$
 waves the lowest BSE order approximation gives $T=0$, and therefore
 for those channels we present $W$ itself.}, together with both
 experimental and theoretical error estimates, for $W_{00}^{-1},
 W_{20}^{-1},$ $W_{02},W_{22}$
 \noindent  and $W_{11}^{-1}$.  Next-to-leading
 results are obtained from Eq.~(\ref{eq:nl}). Central values and
 uncertainties of the needed low energy constants were discussed in
 Subsect.~\ref{sec:num-res-on} (Table~\protect\ref{tab:on-fit} and
 Eqs.~(\ref{eq:l4}),~(\ref{eq:form-res}), (\ref{eq:l12}),
 and~(\ref{eq:l3})). Errors have been propagated via a MonteCarlo
 simulation and statistical and systematic errors have been added in
 quadratures.  Leading results have been obtained from
 Eq.~(\ref{eq:leading}) where the unknown low energy constants
 $C_{00}$, $C_{11}$ and $C_{20}$ have been obtained from three one
 parameter best fits to the experimental data, since at lowest order
 the three channels are independent. For the $s-$wave
 channels we have fitted the same set of data and use the same energy
 cuts as in Table~\protect\ref{tab:on-fit} and
 Fig.~\ref{fig:on-fit}. We have obtained 
\begin{eqnarray}
C_{00} = -0.0273 \pm 0.0004 &\qquad& \chi^2 / {\rm num.\, data} = 31.2
/20 \\\nonumber
&&\\\nonumber
C_{20} = -0.051 \pm 0.014 &\qquad& \chi^2 / {\rm num.\, data} = 21.4
/21
\end{eqnarray}
For these two channels both leading and next-to-leading approximations
provide similar descriptions of the data. For the $\rho-$channel the
situation is different. We have fitted $C_{11}$ to the data of
Ref.~\cite{pa73}. Fits from threshold up to 0.9 GeV give values of
$\chi^2/dof$ of the order of 12, being then highly unlikely that
such a big discrepancy between theory and data is due to statistical
fluctuations, disqualifying any error analysis. We find significantly
lower values of $\chi^2/dof$ for smaller energy cuts, being the optimum
choice obtained with a energy cut of about 690 MeV, for which we find:
\begin{eqnarray}
C_{11} = -0.1275 \pm 0.0006 &\qquad& \chi^2 /dof = 4.1
\end{eqnarray} 
The above error is quite small, due to large value of $\chi^2$. 
Leading and next-to-leading estimates for all three parameters 
$C_{00},C_{20}$  and $C_{11}$ are compatible within two sigmas.

Results from Fig.~\ref{fig:comp} show a good rate of convergence of the
expansion proposed in this work.

\section{The non-perturbative nature of the $C_{IJ}$ parameters. }
\label{sec:c_non}

In this section we link the renormalization $C_{IJ}$ parameters to
some non-perturbative effects of the underlying theory, such as the  
the existence of resonances. 

For simplicity, let us consider the $\rho-$channel ($I=J=1$) for which
the lowest order inverse amplitude, within the on-shell scheme 
(Eq.~(\ref{eq:leading})), reads
\begin{eqnarray}
T_{11}^{-1}(s)-T_{11}^{-1}(-\mu^2) &=& - \left ( {\bar I}_0(s) - {\bar
I}_0(-\mu^2)  \right )
 - 6 f^2 \left ( \frac{1}{s-4m^2}+\frac{1}{\mu^2+4m^2}\right) 
\end{eqnarray}
where we have chosen the subtraction point $s_0 =
-\mu^2 $ . From above, we identify $C_{11}$ by
\begin{eqnarray}
C_{11} = -T_{11}^{-1}(-\mu^2) - {\bar I}_0(-\mu^2) + \frac{6f^2}{\mu^2+4m^2} 
\label{eq:c_pwave}
\end{eqnarray}
which should be independent of the scale $\mu^2$. Thus we have
\begin{eqnarray}
T_{11}^{-1}(s) &=& - \left ( {\bar I}_0(s) + C_{11} \right )
 -  \frac{6 f^2}{s-4m^2} 
\end{eqnarray}
At the resonance, $s=m_\rho^2 $, $ {\rm Re}\, T_{11} (m_\rho^2 )^{-1}
=0$, and thus if we look at the real parts, we have
\begin{eqnarray}
C_{11} &=& -{\rm Re}\,{\bar I}_0(m_\rho^2) - \frac{6f^2}{m_\rho^2 -
4m^2}\nonumber \\ 
&=& -\frac{1}{16\pi^2}\log\frac{m_\rho^2}{m^2} -
\frac{6f^2}{m_\rho^2} + {\cal O}(\frac{m^2}{m_\rho^2})\\\nonumber 
&\approx & -0.110
\end{eqnarray}
in qualitative good agreement with the findings of previous sections
(see, for instance, the $I_0^R$ entries for this channel in
Table~\ref{tab:off-res} or Eq.~(\ref{eq:form-res})).

Similarly for the scalar-isoscalar channel,
\begin{eqnarray}
T_{00}^{-1}(s)-T_{00}^{-1}(-\mu^2) &=& - \left ( {\bar I}_0(s) - {\bar
I}_0(-\mu^2)  \right )
 -  f^2 \left ( \frac{1}{s-m^2/2}+\frac{1}{\mu^2+m^2/2}\right) 
\end{eqnarray}
where we have chosen the subtraction point $s_0 =
-\mu^2 $ and it can be different for each $IJ$ channel since we may take
$ T_{IJ} (-\mu^2 ) $ as the fitting and {\it a priori} unknown parameter.
As before, we can identify $C_{00}$ 
\begin{eqnarray}
C_{00} &=& -T_{00}^{-1}(-\mu^2) - {\bar I}_0(-\mu^2) + \frac{f^2}{\mu^2+m^2/2} 
\label{eq:c_swave}
\end{eqnarray}
From all our previous discussions and fits it is rather clear that in the 
$I=J=0$ channel the $\sigma$ resonance cannot be completely understood without 
inclusion of sub-threshold $K \bar K$ contributions. Nevertheless, let us 
define the point $s=m_\sigma $ as the one  fulfilling  
$ {\rm Re} T_{00} (m_\sigma^2 )^{-1} =0$, yielding 
\begin{eqnarray}
C_{00} &=& -{\rm Re}\,{\bar I}_0(m_\sigma^2) - \frac{f^2}{m_\sigma^2 -
m^2/2}\nonumber \\ 
&=& -\frac{1}{16\pi^2}\log\frac{m_\sigma^2}{m^2} -
\frac{f^2}{m_\sigma^2} + {\cal O}(\frac{m^2}{m_\sigma^2})\\\nonumber 
&\approx & -0.043
\end{eqnarray}
for $m_\sigma = 600$ MeV. This is not completely unreasonable since
the phase shift at about $s=600 {\rm MeV}$ go up to $60^o$, which is
equivalent to take $ {\rm Re}\, T_{00} (m_\sigma^2 )^{-1} $
small. From this discussion, we understand the origin of the
significant difference of size between $C_{11}$ and $C_{00}$. Indeed,
neglecting the logarithmic contributions, 
\begin{equation}
\frac{C_{11}}{C_{00}} \approx 6 \frac{m_\sigma^2}{m_\rho^2}
\end{equation}
which clearly points out that the reason for the different size
between the two $C$ parameters relies on the fact that the potential
in the $\rho-$channel is about six times smaller than that in the
scalar-isoscalar one. This has also been pointed out by the authors of
Ref.~\cite{oo98} who also use this argument to stress the dominant
role played by unitarization in the $\sigma$ channel. The authors of
this same work, also claim that the quite different values of $C_{00}$
and $C_{11}$ could not be understood if the same scale or cutoff, as
they used in a previous work~\cite{O98}, are used for both
channels. Indeed, if one uses dimensional regularization to compute
the divergent integral $I_0(s)$ and in the $\overline {{\rm MS}}$
scheme, one gets the finite piece
\begin{equation}
I^D_0(s) = \frac{1}{16\pi^2}\left\{ -2 + \log \frac{m^2}{\nu^2}\right
\} + {\bar I_0} (s) 
\end{equation}
being $\nu$ some renormalization scale. Thus, one might identify 
\begin{equation}
C = \frac{1}{16\pi^2}\left\{ -2 + \log \frac{m^2}{\nu^2}\right
\} \label{eq:cdim} 
\end{equation}
for all angular momentum and isospin channels. The above equation
gives values for the $C$ parameter of about $-0.04$ when scales, $\nu$,
of the order 1 GeV are used. This agrees reasonably well with the values found
for $C_{00}$, but to get values of about $-0.11$, as it is the case for
$C_{11}$, one needs unrealistic scales or cut-offs of the order of 300
GeV. All confusion, comes from the interpretation of the  $C$ parameter
as a renormalization of the bubble divergent integral $I_0(s)$,
instead of a renormalization of the full amplitude once that a
renormalized potential is iterated by means of the BSE. A direct,
comparison of Eqs.~(\ref{eq:cdim}) and~(\ref{eq:c_pwave})
or~(\ref{eq:c_swave}) shows that the {\it free} parameters $ T_{IJ}
(-\mu^2 ) $ allow for the use of the same value of the scale $\mu$ to
generate such different values for the $C_{IJ}$ parameters in the
scalar-isoscalar and vector-isovector channels. Furthermore, as we
said above, the $C$ parameters defined in Eq.~(\ref{eq:c_pwave})
or~(\ref{eq:c_swave}) do not depend on scale. On the light of this
discussion, the numerical coincidence between the values provided by
Eq.~(\ref{eq:cdim}) and $C_{00}$ might be considered accidental.

From the former expressions and neglecting corrections of the order of
the ${\cal O}(m^2/m_{\rho,\sigma}^2)$ we find simple expressions for
the amplitudes in the two channels considered in this section,
\begin{eqnarray}
T_{11}^{-1}(s) &\approx& -{1\over 16\pi^2} {\rm log}
(-{s\over m_\rho^2} ) - 6 f^2 ( {1\over s}-{1\over m_\rho^2} ) \nonumber\\
T_{00}^{-1}(s) &\approx& -{1\over 16\pi^2} {\rm log}
(-{s\over m_\sigma^2} ) - f^2 ( {1\over s}-{1\over m_\sigma^2} ) 
\end{eqnarray}
where we take $ {\rm log} (-s/m_R^2 ) = {\rm log} (|s|/m_R^2 )
- {\rm i} \pi \Theta (s) $.  The above expressions lead to
approximate expressions for the decay widths (see Eq.~(\ref{eq:width}))
\begin{eqnarray}
\Gamma_\rho &=& {m_\rho^3 \over 96 \pi f^2} {1\over 1- { m_\rho^2
\over 96\pi^2 f^2}} \\ \Gamma_\sigma &=& {m_\sigma^3 \over 16 \pi^2
f^2} {1\over 1- { m_\sigma^2 \over 16\pi^2 f^2}}
\end{eqnarray}
For the $\rho-$channel, this corresponds to a coupling $ g_{\rho
\pi\pi} \approx (m_\rho / \sqrt{2} f) $, the value predicted by the
KSFR relation \cite{KSFR}. KSFR predicts, for $ m_\rho= 770$ MeV a
width of $\Gamma_\rho = 141 {\rm MeV} $, in excellent agreement with
the experimental value. Such a derivation of the KSFR relation is
actually not completely unexpected since this relation is a
consequence of PCAC plus vector meson dominance through current-field
identities; in our model vector meson dominance is realized, i.e. the
real parts of the inverse scattering amplitude in the $I=J=1$ channel
and of the inverse vector form factor vanish simultaneously, as we
pointed out after Eq.~(\ref{eq:fv}).  The width for the $\sigma$ meson
turns out to be very large making this way of determining it
doubtful. Notice that in both cases the conditions for non negative
widths are $ m_\rho^2 < 96\pi^2 f^2 $ and $ m_\sigma^2 < 16 \pi^2 f^2
$ respectively.

\section{Comparison with other Unitary Approaches}
\label{sec:comp}
As we have hopefully pointed out along this work, the unitarization
generated through a solution of the BSE becomes extremely important to 
describe the resonance region and also
``guides'' the theory in the low energy regime, allowing for a reasonably
accurate description of low energy and threshold parameters.
 
In what follows we want to compare the BSE with some other approaches
pursued in the literature for on-shell amplitudes which also
incorporate elastic unitarity. Some of the methods presented here have
already been discussed, \cite{Tr91}, in the chiral limit. 

All unitarization schemes provide the same imaginary part for the
inverse amplitude and they differ in the way the real part is
approximated. Thus, it is not surprising that formal relations among
them can be found.

\subsection{ K-matrix}

A usual unitarization method consists in working with the 
$K-$matrix, defined as
\begin{eqnarray}
K^{-1}_{IJ} &=& {\rm Re}(T^{-1}_{IJ}) = 
 T^{-1}_{IJ} - i \rho  \nonumber\\
\rho &=& - {\rm Im}{\bar I}_0 (s+{\rm i}\epsilon)
= \frac{\lambda^{\frac12}(s,m^2,m^2)}{16\pi s}
= {1\over 16\pi }  \sqrt{1-{4m^2 \over s}}\label{eq:kmatrix} 
\end{eqnarray}
In the bottom graph of Fig.~\ref{fig:pole}, we plot the inverse of the
$K-$matrix for the isovector $p-$wave channel. The presence of a pole 
(zero in the inverse function) in $K_{11}$ limits the use of any
``perturbative'' expansion of $K_{11}$  in the region of the
resonances and this fact prevents the 
use of this unitarization scheme above threshold when there exists 
a resonance.

\subsection{Dispersive Representation of the Inverse of the Partial Wave
Amplitude.}
\label{sec:disp}

Partial wave amplitudes, $T_{IJ}$, have both right and left hand
cuts. Besides, the only poles that are
allowed on the physical $s$ sheet are the bound states poles, which
occur on the positive real axis below $4m^2$. Poles on the real axis
above threshold violate unitarity, and those on the physical sheet
that are off the real axis violate the Mandelstam
hypothesis~\cite{Ma58} of maximum analyticity. Thus, $T_{IJ}$ and
$T^{-1}_{IJ}$ have similar right and left cuts. However, partial wave
amplitudes have zeros, which in turn means that their inverses have poles
on the physical sheets.

A seemingly significant advantage of considering the dispersive
representation of the inverse of a partial wave amplitude is that its
right hand cut discontinuity is just given by phase space in the
elastic region, thanks to unitarity. The relevance of this property
was already emphasized in Ref.~\cite{pb97}. For simplicity, 
let us suppose that $T_{IJ}$, has a zero of
order 1 for $s=s_A$\footnote{For the $s-$waves, these are demanded
by the Adler condition, for the $p-$waves, the zero is kinematic being
at threshold, $s_A = 4m^2$. In both latter cases, the zeros are of
order one, higher partial waves have higher order zeros.}.

If we define,
\begin{equation}
r = \lim_{s\to s_A} (s-s_A) T^{-1}_{IJ}(s)
\end{equation}
assuming that $|T^{-1}_{IJ}(s)| < s $ as $s\to\infty$ then each
inverse amplitude satisfies a once--subtracted dispersion
relation~\cite{pb97}, with $s=s_1$, the subtraction point (dropping
the labels $I,J$ for simplicity)
\begin{eqnarray}
T^{-1}(s)-T^{-1}(s_1) &=& \left(
\frac{r}{(s-s_A)}-\frac{r}{(s_1-s_A)}\right) + \frac{(s-s_1)}{\pi}
\int_{4m^2}^{\infty}
\frac{ds^{\,\prime}}{(s^{\,\prime}-s_1)(s^{\,\prime}-s)}
\,{\rm Im }T^{-1}(s^{\,\prime}\,)\nonumber\\
&&\nonumber\\
&+ & \frac{(s-s_1)}{\pi} \int_{-\infty}^{0}
\frac{ds^{\,\prime}}{(s^{\,\prime}-s_1)(s^{\,\prime}-s)}
\,{\rm Im }T^{-1}(s^{\,\prime}\,)
\end{eqnarray}
Elastic unitarity (Eq.~(\ref{eq:uni})) allows to perform the integral
over the right hand cut, thus one gets
\begin{eqnarray}
T^{-1}(s)-T^{-1}(s_1) &=& \left(
\frac{r}{(s-s_A)}-\frac{r}{(s_1-s_A)}\right) - \left ( {\bar I}_0
(s) - {\bar I}_0 (s_1)\right) \nonumber\\
&&\nonumber\\
&+ & \frac{(s-s_1)}{\pi} \int_{-\infty}^{0}
\frac{ds^{\,\prime}}{(s^{\,\prime}-s_1)(s^{\,\prime}-s)}
\,{\rm Im }T^{-1}(s^{\,\prime}\,)
\label{eq:inv-disp}
\end{eqnarray}
Then one clearly sees that all the dynamics of the process is
contained in both the poles of $T^{-1}$ (zeros of $T$) and in the left
hand cut. This latter contribution can be split as follows:
\begin{equation}
\frac{(s-s_1)}{\pi} \int_{-\infty}^{0}
\frac{ds^{\,\prime}}{(s^{\,\prime}-s_1)(s^{\,\prime}-s)}
\,{\rm Im }T^{-1}(s^{\,\prime}\,) = \frac{1}{{\cal V}(s)}-
\frac{1}{{\cal V}(s_1)}
\end{equation}
where
\begin{equation}
\frac{1}{{\cal V}(s)} =   \frac{1}{\pi} \int_{-\infty}^{0}
\frac{ds^{\,\prime}}{(s^{\,\prime}-s)}
\,{\rm Im }T^{-1}(s^{\,\prime}\,) + P 
\end{equation}
where $P$ is a renormalization constant, though in general might be a
polynomial. Thus, $T^{-1}(s)$ can be written as
\begin{equation}
T^{-1}(s) = - {\bar I}_0(s) - k + \frac{r}{s-s_A} + {\cal V}^{-1}(s)
\end{equation}
being $k$ a constant. The above equation with the obvious replacements
is identical to Eq.~(\ref{eq:defV}).  As a final remark to this
method, we add that the solution to the dispersion relation is not
unique (any pair of solutions with different values of the constant $k$
satisfy the same dispersion relation), since we may include any zeros
to the inverse amplitude, which become poles in the amplitude. In the
particular case of $\pi\pi$ scattering such poles are inadmissible, so
an additional condition on the constant $k$ has to be imposed to
describe the physics.

\subsection{ Blackenbekler Sugar type equations}

It is interesting to see that many of our considerations are strongly
related to a Blackenbekler-Sugar type equation \cite{Bl66}.
Let us consider the BSE in operator form,
\begin{eqnarray}
T = V + V G_0 T
\end{eqnarray}
As we know the ``potential'' $V$ is the two particle irreducible amputated
Green function. Thus, it does not contribute to the $s$-channel unitarity
cut, i.e. $ {\rm Disc} [V_{pk} (s)] = 0 $ for $s>4m^2$. On the other hand,
the $s$-channel two particle reducible diagrams
do contribute to the unitarity cut, but the part of them which does not
contribute to the discontinuity is not uniquely defined. The idea is to
split the two particle propagator into two parts, one   
containing the elastic unitarity cut and the rest. This separation is
ambiguous, since we can sum to the former any function with no
discontinuity. A practical way to do this is as follows. We write
\begin{eqnarray}
G_q (s) = \Delta (q_+) \Delta (q_-) = {\bar G}_q(s) + g_q (s)
\end{eqnarray}
where we impose that
\begin{eqnarray}
{\rm Disc }[G_q (s)] = {\rm Disc }[{\bar G}_q(s)]=
(-2\pi {\rm i}) \delta^+ \Big[ q_+^2-m^2 \Big]
(-2\pi {\rm i}) \delta^+ \Big[ q_-^2-m^2 \Big],\,\, s> 4m^2
\end{eqnarray}
A solution to this discontinuity equation is\footnote{ Notice the
identity $ (-2\pi i)^2 \delta^+ [ q_+^2-m^2 ] \delta^+ [ q_-^2-m^2 ] =
-2 \pi^2 \delta ( P \cdot q ) \delta ( q^2 - m^2 + \frac{s}{4} )$}
\begin{eqnarray}
{\bar G}_q (s) = -i \pi \frac{\delta (P\cdot q)}{q^2-m^2 + s/4 + i\epsilon}
\end{eqnarray}
This, of course, is not
the only choice; we could equally well add a polynomial to $ {\bar
G}_q (s) $ without changing the discontinuity. Proceeding in
perturbation theory we can see that
\begin{eqnarray}
T=V+ V ({\bar G}_0+g) T = t + t {\bar G}_0 T
\end{eqnarray}
where $ t = V +  V g t = V + V g V + V g V g V + \dots $.

Thus we are finally led
to a equation of the form
\begin{eqnarray}
T_P(p,k) &=& t_P(p,k) + \pi\int\frac{d^4
q}{(2\pi)^4}T_P(q,k) {  \delta ( P \cdot q )  \over q^2 - m^2 +
\frac{s}{4}+ i\epsilon} t_P(p,q)\label{eq:bsu}
\end{eqnarray}
This equation looks like the Blackenbecler-Sugar equation \cite{Bl66}
but with the important difference that instead of the potential $ V_P
(p,k) $ we iterate the reduced amplitude $ t_P (p,k) $ which at lowest order
in $1/f^2 $ coincides with the original potential. This equation
satisfies de unitarity condition of Eq.~(\ref{eq:off-uni}).

If we set the on-shell conditions $ P \cdot p = P \cdot k = 0 $ we have
a closed three dimensional equation for the amplitudes
\footnote{ We have the formula $ {\rm d}^4 q = {1\over 2} \sqrt{
 ( \hat P \cdot q )^2 - q^2 } \, {\rm d} q^2 \, {\rm d} (\hat P \cdot q)
\, {\rm d}^2 \hat q $, where $ \hat q^2 = -1 = - \hat P^2 $ and 
 $ P \cdot \hat q = 0 $.}.

\subsection{ Lippmann-Schwinger Equation}

The Lippmann-Schwinger equation has been employed
recently~Ref.\cite{O97} in successfully describing $s-$wave
meson-meson scattering using a coupled channel
formalism\footnote{Actually, their equation is the BSE, but we comply
with their own notation.}. This approach is also related to ours,
although the way they obtain some results is not fully
unproblematic. The authors of that reference 
compute the $q^0$ integral first, asumming that the contour can be
closed at infinity, which for regular potentials and amplitudes may be
acceptable. Then they introduce a three-momentum cut-off, hence
breaking Lorentz and chiral invariance, and obtaining a three
dimensional two body equation in the CM frame. Though this seems
convenient for practical calculations, the justification for choosing
this particular frame to include the cut-off is doubtful. Similarly to
us, they also find that off-shellness may be ignored but it is not
clear why chiral symmetry seems to play no role, particularly because
their cut-off breaks it. In addition, to justify why the off-shellness
can be ignored, they stick just to the one loop approximation.  In our
case we fully rely on the divergence structure of the non-lineal sigma
model, to all orders in the loop expansion, and on its chiral symmetric
structure and make use of the chirally symmetric dimensional
regularization to establish that the off-shellness can effectively be
incorporated as a renormalization of the two particle irreducible amplitude.
 
Finally, in their finite cut-off framework they get the renormalization
constant $C_{00}$ and all higher order low-energy parameter
contribution to this channel, as a specific function of the
cut-off. That function should be the same for all isospin-angular
momentum channels. This introduces undesired correlations between the 
higher order corrections in different channels which violates the
spirit of ChPT. As a matter of fact, to reproduce the $\rho-$resonance
within their framework requires a cut-off about three orders of
magnitude bigger than that used in the $\sigma-$channel. We have
already illustrated this point in Sect.~\ref{sec:c_non} after
Eq.~(\ref{eq:cdim}). There, we compute the $C$ parameter for the
special case of dimensional regularization, but the discussion is
identical if a three dimensional cut-off regularization is considered
(see Eqs. (A3)-(A8) in the second item of Ref.~\cite{O98}).

\subsection{ Inverse Amplitude Method}

Eq.~(\ref{eq:defV}) provides an exact
solution for the $\pi\pi$ elastic scattering amplitude, thus it is not
fully surprising that the IAM of Ref.~\cite{dp93} can be rederived from it.
The essential point is our lack of knowledge on the two particle
irreducible amplitude, $V_{IJ}$, and the type of expansion proposed
for it.  For simplicity, let us consider the
$\rho-$channel ($I=J=1$). In this channel, we showed in
Subsect.~\ref{sec:app-pot} that if the $C_{11}$ parameter is set to
zero, the potential $V_{11}$ has to diverge before the resonance
energy is reached. That makes hopeless a perturbative expansion for
$V_{11}$, but it suggests an expansion for $1/V_{11}$, which should be
small. Thus, expanding $1/V_{IJ}$ (we come back to the general case
$IJ$) in powers of $1/f^2$ in Eq.~(\ref{eq:tinv_fin}), we get
\begin{eqnarray}
T^{-1}_{IJ}(s)|_{\rm IAM} = f^2 [ T_{IJ}^{(2)} (s)]^{-1} - T_{IJ}^{(4)} (s)
[T_{IJ}^{(2)} (s)]^{-2}
+ \cdots = - {\bar I}_0(s) + \frac{1}{V^{\rm IAM}_{IJ}(s)}
\label{eq:iam}
\end{eqnarray}
where the $C_{11}$ constant cancels out to all orders and $ V(s)^{\rm
IAM}_{IJ} = T^{(2)}_{IJ}/( f^2-\tau_{IJ}^{(4)} / T^{(2)}_{IJ} )
$. If $ T^{(2)}_{IJ} $ has a single non-kinematical Adler zero,
like for $s$-wave scattering, then this zero becomes a double one at the
first order of the approximation, since $ T =
(T^{(2)})^2/(f^2 T^{(2)}-T^{(4)}) $. In other words, $ T^{-1}$ has a
double pole, in contradiction with the dispersion relation for the
inverse scattering amplitude, Eq.~(\ref{eq:inv-disp}), as pointed out
in Ref.~\cite{pb97}.  Notice that in
the BSE the order of the Adler zero is always preserved. If in a
particular channel, a resonance does not exist, one should expect $1/V$
not to be necessarily small and therefore the IAM might present
some limitations. In our BSE framework, a chiral expansion of $V$ is
performed.  Thanks to the inclusion of the renormalization parameter
$C$, $V$ might remain reasonably small in a wide region of
energies. Besides, this scenario also allows us to compute form-factors.

\subsection{ Pade Approximants }
In the Pade method~\cite{dht90}, the series expansion in $1/f^2 $ of the 
amplitude, 
\begin{equation}
T_{IJ}(s)=T_{IJ}^{(2)}(s)/f^2+T_{IJ}^{(4)}(s)/f^4+T_{IJ}^{(6)}(s)/f^6+ \dots  
\end{equation}
truncated at some finite order, $2N$, is rewritten as a rational 
representation, 
\begin{equation}
T_{IJ}(s) \approx T_{IJ}^{[2N-2K,2K]}(s)={P_{IJ}^{2N-2K}(s)\over
Q_{IJ}^{2K}(s) }
\end{equation}
where $K=0, \dots, N $, $ P_{IJ}^{2N-2K}(s) $ and $ Q_{IJ}^{2K}(s) $
are polynomials in $1/f^2$ of degrees $2N-2K$ and $2K$ respectively. 
This approximation is enforced to reproduce the series expansion for
the amplitude and also to satisfy the unitarity condition, Eq.~(\ref{eq:uni}). 
For the particular case $2N=4$, the only acceptable Pade approximant 
is the $[2,2]$, and  the IAM method is recovered. As we have already 
mentioned single non-kinematical Adler zeros are transformed at this level of 
approximation into double ones. In the language of the BSE, the Pade method 
is translated  as a Pade approximant for the potential, for instance 
the $[2,2]$ approximant would be 
\begin{equation}
V_{IJ}^{[2,2]}= { \frac{m^2}{f^2} [V^{(2)}_{IJ}(s)]^2 \over 
V^{(2)}_{IJ}(s)- \frac{m^2}{f^2} V^{(4)}_{IJ}(s) }
\end{equation}
which thanks to Eqs.~(\ref{eq:defV}) and~(\ref{eq:approx_v}) exactly
yields to Eq.(\ref{eq:iam}).

\subsection{ N/D Method }
In the N/D method one starts with the dispersion relation for the
partial wave scattering amplitude~\cite{CM60} (dropping the labels
$I,J$ for simplicity), 
\begin{eqnarray}
T (s) = B (s) + {1\over \pi} \int_{4m^2}^\infty  {\rm d} s' { T (s')
\rho (s') T^* (s') \over s-s'}
\end{eqnarray}
where $\rho(s)$ was defined in Eq.~(\ref{eq:kmatrix}) and $ {\rm Disc}
[B(s)] = 0 $ for $ s > 4m^2 $. $B(s)$ presents a discontinuity for $s
< 0$. The method is based on the following assumptions: $ T(s) = N(s)
D (s)^{-1} $ and the discontinuity conditions
\begin{eqnarray} 
{\rm Disc} [N (s)] &=& D (s) \, {\rm Disc} [T (s)] = D (s) \, {\rm Disc} [B
(s)] \, \qquad \phantom{p} s < 0 \nonumber\\ {\rm Disc} [D (s)] &=& N (s)
\, {\rm Disc} [T^{-1} (s)] = - N(s) 2i \rho(s) \, \qquad s > 4 m^2 \label{eq:nd}
\end{eqnarray} 
Dispersion relations for $D(s)$ and $N(s)$ with suitable subtractions
determine the full amplitude solely in terms of the left hand cut
discontinuity. The point here is that the discontinuity conditions in
Eq.~(\ref{eq:nd}) do not determine uniquely the amplitude, as it also
happened in the approach presented in Subsect.~\ref{sec:disp}. Thus,
one should supplement the discontinuity conditions with further
information if possible. For instance the existence and position of
the CDD (Castillejo, Dalitz and Dyson) poles~\cite{cdd56}, as recently
invoked in Ref.~\cite{oo98} for meson-meson scattering, or of the bound
states might help to find a physical solution from the whole family of
solutions which satisfy the above discontinuity conditions. In any
case, however, the multiplicative structure of the left and right hand
cuts implied by the N/D method seems to contradict the additive cut structure
deduced at two loops~\cite{mksf95,bc97,ssf93}. A more detailed study
of the advantages and limitations of a N/D approach in this context
will be presented elsewhere~\cite{ej99-3}.  We note here, that the BSE
preserves that additive left and right hand cuts, and hence a
direct comparison with standard ChPT becomes possible.

\section{ CONCLUSIONS} 
\label{sec:concl}
In this paper we have studied the consequences of chiral symmetry for
$\pi\pi$ scattering and
also for the scalar and vector form factors within an approach based on
the BSE. Besides the automatic incorporation of elastic unitarity, it
becomes clear what is the subset of diagrams which is summed up. This
requires the identification of a {\it potential} which corresponds to
the amputated two particle irreducible Green function.  As such, the
identification is not unique, leaving room for a scale ambiguity, or
equivalently a subtraction constant in the language of dispersion
relations. We have dealt with the issue of renormalizability of the
BSE within the framework of ChPT. We have recognized two alternative
extreme viewpoints to implement the renormalization program which we
have named: off-shell and on-shell schemes.  

In the off-shell scheme, the off-shell dependence of the potential is
kept while solving the BSE. This way of proceeding allows to choose a
renormalization scheme where the finite parts of the ultraviolet power
divergences are not set to zero. This, still, produces more constants
than those allowed by crossing at a given order in the chiral
expansion, and it becomes sensible to fine tune them in a way to
comply as best as we can with the crossing symmetry requirement. In
practice we have found it convenient to implement this as a
restriction on our partial wave amplitudes by matching those to Taylor
expansions of the ChPT ones, including the contribution of the left
hand cut. This off-shell scheme is rather simple at the lowest
approximation level, and at the same time allows for a satisfactory
fit to the phase shift $\pi\pi$ scattering data in the $I=0,1,2$ and
$J=0,1$ channels leading to a prediction of the low energy parameters
$\bar l_{1,2,3,4}$ with varying degree of accuracy~\cite{ej99}.
Conversely, using commonly accepted values for the $\bar l_{1,2,3,4}$
with their error-bars, a satisfactory and compatible prediction of the
scattering data is achieved. In particular, our extrapolation of the
low energy standard ChPT phase shifts in the $I=J=1$ channel to higher
energies predicts the correct $\rho$ meson mass and width with 10\%
and 25\% accuracy respectively. This approach, although very efficient
to describe the data in the different isospin channels, becomes very
cumbersome to pursue at next to leading order. 

Thus, we have also considered the on-shell scheme.  As discussed in
the paper, the fact that we are dealing with an EFT provides
interesting insight into the problem. Given the infinite number of
counter-terms necessary to ensure the renormalizability of the theory
in a broad sense, one can reorganize the summation of diagrams in a
way as to include in the potential the unknown coefficients. As an
important consequence, a dramatic simplification in the solution of
the BSE arises: all the divergences originated by the off-shell
behavior of the potential can be effectively renormalized by
redefining the potential and simultaneously ignoring the off-shellness
in the solution of the BSE.  In particular, on shell information for
the full scattering amplitude can simply be obtained from the on-shell
potential with unknown coefficients. We have found the on-shell scheme
more convenient to take advantage of the ChPT information to one and
two loops than the off-shell scheme. Eventually, both schemes would
become equivalent if the {\it exact} amplitude were considered.

In particular, by using the on-shell scheme we have constructed a
$\pi\pi$ amplitude, which exactly reproduces the tree plus one loop
scattering amplitude of ChPT and embodies exact elastic unitarity to
all orders in the chiral expansion parameter $1/f^2 $. Thus, the
definition of the phase shifts is unambiguous.  In addition, a
prediction for some ChPT two-loop low energy parameters can be also
made. The BSE within the on-shell scheme proves particularly useful
when dealing with the vector and scalar form factors. Watson's theorem
is automatically satisfied without explicitly invoking an Omn\`es
representation.  The requirement of an asymptotic behavior compatible
with a once subtracted dispersion relation yields to a unambiguous
prediction for the form factors, with no more undetermined constants
than those suggested by standard ChPT.  Equipped with all this
formalism we have been able to make a very accurate determination, in
terms of estimated errors, of some low energy parameters.  In
particular, a remarkably accurate prediction has been achieved for the
difference $\bar l_1 - \bar l_2 $ and the parameter ${\bar l}_6$ by
fitting the vector form factor, rather than the $\pi\pi$ scattering
amplitude in the vector-isovector channel. Predictions for the
corresponding $\delta_{11} $ phase shift with the propagated errors
are satisfactory and accurate from threshold up to 1200 MeV. Similar
features are also found for other channels, although there the maximum
energy, for which the predictions with their error-bars apply, ranges
from 600 MeV in the scalar-isoscalar, to 1400 MeV in the
scalar-isotensor channel.  $d-$waves corresponding to the $I=0$ and 2
channels are also describable up to 1000 MeV. In all cases we observe
that the higher the energy the larger the error bars. This is not
surprising since, our unitary amplitudes are generated from low energy
information.

From the point of view of predictive power we have also undertaken a
thorough analysis of the predictions for the ${\cal O}(1/f^6)$
contributions of the amplitudes to the threshold parameters, as
compared to standard ChPT two--loop ones under a resonance saturation
hypothesis~\cite{bc97}. The level of prediction is never less accurate
than in the case of ChPT, and in some cases it is much better. The
total predictions for the effective range parameters are in agreement
with the experimental values within errors.
 
The present calculation can be improved and extended in several ways
which we describe in the following. First of all, the formalism
presented in this paper can be enlarged to include coupled channel
contributions. All of our formulae are valid for a coupled channel
scenario just bearing in mind that now one is dealing with matrices
instead of commuting $\comp-$numbers. This topic has been partially
addressed in Sect.~\ref{sec:pwd}. Secondly, our results, regarding the
accurate description of one loop parameters with the present
unitarization method, advises to implement the next order in our
expansion. In this way we might try to determine, from fits to the
data, more accurately some of the two loop parameters
$b_{1,2,3,4,5,6}$, disposing the physically compelling, but
unmotivated from a standard ChPT viewpoint, resonance saturation
hypothesis. In this regard, the consideration of our formalism for
describing $K_{l_4}$ decays, in conjunction with the electric pion
form factor might prove very fruitful in order to provide more accurate
constraints for the one loop parameters. In neither case, have we
addressed the determination of $\bar l_5$, which is related to the
pion polarizability, and would require a thorough analysis of Compton
scattering on the pion. In this respect, there exist already attempts
in the literature trying to describe the physical process, but never
focused from the point of view of determining this parameter. We leave
these points for future research.

\section*{Acknowledgments}
We would like to acknowledge useful discussions with J.A. Oller, E.
Oset and L.L. Salcedo.  This research was supported by DGES under
contract PB98-1367 and by the Junta de Andaluc\'\i a.

\newpage

\appendix

\section{Details on the off-shell BSE scheme}\label{sec:appea}

\subsection{I=0 $\pi \pi$ scattering}

The ansatz of Eq.~(\ref{eq:i0}) reduces the BSE to the linear algebraic
system of equations

\begin{eqnarray}
A &=& {5m^2-3s \over 2f^2} + {(5m^2-3s) I_0 (s) - 2 I_2 (s)\over 2f^2}
A + {(5m^2-3s) I_2(s) -2 I_4 (s) \over 2f^2}\, B \nonumber\\ 
B &=& -{1\over f^2} - { A I_0 (s) + B I_2 (s) \over f^2}\nonumber \\ 
B &=& -{1\over f^2} + {(5m^2-3s) I_0 (s) - 2 I_2 (s)\over 2f^2} B +
{(5m^2-3s) I_2(s) -2 I_4 (s) \over 2f^2} C \nonumber \\ 
C &=& -{ B I_0(s) + C I_2 (s) \over f^2}\label{eq:sys}
\end{eqnarray}
as we see there are four equations and three unknowns. For the system to
be compatible necessarily one equation ought to be linearly dependent.
This point can be verified by direct solution of the system. The
integrals appearing in the previous system are of the form
\begin{eqnarray}
I_{2n} (s) = {\rm i}
\int \frac{d^4 q}{(2\pi)^4}\frac{(q^2)^n}
{\left[q_-^2-m^2+{\rm i}\epsilon\right]
\left[q_+^2-m^2+{\rm i}\epsilon\right]}\label{eq:i2n}
\end{eqnarray}
$I_0 (s), I_2(s) $ and $I_4 (s) $  are logarithmically, quadratically
and quartically ultraviolet divergent integrals. 
Translational and Lorentz invariance relate the
integrals $I_2(s)$ and $I_4(s)$ with $I_0(s)$ and the divergent
constants $I_2(4m^2)$ and $I_4(4m^2)$ . 
\begin{eqnarray}
I_0 (s) &=&  I_0 (4m^2) + {\bar I}_0 (s)\nonumber\\
I_2 (s) &=& \Big( m^2-s/4 \Big) I_0 (s) + I_2 (4m^2)\nonumber \\
I_4 (s) &=& \Big( m^2-s/4 \Big)^2 I_0 (s) + I_4 (4m^2)\label{eq:integrals}
\end{eqnarray}
Note also that
$I_0(s)$ is only logarithmically divergent and it only requires one
subtraction, i.e., ${\bar I}_0 (s) = I_0(s)-I_0(4m^2)$ is finite ant it
is given by
\begin{eqnarray}
{\bar I}_0 (s) &=& \frac{1}{(4\pi)^2} \sqrt{1-\frac{4m^2}{s}}  \log
\frac{\sqrt{1-\frac{4m^2}{s}}+1 }{\sqrt{1-\frac{4m^2}{s}}-1}
\end{eqnarray}
where the complex phase of the argument of the $\log$ is taken in the
interval $[-\pi,\pi[$. Using the relations of Eq.~(\ref{eq:integrals}) 
in the solution of the linear system of Eq.~(\ref{eq:sys}) we get
\begin{eqnarray}
A(s) &=& {1\over D(s)}
\Big[ { 5m^2 - 3 s \over f^2 } + { I_4 (4m^2)+( m^2-s/4)^2 I_0(s)\over f^4 } \Big] \nonumber\\
B(s) &=& {-1\over D(s)}
\Big[ {1 \over f^2 } + { I_2 (4m^2)+( m^2-s/4)I_0(s) \over f^4 } \Big] \nonumber\\
C(s) &=& {1\over D(s)} {I_0 (s) \over f^4 } \label{eq:i0_1}
\end{eqnarray}
where,
\begin{eqnarray}
D(s) &=& \Big[ 1 + {I_2 (4m^2) \over f^2} \Big]^2 
 +  I_0(s)  \Big( { 2s-m^2 \over 2f^2} 
-            {        (s-4m^2) I_2 (4m^2) +2 I_4 (4m^2) \over
2f^4} \Big)\label{eq:i0_2}
\end{eqnarray}
Eqs.~(\ref{eq:i0_1}) and~(\ref{eq:i0_2}) require renormalization, we
will address this issue in Subsect.~\ref{sec:reno}. 

\subsection{I=1 $\pi \pi$ scattering}

The full off-shell scattering amplitude solution of the BSE in this
channel is given by the ansatz of Eq.~(\ref{eq:i1ans}) with the
functions $M$ and $N$ given by
\begin{eqnarray}
&&\nonumber\\
M(s) &=& {2\over f^2} \left ( 1 - {2 I_2 (4m^2) +
(4m^2-s) I_0 (s) \over 6f^2} \right )^{-1}\nonumber\\
&&\nonumber\\
s N(s) &=& {M(s)\over 6 f^2} { 4 I_2 (4m^2) - (4m^2-s) I_0(s) \over 1
- I_2(4m^2)/ f^2} \label{eq:i1_1}
\end{eqnarray}
To obtain Eq.~(\ref{eq:i1_1}) we have used that
\begin{eqnarray}
I^{\mu\nu} (s) &=& {\rm i}
\int \frac{d^4 q}{(2\pi)^4} q^\mu q^\nu \Delta(q_+)\Delta(q_-) \nonumber\\
&=& {1\over 3} \Big( g^{\mu\nu}-{P^\mu P^\nu \over s} \Big)
\Big[ (m^2-s/4) I_0(s) - I_2(4m^2) \Big] + {1\over 2} g^{\mu\nu} I_2(4m^2)
\end{eqnarray}

\subsection{I=2 $\pi \pi$ scattering}

The functions $A$,$B$ and $C$ entering in the amplitude of
Eq.~(\ref{eq:i2ans}) can be determined by solving the BSE in this
channel, and thus we find
\begin{eqnarray}
A(s) &=& {1\over D(s)}
\Big[ { m^2 \over f^2 } + { I_4 (4m^2) + (m^2-s/4)^2 I_0(s) 
\over f^4 } \Big] \nonumber\\
B(s) &=& {-1\over D(s)}
\Big[ {1 \over f^2 } + { I_2 (4m^2)+(m^2-s/4)I_0(s) 
\over f^4 } \Big] \nonumber\\
C(s) &=& {1\over D(s)} {I_0 (s) \over f^4 }\label{eq:i2_1}
\end{eqnarray}
where
\begin{eqnarray}
D(s) &=& \Big[ 1 + {I_2 (4m^2) \over f^2 } \Big]^2 +
   I_0(s) \Big[
{2 m^2-s\over 2 f^2 }
-  {         ( s -2m^2 ) I_2 (4m^2) +2 I_4 (4m^2) \over
2 f^4} \Big]
\label{eq:i2_2}
\end{eqnarray}
Once again Eqs.~(\ref{eq:i2_1}) and~(\ref{eq:i2_2}) 
require renormalization, we 
will get back to this point in Subsect.~\ref{sec:reno}. 

\subsection{On-shell and off-shell unitarity}

Off-shell unitarity (Eq.~(\ref{eq:off-uni})) imposes a series of
conditions to be satisfied by the amplitudes obtained in the previous
subsections. Those read for the $I=1$ case
\begin{eqnarray}
{\rm Disc}\, [N(s)] &=& { (s-4m^2) \over 12 } | N(s) |^2 {\rm
Disc}\,[I_0 (s)] \nonumber\\ 
{\rm Disc}\, [M(s)] &=& -{{\rm Disc}\,
[N(s)]\over s} \label{eq:dis_1}
\end{eqnarray}
and for the $I=0,2$ cases, they are
\begin{eqnarray}
{\rm Disc}\, [A(s)] &=& - \Big| A(s)+ B(s) ( m^2 - s/4) \Big|^2
 {\rm Disc}\,[I_0 (s)] \nonumber\\
{\rm Disc}\, [C(s)] &=& - \Big| B(s)+ C(s) ( m^2 - s/4) \Big|^2
 {\rm Disc}\,[I_0 (s)] \nonumber\\
{\rm Disc}\, [B(s)] &=& \left ( (m^2-s/4)B(s) + A(s)\right )^* 
\left ( (m^2-s/4)C(s) + B(s)\right ){\rm Disc} \,[I_0 (s)] \label{eq:dis_2}
\end{eqnarray}
where ${\rm Disc}\,[f(s)] \equiv f(s+i\epsilon)-f(s-i\epsilon),\,s>
4m^2 $.

On the other hand and thanks to the Schwartz's Reflex ion Principle,
${\rm Disc} \,[I_0 (s)]$ is given by
\begin{eqnarray}
{\rm Disc} \,[I_0 (s)] &\equiv& I_0(s+i\epsilon) - I_0(s-i\epsilon)= 
2i {\rm Im} {\bar I}_0(s+i\epsilon)\nonumber\\ 
&=& -{\rm i} (2\pi)^2\int\frac{d^4 q}{(2\pi)^4} 
\delta^+ \left( q_+^2-m^2 \right)
                        \delta^+ \left( q_-^2-m^2 \right) = 
-{{\rm i}\over 8\pi }  \sqrt{1-{4m^2 \over s}},\, s>4m^2
\end{eqnarray}

\subsection{Renormalization method, crossing symmetry and 
lagrangian counter-terms}\label{sec:appea1}

The method of subtraction integrals makes any amplitude finite, by
definition. In customary renormalizable theories or EFT's, one can
prove in perturbation theory (where crossing is preserved order by
order) that this method has a counter-term interpretation. This is
traditionally considered a test for a local theory, from which
microcausality follows. Since we are violating crossing it is not
clear whether or not our renormalization method admits a Lagrangian
interpretation beyond the actual level of approximation. These are in
fact a sort of integrability conditions; the renormalized amplitudes
should be indeed functional derivatives of the renormalized
Lagrangian. It is actually very simple to see that these conditions are
violated by our solution. In terms of the generating functional $Z[J]$
the four point renormalized Green's function is defined as
\begin{eqnarray}
& & \langle 0 | T\{ \pi_a (x_1) \pi_b (x_2) \pi_c (x_3) \pi_d (x_4)\} | 0
\rangle =
{1\over Z [ J ]} { \delta^4 Z [ J] \over
\delta J_a (x_1)
\delta J_b (x_2)
\delta J_c (x_3)
\delta J_d (x_4) } \Big |_{J=0} = \nonumber\\
&& \int
{{\rm d}^4 k \over (2\pi)^4 }
{{\rm d}^4 p \over (2\pi)^4 }
{{\rm d}^4 P \over (2\pi)^4 }
{{\rm d}^4 P' \over (2\pi)^4 }
e^{{\rm i} p (x_1-x_2)}
e^{{\rm i} k (x_4-x_3)}
e^{{\rm i} P (x_1+x_2)/2}
e^{-{\rm i} P' (x_3+x_4)} 
 \Delta_{aa'} (p_+) \nonumber\\ &\times &\Delta_{bb'} (p_-)
 (-{\rm i})T_P (p,k)_{a'b';c'd'} \delta^4 (P-P')
 \Delta_{cc'} (k_+) \Delta_{dd'} (k_-)
\end{eqnarray}
The integrability conditions are simply the equivalence between
crossed derivatives which, as one clearly sees, implies in particular 
the crossing condition for the scattering amplitude. Since crossing is 
violated, the integrability conditions are not fulfilled and hence our 
renormalized amplitude does not derive from a renormalized Lagrangian.

Crossing symmetry can be restored, in certain approximation, as it is
discussed in Subsect.~\ref{sec:csr}, by imposing suitable constraints 
between the subtraction constants  which appear in the renormalization
scheme described in Subsect.~\ref{sec:reno}. These conditions reads:
\begin{eqnarray}
75 I^{R,I=0}_2/2m^2 + 8I^{R,I=1}_0+
33I^{R,I=0}_0+5I^{R,I=1}_2/m^2+\frac{10157}{1920\pi^2} &=& 0 \nonumber\\
&&\nonumber\\
I^{R,I=2}_0 - \frac{4I^{R,I=1}_0}{15}-\frac{8I^{R,I=0}_0}{5}- 
\frac{719}{7200\pi^2} &=& 0 \nonumber\\
\frac{I^{R,I=2}_2}{4m^2}+ \frac{4\left (I^{R,I=1}_0 + I^{R,I=0}_0
\right )}{25} + \frac{I^{R,I=1}_2}{12m^2} + 
\frac{887}{36000\pi^2} &=& 0 \nonumber\\
&&\nonumber\\
\frac{I^{R,I=2}_4}{16m^4}+ \frac{17 I^{R,I=1}_0}{500} + 
\frac{111 I^{R,I=0}_0}{4000} + \frac{3I^{R,I=1}_2}{200m^2}-  
\frac{I^{R,I=0}_4}{40m^4} + 
\frac{2159}{480000\pi^2} &=& 0
\label{eq:cons}
\end{eqnarray}
The first of these relations was obtained in Ref.~\cite{ej99}. 

Finally and for the sake of
clarity we quote here the result of inverting the 
Eq.~(18) of Ref.~\cite{ej99},
\begin{eqnarray}
I^{R,I=1}_0 & = & - \frac{1}{16\pi^2} \left (2({\bar l}_2-{\bar
l}_1)+97/60 \right ) \nonumber \\ 
I^{R,I=1}_2 & = & \frac{m^2}{8\pi^2}
\left ((2({\bar l}_2-{\bar l}_1)+3{\bar l}_4-65/24\right )\nonumber \\
I^{R,I=0}_0 & = & -\frac{1}{576\pi^2}\left (22 {\bar l}_1+28 {\bar
l}_2+31/2\right )\nonumber 
\\ I^{R,I=0}_4 & = &
\frac{m^4}{7680\pi^2}\left (-172 {\bar l}_1-568 {\bar l}_2+600 {\bar l}_3 -672
{\bar l}_4 + 1057\right)
\label{eq:li}
\end{eqnarray}

\subsection{ Remarks on the choice of the pion field} \label{sec:parapi}

The solution of the BSE requires an ansatz for the off-shell potential
$V_P(p,k)$. The non-linear realization of chiral symmetry through the
non-linear $\sigma$-model posses an additional difficulty. The pion
field, $\vec \phi$, is encoded in the unitary space-time matrix
$U(x)$, $U^\dagger U =1$, so it lives in the three sphere of radius
$f$. However, the particular set of coordinates is not unique. For
instance, one may have a {\it polar} representation $U(x)=e^{{\rm i\,}\vec
\phi \cdot \vec \tau/f }$, being $\vec{\tau}$ the Pauli matrices , a
{\it cartesian} representation $U(x)= \sqrt{1-\vec\phi~^2/f^2 } + {\rm
i}\vec{\phi} \cdot \vec{\tau} /f $, or a {\it stereographic} representation
$U(x) = ( 1 + {\rm i}\vec \phi \cdot \vec \tau/2f ) / ( 1 - {\rm
i}\vec \phi \cdot \vec \tau/2f ) $. It is well known that any
representation yields to the same kinetic and mass terms, but
different pion interaction terms. Indeed, up to second order in the
chiral expansion, the effective lagrangian reads
\begin{eqnarray}
{\cal L} = {f^2 \over 4} {\rm tr} \Big( \partial^\mu U^\dagger
\partial_\mu U \Big)  + {f^2\, m^2 \over 4} {\rm tr} \Big( U + U^\dagger
-2 \Big)
\end{eqnarray}
which up to fourth order in the pion field becomes 
\begin{eqnarray}
{\cal L} = \frac12 (\partial_\mu \vec \phi)^2  - 
\frac{m^2}{2} \vec \phi~^2 - \frac{\alpha}{4f^2} (\partial_\mu \vec \phi)^2
\vec \phi~^2 + \frac{1-\alpha}{2f^2} (\vec \phi\,\partial_\mu \vec
\phi)^2 + (\alpha -\frac12)\frac{m^2}{4f^2} \vec \phi~^4   
\end{eqnarray}
where the unfixed $\alpha$ parameter has its origin in the
arbitrariness on the form of the $U-$matrix. Thus, the polar and
cartesian representations correspond to $\alpha = 0$ whereas the
stereographic representation leads to $\alpha = 1$. The $\alpha$
dependence disappears if one uses the equation of motion.  The above
Lagrangian gives
\begin{equation}
A_P(p,k) = -\frac{1}{f^2} \left \{ s\left(1-\frac{\alpha}{2} \right ) 
-\alpha \left ( p^2+k^2   \right ) + m^2 \left ( 2\alpha -1 \right ) \right \}
\end{equation}
which by means of the symmetry properties discussed in
Sect.~\ref{sec:bse} completely defines the two identical isovector
(off-- or on--shell) meson scattering amplitude ($T_P (p,k)_{ab;cd}$)
for the process $ (P/2+p,a)+(P/2-p,b) \to (P/2+k,c)+(P/2-k,d) $,
defined in Eq.~(\ref{eq:defTabcd}). On the mass-shell the dependence
on $\alpha$ of the amplitude disappears. Off the mass shell an
explicit $\alpha$ dependence is exhibited.
Thus, the BSE kernel $V_P(p,k)$ will depend on $\alpha$ and might generate
also an $\alpha-$dependence in the solutions of the BSE. For on-shell
pion scattering and when the exact potential $V$
and propagator $\Delta$ are used the $\alpha$ dependence drops
out. However, when  approximated $V$
and $\Delta$ are inserted in the BSE, the solution of the equation
might display an $\alpha$ dependence, even for on--shell pions. 
This is the case for  the lowest order $I=0$ and $I=2$ on--shell solutions, 
\begin{eqnarray}
T^{-1}_{00}(s) &=& -I_0(s) +
\frac{2\left(f^2+(1-5\alpha/2)I_2(4m^2)\right)^2}{(m^2-2s)f^2
+ (1-5\alpha/2)^2\left(2I_4(4m^2)+(s-4m^2)I_2(4m^2)\right ) } \label{eq:al0}\\
&&\nonumber\\
T^{-1}_{20}(s) &=& -I_0(s) +
\frac{2\left(f^2+(1-\alpha)I_2(4m^2)\right)^2}{(s-2m^2)f^2
+ (1-\alpha)^2\left(2I_4(4m^2)+(s-4m^2)I_2(4m^2)\right ) } \label{eq:al2}
\end{eqnarray}
where both amplitudes above require renormalization and reduce to
those given in Eqs.~(\ref{eq:i_0}) and Eqs.~(\ref{eq:i_2}) for $\alpha
= 0.$ The lowest order $I=1$ BSE solution turns out to be independent
of $\alpha$. 

It is worth noticing that the energy dependence encoded
in the amplitudes above is not changed by this $\alpha$ dependence,
and they can generically be written as
\begin{eqnarray}
T^{-1}_{00\,(20)}(s) &=& -{\bar I}_0(s) -c + \frac{1}{a+b s}
\end{eqnarray} 
with $a$, $b$ and $c$ suitable constants. So there is redundancy of
parameters, since we have four parameters $I_0 (4m^2) $, $I_2 (4m^2)
$, $I_4 (4m^2) $ and $\alpha$, mapped into three parameters $a$,$b$
and $c$.  Obviously, the particular value of $\alpha$ would be
irrelevant, when a fit of these constants to data is performed, as it
was done in Ref.~\cite{ej99}. The trouble might appear, as we will
see, when matching these amplitudes to those deduced in ChPT, which of
course do not present any dependence on $\alpha$ for pions on the mass
shell.

The $\alpha$ dependence shown in Eqs.~(\ref{eq:al0}) and~(\ref{eq:al2}) is
undesirable because we are dealing to on--shell amplitudes. It appears
because we have not generated loops in the $t-$ and $u-$channels,
i.e. the amplitudes do not satisfy exact crossing. On the light of
this discussion, we envisage two alternative procedures linked to
different renormalization schemes:
\begin{itemize}
\item Within the renormalization scheme presented in
Sect.~\ref{sec:reno}, we can match the $1/f^2$ expansion of the BSE
amplitude up to ${\cal O}(1/f^4)$, to the threshold expansion, up to
order $(s-4m^2)^2$, of the Gasser-Leutwyler amplitudes, as it has been
done in Sect.~\ref{sec:appea1}. This procedure modifies the last of
Eqs.~(\ref{eq:li}) and all of the constraints of Eq.~(\ref{eq:cons}),
and ensures the $\alpha$ independence of the matched terms, but there
remains a residual $\alpha$ dependence in the renormalized on--shell
BSE amplitude, starting at order ${\cal O}(1/f^6)$. Actually, this
matching procedure does not work if $\alpha=2 $ or $\alpha=-1$,
because in this cases the BSE coefficients of $(s-4m^2)/f^4$ and
$(s-4m^2)^2/f^4$ pieces are not independent for $I=0$ and $I=2$
respectively, making then the matching to the Gasser-Leutwyler
amplitudes overdetermined\footnote{That would require a restriction
among the ${\bar l}$'s.}. In addition, we do not have any restriction
on the possible $\alpha$ values. This is way we find more appropriate
the following scheme.

\item From the renormalization discussion presented in
Sect.~\ref{sec:reno} it is clear, as it is also the case in standard
ChPT, that to achieve renormalization of the amplitudes an infinite
number of counter-terms is required. Being then the finite parts, order
by order in the perturbative expansion, undetermined and have to be
either fitted to data or determined from the underlying QCD
dynamics. This implies in particular that the divergent integrals
appearing in the numerator and denominator in 
Eqs.~(\ref{eq:al0}) and~(\ref{eq:al2}) may be chosen to be independent
of each other. This can be  used advantageously to eliminate the
explicit $\alpha$ dependence, by simply redefining the integrals, $I_2
(4m^2 )(1-5\alpha/2) \to I_2(4m^2) $, $I_2 (4m^2)(1-\alpha) \to
I_2(4m^2) $ in the numerators of Eqs.~(\ref{eq:al0}) and~(\ref{eq:al2}),
and $I_{2,4} (4m^2) (1-5\alpha/2)^2 \to I_{2,4} (4m^2)  $, $I_2
(4m^2)(1-\alpha)^2  \to
I_2(4m^2)  $ in the denominators of Eqs.~(\ref{eq:al0})
and~(\ref{eq:al2}), respectively, since we know that the total amplitude
must be $\alpha$ independent. 
Even though this choice might appear more arbitrary than the scheme
presented in Sect.~\ref{sec:reno}, it is still more restrictive than
what the renormalization of an EFT allows for.
\end{itemize}

From the discussion above, we prefer to work within the second scenario where 
the unphysical $\alpha-$dependence can be ignored, when looking at
on-shell amplitudes, and the particular value  $\alpha=0$
can be considered, as we have done along Sect.~\ref{sec:off}. Within
this renormalization scheme any value of $\alpha$ would lead to the
same results presented in that section.

\section{Leading and next-to-leading elastic $\pi\pi$ scattering
amplitudes in ChPT} \label{sec:appeb}

The   $ {\cal O}(p^2)+{\cal O}(p^4)$ on-shell $SU(2)-$ ChPT elastic 
$\pi \pi$ amplitudes, expressed in terms of the
renormalization  invariant parameters ${\bar l}_{1,2,3,4}$,
are given\footnote{Note that the function $ {\bar I}_0 (s) $, 
fulfilling ${\bar I}_0 (4m^2)=0$ is
related to that of \cite{GL84} fulfilling $ {\bar J}(0)=0$
by
\begin{eqnarray}
{\bar I}_0 (s)=-{\bar J}(s)+{1\over 8\pi^2}\nonumber
\end{eqnarray}}
 by~\cite{GL84}
\begin{eqnarray}
T_{IJ}(s)&=&  T_{IJ}^{(2)}(s) /f^2  +  T_{IJ}^{(4)}(s) /f^4 
\\
&&\nonumber\\
T_{IJ}^{(4)}(s)&=&  \tau_{IJ}^{(4)}(s)  +  {\bar I}_0
(s)\times [\,T_{IJ}^{(2)}(s)\,]^2
\label{eq:tau}\\
&&\nonumber\\
T_{IJ}^{(2)}(s)&=&\left\{\begin{array}{ll}\frac{m^2-2s}{2}&I=0\,;\,J=0 \\
&\\
0 &I=0\,;\,J=2 \\
&\\
\frac{4m^2-s}{6}&I=1\,;\,J=1 \\
&\\
\frac{s-2m^2}{2}&I=2\,;\,J=0\\
&\\
0 & I=2\,;\,J=2 \end{array}\right. \\
&&\nonumber\\
\tau_{IJ}^{(4)}(s)&=&-\frac{1}{192\pi^2}g_{IJ}+
\frac{1}{12}h_{IJ} \label{eq:gl}\\
&&\nonumber\\
h_{IJ}&=& \frac12\int_{-1}^1 d(\cos\theta )\left (
f_{I}(t){\bar I}_0 (t) + f_{I}(u){\bar I}_0 (u)\right )
P_J(\cos\theta )\nonumber\\
&&\nonumber\\
&=& \left\{\begin{array}{ll} \frac{5m^4}{4\pi^2}+ \frac{101 m^2
(s-4m^2)}{96\pi^2} + \frac{191 (s-4m^2)^2}{288\pi^2}+ \cdots
&I=0\,;\,J=0 \\ &\\ 
{287 (s-4m^2)^2 \over 2880 \pi^2 } + {481 (s-4m^2)^3 \over 67200 m^2 \pi^2
} \cdots &I=0\,;\,J=2 \\ &\\ 
\frac{89 m^2 (s-4m^2)}{288\pi^2} - \frac{37 (s-4m^2)^2}{2880\pi^2}+ \cdots
&I=1\,;\,J=1\\ &\\
\frac{11m^4}{4\pi^2}+ \frac{179 m^2
(s-4m^2)}{96\pi^2} + \frac{769 (s-4m^2)^2}{1440\pi^2}+ \cdots
&I=2\,;\,J=0 \\ &\\
{529 (s-4m^2)^2 \over 14400 \pi^2  }+ {277 (s-4m^2)^3 \over 67200 m^2 \pi^2
} + \cdots &I=2\,;\,J=2 
\\ &\\\end{array}\right. \label{eq:hij_exp}
\end{eqnarray}
where $t= -2(\frac{s}{4}-m^2)(1-\cos\theta )$, 
$u= -2(\frac{s}{4}-m^2)(1+\cos\theta
)$ and
\begin{eqnarray}
f_{I}(x)&=&\left\{\begin{array}{ll} 10x^2 + x(2s-32m^2)+37m^4-8sm^2
&I=0 \\
2x^2 + x(s+2m^2)-m^4-4sm^2 &I=1\\ 
4x^2 - x(s+2m^2)-5m^4+4sm^2
&I=2\end{array}\right. \\
& & \nonumber\\
& & \nonumber\\
g_{IJ}&=&\left\{\begin{array}{ll} m^4\left(40{\bar l}_1+ 80{\bar l}_2-
15{\bar l}_3+84{\bar l}_4+125\right) + m^2(s-4m^2)\times &\\
\left(32{\bar l}_1+ 48{\bar l}_2+24{\bar l}_4+\frac{232}{3}\right)+
(s-4m^2)^2\left(\frac{22}{3}{\bar l}_1+ \frac{28}{3}{\bar l}_2+
\frac{142}{9}\right)&I=0\,;\,J=0 \\ & \\
\frac{2(s-4m^2)^2}{15}\left({\bar l}_1+4{\bar l}_2+\frac{16}{3}
\right)&I=0\,;\,J=2 \\ & \\
\frac{s-4m^2}{3}\left [ 4m^2 \left(-2{\bar l}_1+2{\bar l}_2
+3{\bar l}_4+1 \right)+ (s-4m^2)\left(-2{\bar l}_1+2{\bar l}_2+1
\right)\right]&I=1\,;\,J=1 \\ & \\
m^4\left(16{\bar l}_1+ 32{\bar l}_2-
6{\bar l}_3-24{\bar l}_4+50\right) + m^2(s-4m^2)\times &\\
\left(8{\bar l}_1+ 24{\bar l}_2-12{\bar l}_4+\frac{100}{3}\right)+
(s-4m^2)^2\left(\frac{4}{3}{\bar l}_1+ \frac{16}{3}{\bar l}_2+
\frac{64}{9}\right)&I=2\,;\,J=0 \\ & \\
\frac{2(s-4m^2)^2}{15}\left({\bar l}_1+{\bar l}_2+\frac{11}{6}
\right)&I=2\,;\,J=2 \\&\\\end{array}\right.
\end{eqnarray}

The integrals which appear in the definition of $h_{IJ}$ in
Eq.~(\ref{eq:gl}) can be done analytically by means of the change of
variables: 
\begin{eqnarray}
\cos\theta \to x = 
\frac{\sqrt{1-\frac{4m^2}{\xi(\cos\theta)}}+1
}{\sqrt{1-\frac{4m^2}{\xi(\cos\theta)}}-1}
\end{eqnarray}
being $\xi(\cos\theta)= t$ or $u$. Thus we get,
\begin{equation}
h_{IJ}(s)= a_{IJ}(s) + b_{IJ}(s) L(s) + c_{IJ} (s) L(s)^2
\end{equation}
where
\begin{equation}
L(s)= {\rm log} \Big( { 1+\sqrt{1-{4m^2\over s}} \over
                        1-\sqrt{1-{4m^2\over s}} } \Big)
\end{equation}
and $s>4m^2 $. In the following we list some special cases of
interest,
\begin{eqnarray}
a_{00} &=&  { -506m^4+130m^2 s - 11 s^2 \over 144\pi^2 } \nonumber\\
b_{00} &=&  {75m^4 - 40m^2 s + 7s^2\over 24\pi^2  }
\sqrt{s\over  s-4m^2}  \nonumber\\
c_{00} &=&  { m^4 ( 25m^2-6s ) \over 8\pi^2  (s-4m^4)}  \\
&& \nonumber\\
a_{11} &=&  {120m^6-149m^4s+37m^2s^2+s^3 \over 144\pi^2  (s-4m^2) }
\nonumber\\ 
b_{11} &=&  {36m^6 -72m^4 s + 16m^2s^2-s^3 \over 48\pi^2  (s-4m^2) }
\sqrt{s\over  s-4m^2}  \nonumber\\
c_{11} &=& { m^4 (6m^4+13m^2 s - 3s^2 ) \over 8\pi^2  (s-4m^2)^2 } \\
&& \nonumber\\
a_{20} &=& { 308m^4-58m^2 s -25s^2 \over 288\pi^2  }  \nonumber\\
b_{20} &=& { 6m^4 -32m^2 s + 11 s^2 \over 48\pi^2  }
\sqrt{s\over  s-4m^2}  \nonumber\\
c_{20} &=& { m^4 ( m^2 + 3 s) \over 8\pi^2  (s-4m^2) } \\
&& \nonumber\\
a_{02} &=& {-31936 m^8+27646 m^6 s-4371 m^4 s^2+61 m^2  s^3  + 14 s^4
\over 720\pi^2  (s-4m^2)^2 }  \nonumber\\
b_{02} &=& {  516m^8 -760m^6 s + 180 m^4  s^2  - 18 m^2  s^3  + s^4
\over 24 \pi^2   (s-4m^2)^2 } \sqrt{s\over  s-4m^2}  \nonumber\\
c_{02} &=& { m^4 (172 m^6- 98 m^4 s + 49 m^2  s^2  - 6 s^3 ) \over
8\pi^2  (s-4m^2)^3 } \\
& &  \nonumber\\
a_{22} &=& {124672 m^8-48832m^6 s+53592m^4 s^2 -5182 m^2 s^3+ 247 s^4
\over   14400 \pi^2   (s-4m^2)^2 }  \nonumber\\
b_{22} &=& { -480 m^8- 980 m^6 s + 117 m^4 s^2 -36 m^2 s^3+ 2 s^4 \over
120 \pi^2  (s-4m^2)^2 }
\sqrt{s\over  s-4m^2}  \nonumber\\
c_{22} &=&  -{m^4 (32 m^6 -76 m^4 s + 11 m^2 s^2 - 3 s^3 ) \over
8\pi^2  (s-4m^2)^3 }
\end{eqnarray}

\end{document}